\documentclass[aps,prb,showpacs,superscriptaddress,twocolumn]{revtex4-1}
\usepackage{amsfonts}

\usepackage{amssymb}
\usepackage{amsmath}
\usepackage{graphicx}
\usepackage{hyperref}
\usepackage{siunitx}
\usepackage[dvipsnames]{xcolor}
\usepackage{nicematrix}
\usepackage[normalem]{ulem}
\usepackage[version=4]{mhchem}

\hypersetup{colorlinks=true,
            linkcolor=blue,
            citecolor=blue,
            urlcolor=orange}

\newcommand{\I}{{\rm i}}

\begin{document}

\title{Antiferromagnetic and spin {spiral} correlations in the doped 
\\two-dimensional Hubbard model:
gauge symmetry, Ward identities, \\and dynamical mean-field theory analysis}
\author{I. A. Goremykin}
\affiliation{Center for Photonics and 2D Materials, Moscow Institute of Physics and Technology, Institutsky lane 9, Dolgoprudny
141700, Moscow region, Russia}
\author{A. A. Katanin}
\affiliation{Center for Photonics and 2D Materials, Moscow Institute of Physics and Technology, Institutsky lane 9, Dolgoprudny
141700, Moscow region, Russia}
\affiliation{M. N. Mikheev Institute of Metal Physics, Kovalevskaya Street 18,
Ekaterinburg {620219}, Russia.}

\begin{abstract}
We reconsider derivation of Ward identities for spin stiffnesses, which
determine the non-linear sigma model of magnetic degrees of freedom of
interacting electrons in the presence of antiferromagnetic or incommensurate
correlations. We emphasize that in the approaches, which do not break
explicitly spin symmetry of the action, 
the spatial components of gauge kernel, which is used to obtain spin stiffnesses, remain gauge
invariant even in case of spontaneous spin symmetry breaking. We derive the corrected 
Ward identities, 
which account for this gauge invariance. 
We find that 
the frequency dependence of temporal spin stiffnesses is not
fixed by the obtained identities, and show that the infinitesimally small
external staggered field is crucially important to obtain finite static uniform
transverse susceptibility. On the other hand, the spatial spin stiffnesses are determined by the gauge kernel of the Legendre transformed theory, which is in general {\it different} from the gauge kernel of the original theory 
and obtain an explicit expressions for spatial spin stiffnesses  
through susceptibilities and current correlation functions. We verify numerically
the obtained results within dynamic mean field theory, and obtain
doping dependencies of the resulting spin stiffnesses 
for antiferromagnetic and incommensurate phase.
%within the
%dynamical mean field theory 
%(e.g. those not performing
%Hubbard-Stratonovich transformation with the subsequent mean-field
%approximation), 
%tions to previously proposed
%[P. M. Bonetti, Phys. Rev. B \textbf{106},155105 (2022)] 
%.  the contradiction occurs because previously derived Ward identities relied on incorrect assumptions regarding performed Legendre transformations
%derive the corrected form of Ward identities, and show that they result
%in a different (matrix) form of the frequency dependent parts of gauge
%kernel, with 
%The obtained correction terms compensate contribution of Golds
%. We find that previously derived results for the  compensation of various correction
%terms, and 
%reproduce
%previous results, 
%with obtaining
%which 
%is however due to 
%on the basis of the obtained
%expressed 
\end{abstract}

\maketitle

%correlations as a driving force \\of the pseudogap phase of the 2D Hubbard model}

\affiliation{Center for Photonics and 2D Materials, Moscow Institute of Physics and Technology, Institutsky lane 9, Dolgoprudny
141700, Moscow region, Russia}

\affiliation{Center for Photonics and 2D Materials, Moscow Institute of Physics and Technology, Institutsky lane 9, Dolgoprudny
141700, Moscow region, Russia} 
\affiliation{M. N. Mikheev Institute of Metal Physics, Kovalevskaya Street 18,
Ekaterinburg {620219}, Russia.}

\section{Introduction}

%Cuprates
The properties of antiferromagnetic and incommensurate magnetic correlations
near half filling remain an actively discussed topic, in particular in
connection with the physics of high-$T_c$ compounds. In these compounds the
commensurate long-range magnetic order is destroyed already at small doping
(see, e.g., Refs. \onlinecite{Plakida,StaticOrd}), but the short-range magnetic
order 
%dominating at the wave vector ${\bf Q}=(Q,Q)$ at low doping and ${\bf Q}=(Q,\pi)$ at larger doping, 
is present at higher doping. The observed short-range magnetic correlations
are incommensurate and characterized by the wave vector $\mathbf{Q}%
=(\pi,\pi-\delta)$ \cite{ExpQ1,ExpQ2,ExpQ3,StaticOrd,Plakida}. These
short-range magnetic correlations are considered {to be} one of the viable
scenarios {for} pseudogap formation \cite%
{PseudogapSF,PseudogapSad,PseudogapOnufr,PseudogapToschi,PseudogapMag1,PaseudogapMag2,PseudogapMag3,Sachdev1,Sachdev2,BonettiMetzner}%
. 
%The incommensurate magnetic order appears in two dimensions as a result of geometric frustration and/or doping.  

%In particular, these correlations are considered as one of the viable scenarios of pseudogap formation in cuprates. 

The low energy spin excitations of magnetic systems can be described by 
%widely used model, which describes low-energy magnetic degrees of freedom,
%is 
the nonlinear sigma model (NLSM), see, e.g., Ref. \onlinecite%
{Auerbach}. The key parameters of this model are the temporal and spatial spin stiffnesses, which are determined from the microscopic analysis. The classical version of the $O(3)$ non-linear sigma model was
derived in the continuum limit of the commensurate antiferromagnetic ordered classical
Heisenberg model \cite{ZinnJustin,NelsonPelkovits} and later generalized to
the quantum case \cite{Haldane,Chakravarty}. The  nonlinear sigma
model in the $O(3)/O(2)$ manifold was considered as a continuum limit of frustrated quantum
antiferromagnets with spin spiral ground state \cite{DombreRead,Mouhanna}.

For itinerant (collinear) antiferromagnets the derivation of the NLSM, which also allows obtaining the respective spin stiffnesses, was
proposed in Refs. \onlinecite{Weng,Schulz,Dupuis,Dupuis1,Sachdev,Sachdev1}. This derivation uses
Hubbard-Stratonovich transformation, and therefore introduces in the action the effective
fluctuating field, corresponding to the order parameter, which is then fixed at its mean field value.
This approach explicitly breaks $SU(2)$ gauge invariance of the theory,
although the $O(3)$ invariance of the resulting bosonic action remains
unbroken. Breaking of the $SU(2)$ gauge symmetry can be represented as a
condensation of the Higgs field \cite{Sachdev,Sachdev1}. The derivation of
the corresponding bosonic action from the microscopic models (e. g., Hubbard
model), relies within this strategy on the average over the fermionic fields
in the presence of the gauge field, which can be performed only in some
approximate way. This approach is therefore difficult to generalize beyond
the mean field approach for fermionic (also referred as chargon) degrees of freedom.

Recently, using the SU(2)-symmetric gauge theory, which does not introduce symmetry
breaking terms or condensation of gauge fields in the action, was proposed for derivation of the nonlinear sigma model and obtaining respective spin stiffnesses
\cite{Bonetti,BonettiThesis,BonettiMetzner}. In this approach the symmetry is broken
only at the level of dressed single- and two-particle Green's functions via
spin-asymmetric self-energy and the resulting renormalized interaction
vertices. This approach is more convenient for using in combination with the
many-body techniques, since it does not introduce mean fields in the action. 
In view of that, it is also expected to preserve gauge invariance. 
This invariance
%. The
%unbroken gauge invariance 
provides, however, the difficulty, since the respective spin stiffnesses, obtained by expansion of the action in gauge fields are expected to vanish in the gauge-invariant approach. At the same time, finite spin stiffnesses were obtained in Refs. \onlinecite{Bonetti,BonettiThesis,BonettiMetzner}, and
in Refs. \onlinecite{Bonetti,BonettiThesis} the Ward identities, which relate the second
derivatives of the gauge kernel to the spin stiffnesses were proposed for
the same SU(2) gauge symmetric action. The applicability of previously obtained results, and in particular respective Ward
identities requires therefore further investigation in view of the above discussed argument of
gauge invariance of the considered theory, and the respective vanishing of
second derivatives of the gauge kernel.

In the present paper we show explicitly that in the absence of the external magnetic fields the gauge invariance of the spatial part
of the gauge kernel is preserved in the approach of Refs. \onlinecite%
{Bonetti,BonettiMetzner,BonettiThesis}. 
%which therefore can not be used to derive non-linear sigma model by expansion in the gauge fields, since the corresponding derivatives identically vanish. The gauge invariance becomes however broken by Legendre transformation, which was previously used to derive Ward identities.
We emphasize therefore that the corresponding gauge kernel in general can not  be used
for determination of spin stiffnesses, since the corresponding contributions
to the stiffnesses of the non-linear sigma model, which are proportional to
the derivatives of the gauge kernel, vanish. 
Instead, we show that spatial spin stiffnesses are determined by the
second derivatives of the effective Legendre transformed action over gauge field,
which are in general \textit{not} identical to the gauge kernel of original theory, and derive
the corresponding corrections to the Ward identities,
%of Ref. \onlinecite{Bonetti},
%originating from the corrected treatment of Legendre transformation,
which necessarily account for the dependence of the source terms on the
gauge field at fixed order parameter. 
On the other hand, temporal components of the spin stiffnesses are determined by the Ward identities
for the original functional $W$ before the Legendre transformation, which yield, however, the off diagonal
components of the susceptibilities and take as an input the uniform
susceptibilities in small staggered magnetic field. 

We furthermore show that the obtained correction terms in Ward identities compensate contributions of Goldstone modes in spatial spin stiffnesses. 
%, which
%was omitted in Refs. \onlinecite{BonettiMetzner,Bonetti,BonettiThesis}. 
%This implies that the gauge
%fixing conditions were implicitly introduced in Refs. \onlinecite%
%{Bonetti,BonettiMetzner,BonettiThesis} by omitting the
%contribution of Goldstone modes on the basis of vanishing of the
%corresponding vertices in the long-wave limit. 
Although the argument  
of vanishing 
of the contribution of Goldstone modes on the basis of vanishing of the
corresponding vertices in the long-wave limit  was proposed in Refs.  \onlinecite{Bonetti,BonettiMetzner,BonettiThesis},
it is in general
not applicable, since
at the same time the corresponding contributions are potentially singular 
%in the long wave limit 
due to gapless Goldstone excitations. This singularity can be avoided by introducing small external staggered magnetic field, which introduces a gap in the Goldstone excitations, which is however inconvenient for practical calculations beyond mean field approach. The compensation of the contribution of Goldstone excitations by correction terms, which originate from the proper treatment of Legendre transform, is irrespective of the presence of the external magnetic fields and on one hand shows the universality of the resulting spin stiffnesses, but on the other hand allows performing calculations in zero external magnetic field. 
The obtained corrections also correspond to a certain gauge fixing in the (properly treated) Legendre transform. 
%These contributions compensate contributions of the Goldstone modes, which
%was omitted in Ref. \onlinecite{Bonetti}. 
%, which is in general finite.%
%This difficulty occurs because of the absence of gauge-fixing contributions in the action of Ref. \cite{BonettiMetzner}. 
%The latter dependence occurs because of fixing the arguments $\phi$ of the Legendre functional $\Gamma[\phi]$ as independent variables, which yields additional important contributions. 
%Yet, the obtained correct terms mostly compensate the contribution of Goldstone modes. The compensation is however not complete, and yields additional contributions to the spatial components of spin stiffness in the incommensurate phase. 
%We also we provide
%modified derivation of the non-linear sigma model, which does not reply on
%Hubbard startonovich transformation with the subsequent men field
%approximation, and, at the same time, does not suffer from the difficulties
%of previous approach of Ref. \cite{BonettiMetzner}.

Finally, we apply the developed approach to calculation of spin stiffnesses of doped two-dimansional Hubbard model with hopping between nearest- and next nearest neighbours within the recently proposed dynamical mean field theory (DMFT) for incommensurate long range order \cite{OurFirst}. We verify obtained Ward identities and 
%the obtained results using the recently proposed dynamical mean field theory for incommensurate
%long range magnetic order. We 
determine the doping dependence of the
respective spin stiffnesses, which are
used for construction of the
non-linear sigma model. These stiffnesses can be further used for the
analysis of the magnetic properties of the model in various temperature- and
doping regimes.

The plan of the paper is the following. In Sect. II we introduce the model,
the respective gauge transformations, and present the modified 
Ward identities (the details of their derivation can be found in Appendixes \ref{AppGeneral}, \ref{AppWard}). We also discuss in detail the differences to the previous form
of Ward identities, and identify their sources. In Sect. III we provide
analytical results for the susceptibilities at the momenta close to ${\mathbf{q}}={%
\mathbf{0},\mathbf{Q}}$, obtained from the modified form of Ward identities, and
reveal their important matrix structure. We show that the correct form of
the transverse uniform susceptibility can be obtained only in the presence
of (infinitesimally) small staggered magnetic field, without which the
respective susceptibility vanishes in accordance with spin conservation.
%We also derive the momentum dependencies of the susceptibilities close to
%momenta $q={\mathbf{0},\mathbf{Q}}$. 
%In Sect. IV we consider the modified
%derivation of non-linear sigma model. 
In Sect. IV we present numerical
results for the doped Hubbard model within 
%static and 
dynamic mean field theory, which confirm our
analytical results, and obtain the respective temporal and spatial spin stiffnesses. Finally, in Sect. V we present conclusions.
%\vspace{-0.2cm}
\section{The model and Ward identities}

\subsection{The model and gauge transformation}

We consider Hubbard model  
\begin{equation}
H = - \sum_{ {i, j, \sigma }} t_{ij}{\hat{c}_{i\sigma}^\dagger \hat{c}%
_{j\sigma}}+U\sum_i{\hat{n}_{i\uparrow}\hat{n}_{i\downarrow}},   \label{H}
\end{equation}
where $\hat{c}_{i\sigma}^\dagger$ and $\hat{c}_{i\sigma}$ are creation and
destruction operators of electron at site $i$, spin $\sigma$, $\hat{n}%
_{i\sigma}=\hat{c}_{i\sigma}^\dagger \hat{c}_{i\sigma}$. To describe magnetic excitations, we introduce
non-abelian SU(2) gauge field by performing rotations of fermionic operators $%
\hat{c}_i\rightarrow \mathcal{R}_i \hat{c}_i$, $\mathcal{R}_i$ is the coordinate-
and time-dependent SU(2) rotation matrix. The generating functional of the
model in the rotated reference frame reads 
\begin{equation}
W[\mathcal{R}] = - \ln \left[ \int \mathcal{D}\left[c,c^+\right]
\exp\left( -S[c,c^+,\mathcal{R}] \right) \right],
\end{equation}
where $c,c^+$ are Grassmann variables, $S[c,c^+,\mathcal{R}]$ is the
fermionic action in the rotated reference frame. 
%$\mathcal{J}$ are the source fields, 
We note that the fields $%
c_i$, obtained after the rotation of the reference frame are sometimes
refered as ``chargons", while the spin fields, corresponding to the rotation
matrices $\mathcal{R}$ are called ``spinons" (see, e.g., Ref. \onlinecite%
{BonettiMetzner}). This separation should be supplemented however by the gauge fixing, which will be discussed below in Sect. \ref{GammaSect}. The long-range magnetic order, which is present in the
chargon sector (i.e. related to the fields $c,c^+$) does not necessarily
imply long-range order of the spinon sector after considering the
fluctuations of the metric $\mathcal{R}$.

%For completeness we start from a short review of gauge transformation,
%considered in Refs. . 
The dependence of the action on the spinon fields can be reduced to the
dependence on four $SU(2)$ gauge fields $A_{\mu }$, which are defined by 
\begin{equation}
A_{\mu i}=\mathrm{i}\mathcal{R}_{i}^{+}\partial _{\mu }\mathcal{R}_i,
\label{AR}
\end{equation}%
%\begin{align}
%    A_{0 i} &= i \mathcal{R}^+_i (\tau) \frac{\partial}{\partial \tau} \mathcal{R}_i (\tau) \\
%    A_{m i} &= i \mathcal{R}^+_i (\tau) \partial_m \left( \mathcal{R}_i (\tau) \right)
%\end{align}
where we use the 4-derivative $\partial _{\mu }=(\partial _{\tau },\nabla )$%
. The corresponding action takes the form 
\begin{align}
&S[c,c^{+},\mathcal{R}]=\sum_{ij}\int\limits_{0}^{\beta }d\tau \;c_{i}^{+}%
\left[ \left( \frac{\partial }{\partial \tau }-\mu -\mathrm{i}A_{0i}\right)
\delta _{ij}\right.\\
&\left.-t_{ij}\exp \left( -\vec{r}_{ji}(\nabla -\mathrm{i}\vec{A}%
_{i})\right) \right] c_{i}+U\sum_{i}\int_{0}^{\beta }d\tau \;n_{i\uparrow
}n_{i\downarrow },\notag
\end{align}%
where $\mu $ is the chemical potential, $\beta$ is inverse temperature (in energy units). We note that the external (non-uniform) magnetic field can be absorbed into the $A_{0i}$ component of the gauge field, as we assume in the following. The fields ${A}%
_{\mu i}(\tau )$ 
%, as well as the external $SU(2)$ source (magnetic) fields $J_i(\tau)$ 
can be expanded in Pauli matrices 
%поле  (пока что для общности везде оставляем компоненту с ненулевым следом):
\begin{equation}
A_{\mu i}=\frac{1}{2}\sum\limits_{a=0,x,y,z}A_{\mu i}^{a}{\sigma ^{a}},
\end{equation}%
such that $A_{\mu }^{a}=Tr\left[ \sigma ^{a}A_{\mu }\right] $. 
%, $\mathcal{J}^a = Tr \left[ \sigma^a \mathcal{J} \right]$. 
The corresponding order parameter can be written as a vector 
\begin{equation}
m _{i}^{a}=i\frac{\delta W}{\delta A_{0,i}^{a}(0)}=\frac{1}{2}\left\langle
c_{i}^{+}\sigma ^{a}c_{i}\right\rangle .
\end{equation}%
%It is important for the following that the components $-iA^a_{0i}$ represent an effect of the internal magnetic fields, which are induced by the considered gauge transformation. Since they enter the action in the same way, as the source fields $J^a$, we absorb all the external fields into $A_0$, except those, which are necessary to keep an order parameter fixed during the Legendre transform, see Sect. \ref{GammaSect} for further discussion. 

We consider the gauge
transformations $c_{i}\rightarrow \mathcal{V}_{i}c_{i},\,\,\,c_{i}^{+}%
\rightarrow c_{i}^{+}\mathcal{V}_{i}^{+}$ which imply $\mathcal{R}_{i}\rightarrow \mathcal{R}_{i}\mathcal{V}_{i}^{+},\,\,\,%
\mathcal{R}_{i}^{+}\rightarrow \mathcal{V}_{i}\mathcal{R}_{i}^{+}$. %
% \comIG{Возможно имеет смысл рассматривать калибровочное преобразование более общего вида, реализуемое посредством унитарного поля из U(2). Это позволит включить в разложение (\ref{V_decomposition}) единичную матрицу $\sigma^0$. Не уверен, не придётся ли во всех тождествах рассматривать случай $a=0$ отдельно.}
Expanding the field $\mathcal{V}_{i}(\tau )$ into components 
\begin{equation}
\mathcal{V}_{i}=\exp \left( -\mathrm{i}\sum\limits_{a=x,y,z}\mathcal{V}%
_{i}^{a}\frac{\sigma ^{a}}{2}\right)   \label{V_decomposition}
\end{equation}%
yields the following rules of infinitesimal transformation: 
\begin{align}
A_{\mu }^{0}&\rightarrow A_{\mu }^{0}+o(\vec{\mathcal{V}}),\,\,\,\notag\\
A_{\mu
}^{a}&\rightarrow A_{\mu }^{a}-\partial _{\mu }\mathcal{V}^{a}-\varepsilon
_{abc}A_{\mu }^{b}\mathcal{V}^{c}+o(\vec{\mathcal{V}})
\label{TransformA}
\end{align}%
(here and hereafter the summation over repeated indexes is assumed). Thus,
under infinitesimal spin gauge transformations, the zeroth (charge)
components of the fields ${A}_{\mu i}^{0}(\tau )$ 
%and $J^0_i(\tau)$ 
are not transformed.  
%, and therefore they are put to zero in the following, restricting the summation in Eqs. (\ref{Aex}), (\ref{Jex}) to spatial components only. 
%\begin{align}
%    A_{\mu i} &= \sum\limits_a A^{a}_{\mu i} \frac{\sigma^a}{2} \\
%    \mathcal{J}_i &= \sum\limits_{a} \mathcal{J}^a_i \frac{\sigma^a}{2}
%    \label{A_decomposition}
%\end{align}
%где под $\sum_a$ подразумевается $\sum_{a = x,y,z}$.
%\vspace{-0.5cm}
\subsection{Ward identities for the gauge kernel}
\label{WardW1}

In the following we introduce the gauge kernel $K_{\mu
,x;\nu ,x^{\prime }}^{ad}={\delta ^{2}W}/({\delta A_{\mu ,x}^{a}\delta
A_{\nu ,x^{\prime }}^{d}})$ ($a,d=0,x,y,z$), which explicit form is obtained
in Appendix \ref{AppGeneral}. The derivation of Ward identities for the gauge kernel, obtained from the invariance of the functional $W$ under gauge transformation (\ref{TransformA}), is presented in Appendix \ref{WardW}. For the components of the kernel, which contain at least one spatial index $n>0$, we find the identity  
\begin{equation}
\left( \partial _{\mu ,x}\delta _{ab}+{\epsilon _{acb}A_{\mu ,x}^{c}}\right)
K_{\mu ,x;n,x^{\prime }}^{bd}=0.  \label{eqq}
\end{equation}%
For vanishing fields $A_{\mu,x}$ this is the condition of spin conservation, which is similar to charge conservation in the U(1) case (see, e.g., the discussion in Ref. \onlinecite{KatsLicht}). The conjugated condition,
obtained by interchange $\mu \leftrightarrow n$, $x\leftrightarrow x^{\prime
}$, $a\leftrightarrow d$ (also in the presence of the fields $A_\mu$) corresponds to the condition of gauge invariance,
which is in accordance with the Eq. (\ref{TransformA}). 
We stress that in the presence of long-range magnetic order in the chargon
sector, the spin conservation in zero external fields and gauge invariance of the spatial components
of the gauge kernel  
%(which is {\it identical} to that considered in Refs. \cite{Bonetti,BonettiMetzner,BonettiThesis}) 
remains unbroken. %\comAKK{check that the r.h.s. vanishes at A->0} 

For the temporal and mixed temporal-spatial components of the gauge kernel we find the identity (see Appendix \ref{WardW})
\begin{equation}
\partial_{\mu,x}K_{\mu,x;0,x^{\prime }}^{ad} + {\epsilon_{acb}A_{0,x}^{c}%
\chi^{bd}_{xx^{\prime }}} = i \varepsilon_{a d b} m^b_x \delta_{x,x^{\prime
}},   \label{Wnu0}
\end{equation}
here and in the following we assume summations over repeated indexes, $\chi^{b d}_{xx^{\prime }}=  K^{bd}_{0,x;0,x^{\prime }}$ is the tensor
of the non-uniform dynamic susceptibility of the chargon sector. {For
vanishing temporal component $A_{0,x}=0$ (including also the external
magnetic field), Eq. (\ref{Wnu0}) in the uniform limit implies \textit{%
absence} of time dependence of diagonal components of the uniform dynamic 
chargon spin susceptibility $\sum_{{\mathbf{x}}{\mathbf{x}}^{\prime
}}\chi^{aa}_{xx^{\prime }}$, which is another consequence of spin
conservation in the system. We note again that internal magnetic fields
(such as staggered magnetization in the chargon sector) do not prevent spin
conservation. }

It was suggested in Ref. \onlinecite{CSS} that the inter-band contribution to the
uniform susceptibility determines the spin-wave velocity. Since this
contribution can be selected by considering the $\omega\rightarrow 0$ limit
of the susceptibility taken \textit{after} the uniform ${\mathbf{q}}\rightarrow 0$ limit
(see the discussion in Ref. \onlinecite{BonettiThesis}), vanishing of the dynamic
uniform susceptibility, obtained from the Eq. (\ref{Wnu0}), looks contradicting the suggestion of identifying
dynamic susceptibility with the inter-band contributions to the uniform
susceptibility. As we argue below in Sect. \ref{chiQ}, the reason of this contradiction is in the necessity of introducing infinitesimally small external
staggered magnetic field, which is switched off only \textit{after} the
uniform and static limits are taken. As we show below, this external
staggered magnetic field, absent in previous consideration of Ward
identities \cite{Bonetti,BonettiThesis}, 
%In the following we show that keeping the non-vanishing external field in the right hand side of the Eq. (\ref{Wnu0})  
is crucial to obtain \textit{finite} interband contribution to the uniform susceptibility. 
%Physically this
%corresponds to account of only part of the interband contributions, as it is
%discussed in details in Sect. \ref{chiQ} below. %These are the contributions,
%which yield finite static uniform transverse susceptibility in the
%Heisenberg model, they were also obtained in the derivation of the
%non-linear sigma model in Ref. \cite{Dupuis}. In the following we call them
%static interband contributions to the transverse uniform susceptibility. 
%From transverse components $\sigma^{\pm} = \sigma^x \pm i \sigma^y$ in AFM case (only $m^z \neq 0$) at $A=0$ it means that
%\begin{equation}
%    \partial_{\mu,x} \left( \frac{\delta^2 W}{\delta A^+_{\mu,x} \delta A^-_{0,x'}} \right) = 2 i m^z_x \delta_{x,x'}
%\end{equation}
%\comAK{\it And this is correct at ${\bf q}=0$, at least in the large U, small w<<U limit. See Eq. (29) of Chubukov's paper. Please also note that at arbutrary w the frequency dependence is much more complicated. It is also correct for ferromagnet (arbitrary w)}

While in the theories using the Hubbard-Stratonovich transformation \cite%
{Weng,Schulz,Dupuis,Dupuis1,Sachdev,Sachdev1}, the gauge kernel (which
appears after expanding the action to the second order in gauge fields) is
used for determination of spin stiffnesses, the above consideration implies
that for zero external field the gauge kernel itself (without performing
Hubbard-Stratonovich transformation and/or introducing mean field) does not determine spatial spin stiffnesses, since the corresponding
contributions to the non-linear sigma model, which are proportional to the spatial 
derivatives of the gauge kernel, vanish according to the Eq. (\ref{eqq}). 
%, in contrast to the consideration of Ref. \cite{BonettiMetzner}. 
This difficulty occurs because of the absence of gauge-fixing contributions
in the considered action, which coincides with that of Refs. \onlinecite%
{Bonetti,BonettiThesis,BonettiMetzner}. Presence of the symmetry breaking in
the chargon sector (e.g. spin asymmetric self-energies of fermionic degrees
of freedom) does not resolve this difficulty, since the symmetry is \textit{%
spontaneously} broken at the level of the particular solution (e.g., the
mean-field or dynamic mean field approximation, considered below in Sect. %
\ref{Numer}), and formally the action is still rotation invariant. As we
show below, in Sect. \ref{GammaSect}, the spin stiffness is in fact
determined by the derivatives of the Legendre transformed action, which are
in general \textit{different} from the gauge kernel. 
%\comAKK{The gauge invariance is broken in the functional $\Gamma$, discussed below?}.}

%We note that in the equilibrium ($J=A_m=0$, $m=x,y,z$) the considered theory describes long-range magnetic order in the itinerant electron system. 

%\vspace{-0.5cm}
\subsection{Legendre transformation and Ward identities for the modified gauge kernel}
\label{GammaSect} %\par\noindent\rule{\textwidth}{0.5pt}
%{}
%obtain the momentum dependence of spin susceptibilities $\chi^{ad}_{xx'}$ of the chargon sector, 
To extract spatial spin stiffnesses, we perform Legendre transformation, following
Refs. \onlinecite{Bonetti,BonettiThesis}. To this end we explicitly pick out the part of the
external magnetic field, which regulates the order parameter, by performing the
shift $A_{0}\rightarrow A_{0}+iJ$, and
introduce the functional 
\begin{align}
& \Gamma \lbrack \phi ,A]=W[J,A]-\phi _{x}^{a}J_{x}^{a},  \label{GammaDef} \\
& J_{x}^{a}=-\frac{\delta \Gamma }{\delta \phi _{x}^{a}},\,\,\,\phi _{x}^{a}=%
\frac{\delta W}{\delta J_{x}^{a}}=i\frac{\delta W}{\delta A_{0,x}^{a}}, 
\notag
\end{align}%
which is identical to that considered in Refs. \onlinecite{Bonetti,BonettiThesis}.
%\begin{equation}
%    \phi^a_i(\tau) = \frac{\delta W [J, %A]}{\delta J_i^a(\tau) }
%\end{equation}
The second derivatives of the transformed functional $\kappa _{xx^{\prime
}}^{ab}= {\delta ^{2}\Gamma }/({\delta \phi _{x}^{a}\delta \phi _{x^{\prime
}}^{b}})$ determine the inverse susceptibilities of the chargon sector
according to the standard relation 
\begin{equation}
\kappa _{xx^{\prime
}}^{ac}%
\frac{\delta ^{2}W}{\delta J_{x^{\prime }}^{c}\delta J_{x^{\prime \prime
}}^{b}}=-\kappa _{xx^{\prime
}}^{ac}\frac{\delta ^{2}W}{\delta A_{0,x^{\prime }}^{c}\delta
A_{0,x^{\prime \prime }}^{b}}=-\delta _{a,b}\delta _{xx^{\prime \prime }}.
\end{equation}

%It is crucial that either $\phi$ or $J$ depend implicitly on the remaining gauge fields $A$. 
%This dependence, being quite important, was missed in Refs. \cite{Bonetti,BonettiThesis,BonettiMetzner}. 
%In the following we consider $\phi$ as independent variables, but take into account the dependence $J[A]$. Taking into account that $W[J,A]=W[J-i A_0,A_m]$ ($m=x,y$), we find
%\begin{equation}
%\Gamma[\phi,A]=W[\tilde J,A_m]-\phi_x^a \tilde J_x^a-i \phi_x^a A^a_{0,x}. 
%\label{Gamma_new}
%\end{equation}
%where $\tilde J=J-i A_0$. Therefore, 
%$A_0$ is the ``redundant"\, 
% variable for $\Gamma$, and  
%. According to the Eq. (\ref{Gamma_new}) 
%For the derivatives of the functional $\Gamma$ we obtain
One can then obtain the following relations between the functional
derivatives 
\begin{align}
\frac{\delta \Gamma }{\delta A_{m}^{a}}& =\frac{\delta W}{\delta A_{m}^{a}},
\label{DG1} \\
\frac{\delta \Gamma }{\delta A_{0,x}^{a}}& =\frac{\delta W}{\delta
A_{0,x}^{a}}=-i\phi _{x}^{a}, \\
\frac{\delta ^{2}\Gamma }{\delta A_{\mu ,x}^{a}\delta A_{0,x^{\prime }}^{b}}%
& =0.  \label{Gamma_vanishpart}
\end{align}%
The Eq. (\ref{Gamma_vanishpart}) implies that second derivatives of the functional $\Gamma $ over the
gauge fields, involving at least one $A_{0}$ component, vanish, and $\Gamma $
depends only \textit{linearly} on the field $A_{0}$. This linear dependence
directly follows from the Eq. (\ref{GammaDef}) at fixed $\phi $, since the
change of $A_{0}$ necessarily yields the opposite change of $iJ$ to keep $%
\phi $ fixed. Notably, although our functional $\Gamma $ coincides with that
considered in Refs. \onlinecite{Bonetti,BonettiThesis}, the second
derivative $\delta ^{2}\Gamma /\delta A_{0}^{2}$ was suggested in these studies to determine
the temporal components of spin stiffness (i.e. the transverse
susceptibilities). Our derivation of temporal components of spin stiffness
is presented below.

First, we emphasize that the components of the modified gauge kernel $M_{\mu ,x;\nu ,x^{\prime }}^{ab}\equiv {\delta ^{2}\Gamma }/({\delta
A_{\mu }^{a}\delta A_{\nu }^{b}})$ are in general different from the gauge kernel $K_{\mu ,x;\nu ,x^{\prime }}^{ab}$.  It is crucial that at fixed fields $%
\phi $ the sources $J$ depend implicitly on the gauge fields $A$. 
%This dependence, being quite important, was missed in Refs. \cite{Bonetti,BonettiThesis,BonettiMetzner}. 
%In the following we consider $\phi$ as independent variables, but take into account the dependence $J[A]$.
%At the same time, taking into account the dependence $J[A]$ at fixed $\phi$. 
We have 
\begin{align}
\frac{\delta J_{x}^{a}}{\delta A_{\mu ,x^{\prime }}^{b}}&=-\frac{\delta
^{2}\Gamma }{\delta {\phi _{x}^{a}}\delta A_{\mu ,x^{\prime }}^{b}}=-\frac{%
\delta }{\delta \phi _{x}^{a}}\frac{\delta W}{\delta A_{\mu ,x^{\prime }}^{b}%
}\notag\\
&=-\frac{\delta J_{x^{\prime \prime }}^{c}}{\delta \phi _{x}^{a}}\frac{%
\delta ^{2}W}{\delta J_{x^{\prime \prime }}^{c}\delta A_{\mu ,x^{\prime
}}^{b}}=\kappa _{xx^{\prime \prime }}^{ac}\frac{\delta ^{2}W}{\delta
J_{x^{\prime \prime }}^{c}\delta A_{\mu ,x^{\prime }}^{b}}.
\label{dJA}
\end{align}%
Therefore, we obtain
\begin{align}
M_{\mu ,x;\nu ,x^{\prime }}^{ab}
%\equiv \frac{\delta ^{2}\Gamma }{\delta
%A_{\mu }^{a}\delta A_{\nu }^{b}}& 
&=\frac{\delta ^{2}W}{\delta A_{\mu
,x}^{a}\delta A_{\nu ,x^{\prime }}^{b}}+\frac{\delta ^{2}W}{\delta A_{\mu
,x}^{a}\delta J_{x^{\prime \prime }}^{c}}
\frac{\delta J_{x''}^{c}}{\delta A_{\nu ,x^{\prime }}^{b}}\notag\\
%\kappa _{x^{\prime \prime
%}x^{\prime \prime \prime }}^{cd}\frac{\delta ^{2}W}{\delta J_{x^{\prime
%\prime \prime }}^{d}\delta A_{\nu ,x^{\prime }}^{b}}\notag\\ 
%\end{align}
%begin{align}
%M_{\mu ,x;\nu ,x^{\prime }}^{ab}&\equiv \frac{\delta ^{2}\Gamma }{\delta
%A_{\mu }^{a}\delta A_{\nu }^{b}}
%\notag\\&
&=K_{\mu ,x;\nu ,x^{\prime }}^{ab}-K_{\mu
,x;0,x^{\prime \prime }}^{ac}\kappa _{x^{\prime \prime }x^{\prime \prime
\prime }}^{cd}K_{0,x^{\prime \prime \prime };\nu ,x^{\prime }}^{db}.  \label{most_important}
\end{align}%
%are in general different from the gauge kernel $K_{\mu ,x;\nu ,x^{\prime }}^{ab}$ (see the proof of Eq. (\ref{most_important}) in Appendix \ref{WardGammaSect}). 
The second term in the right hand side of Eq. (\ref{most_important}) is crucial to fulfill the relation (%
\ref{Gamma_vanishpart}) and obtain the correct form of the Ward identities
for the functional $\Gamma $. In particular, for $\mu =0$ or $\nu =0$ it
yields $M_{\mu \nu }=0$ in agreement with the Eq. (\ref{Gamma_vanishpart}). In Sect. \ref{Explicit} we show that the second term in the right hand side
of Eq. (\ref{most_important}) cancels contribution of Goldstone modes in
the first term. This is physically justifiable, since Goldstone modes, being
the low energy excitations, should not give contribution to the spin
stiffness of the non-linear sigma model. On the other hand, taking into account the dependence $J[A]$ at fixed $\phi$, represented by Eq. (\ref{dJA}), implies gauge fixing in the Legendre transform (\ref{GammaDef}) at fixed equilibrium $\phi$, which is characterized by $J[0]=0$.

%, which depends explicitly on the order parameter $\phi$.
%\subsubsection{Second order Ward identity for $\Gamma$}

The respective Ward
identity for the second derivatives $M^{ab}_{\mu,x;\nu,x^{\prime }}$ takes the form (see Appendix \ref{WardGammaSect})
\begin{align}
\partial_{m,x} \partial_{n,x^{\prime }}& M^{ab}_{m,x;n,x^{\prime }} =- i\varepsilon_{a d b} \phi^d_x \partial_{0,x^{\prime }} \delta_{x,x^{\prime
}}
\notag\\
+&\varepsilon_{b \widetilde{d} \widetilde{c}} \phi^{\widetilde{d}}_{x^{\prime
}}\left( \varepsilon_{a d c} \phi^d_x \kappa^{c \widetilde{c}}_{xx^{\prime
}} \right.\left.{+i\varepsilon_{a \widetilde{c} c} A_{0,x}^c\delta_{xx^{\prime }}}\right).  \label{Gamma_Ward1}
\end{align}
{The first term in the right hand side of Eq. (\ref{Gamma_Ward1}) vanishes for almost uniform
components of $M^{ab}_{m,x;n x^{\prime }}$ since it contains an order
parameter vector, oscillating in space for non-zero wave vector ${\mathbf{Q}}
$. The equation (\ref{Gamma_Ward1}) is \textit{different} from that derived
in Refs. \onlinecite{Bonetti,BonettiThesis} by the %last term in the right hand side and 
the derivatives in the left hand side acting on the space components only.} We also note that if the gauge kernel $%
K^{ab}$ would be equal to $M^{ab}$, as it was assumed in Refs. \onlinecite%
{Bonetti,BonettiThesis,BonettiMetzner}, the left-hand side of Eq. (\ref%
{Gamma_Ward1}) would vanish in the zero external field according to the Eqs.
(\ref{eqq}), (\ref{Gamma_vanishpart}), while the right-hand side is finite.
{\ For diagonal components we obtain the identity, which allows us to
evaluate the spatial components of the spin stiffness, 
\begin{align}
\partial_{m,x} \partial_{n,x^{\prime }}& M^{aa}_{m,x;n,x^{\prime }} \notag\\&=
\varepsilon_{a \widetilde{d} \widetilde{c}} \phi^{\widetilde{d}}_{x^{\prime
}}\left( \varepsilon_{a d c} \phi^d_x \kappa^{c \widetilde{c}}_{xx^{\prime
}} {+i\varepsilon_{a \widetilde{c} c} A_{0,x}^c\delta_{xx^{\prime }}}\right).
\label{Gamma_Ward2}
\end{align}
One can see that the corresponding spin stiffnesses are proportional to the
derivatives $M^{aa}_{mn}$ of the functional $\Gamma$ over gauge fields,
which are in general \textit{different} from the gauge kernel, see Eq. (\ref%
{most_important}). Since the temporal parts $M_{\mu\nu}$ with either $\mu=0$
or $\nu=0$ vanish according to the Eq. (\ref{Gamma_vanishpart}), we can
formally add the respective time derivatives to the left hand sides of Eq. (%
\ref{Gamma_Ward2}), but they do not provide actual contribution, in contrast
to Refs. \onlinecite{Bonetti,BonettiThesis}. This implies that the
respective (diagonal) components of the inverse susceptibilities $\kappa$
are frequency-independent. 
%(see explicit expressions in Sect. \ref{chiQ} below).  
At the same time, the temporal components of the spin stiffnesses are
determined by the off-diagonal components of inverse susceptibility $\kappa$,
see explicit calculation below, in Sect. \ref{chiQ}. 
%Теперь воспользуемся тем, что в состоянии термодинамического равновесия
%
%\begin{equation}
%    A^a_\mu = 0
%\end{equation}
%
%\begin{equation}
%    \frac{\delta W}{ \delta A_\mu^a} = \frac{\delta \Gamma}{\delta A^a_\mu} = 0
%\end{equation}
}

%\begin{equation}
%    -J^a = \frac{\delta \Gamma}{\delta \phi^a} = 0
%\end{equation}

%Спонтанное нарушение симметрии сооветствует случаю, когда

%\begin{equation}
%    \phi^a \neq 0
%\end{equation}

%Тогда получим следующие следствия из более общего утверждения, справедливые для термодинамического равновесия
%
%\begin{equation}
%    \partial_\mu \left( \frac{\delta W }{\delta A^a_\mu} \right) = 0
%\end{equation}
Let us compare the obtained relations to the Ward identity of Refs. \onlinecite%
{Bonetti,BonettiThesis} in zero external fields: 
\begin{align}
\partial_{\mu,x} \partial_{\nu,x^{\prime }} K_{\mu,x; \nu,x^{\prime }}^{ab}
=\partial_{\mu,x} \partial_{\nu,x^{\prime }}& M_{\mu,x; \nu,x^{\prime }}^{ab}\notag\\
&=  \varepsilon_{a d c} \phi^c \varepsilon_{b p l} \phi^l
\kappa^{dp}_{xx^{\prime }}.   \label{WardOld}
\end{align}
{As we discuss above, the spatial derivatives in the
left-hand side of Eq. (\ref{WardOld}) for zero external field in fact yield
identically zero according to the gauge invariance condition, Eq. (\ref{eqq}%
). While the temporal part $M^{ab}_{0,x;0,x^{\prime }}$ vanishes too, the
spatial part $M^{ab}_{m,x;n,x^{\prime }}$ (and its derivatives) do not
vanish, which readily shows that the left equality in the relation (\ref%
{WardOld}) can not be correct. This contradiction is resolved in our
approach by account for the additional term
in the Eq. (\ref{most_important}). The second equality in Eq. (\ref{WardOld}%
) is fulfilled in our approach as well, with the important difference that
the contribution of the time components in the summation over $\mu$, $\nu$
in the middle part of Eq. (\ref{WardOld}) vanishes, implying vanishing of the respective staggered components of the inverse susceptibility $\kappa_{xx'}$ in the absence of the external magnetic field, while
%Although this makes also
%the left equality trivially correct for %${\mathbf{q}}\rightarrow 0$ (and
%only in this limit), since both, the left-hand and middle part of Eq. (\ref{WardOld}) vanish
%in that limit, this is also in contradiction with Refs. \cite{Bonetti,BonettiThesis}, which
%considered the 
these components, together with $M^{ab}_{0,x;0,x^{\prime }}$ (and its time
derivatives) were considered finite in Refs. \onlinecite{Bonetti,BonettiThesis}. In our approach this contradiction is resolved by
including off-diagonal spin components of $\kappa^{ab}$ and performing its
inversion as a matrix, as discussed below, in Sect. \ref{chiQ}}. %\newpage

%\textcolor{red}{Our results are not Lorentz-invariant because of the
%off-diagonal terms in the dynamic part. Can they be put in the
%Lorentz-invariant form? Can the second term in the r.h.s of (45) be
%considered as appearing from the matrix inversion?}
%\vspace{-0.5cm}
\section{Analytical results for the chargon susceptibilities near ${\mathbf{q%
}}={\mathbf{0}},{\mathbf{Q}}$}

\label{chiQ}

%\subsection{Explicit form of the Ward identity for the inverse susceptibilities}

To calculate the chargon susceptibilities we introduce the Fourier
components in the global reference frame (we consider only diagonal in
momentum components of the gauge kernel) 
\begin{align}
&M^{ab}_{q;\mu \nu}=\int d\tau \sum_{\mathbf{xx}^{\prime }}e^{-i\mathbf{q}(%
\mathbf{x}-\mathbf{x}^{\prime })+i\omega (\tau-\tau^{\prime
})}M^{ab}_{\mu,x;\nu,x^{\prime }} \\
&\kappa^{ab}_{\mathbf{qq}^{\prime },\omega}=\int d\tau \sum_{\mathbf{x}%
\mathbf{x}^{\prime }}e^{-i\mathbf{q}\mathbf{x}+i\mathbf{q}^{\prime }\mathbf{x%
}^{\prime }+i\omega (\tau-\tau^{\prime })}\kappa^{ab}_{xx^{\prime }}.
\end{align}
and $\chi_{{\mathbf q}{\mathbf q}',\omega}$ defined similarly to $\kappa_{{\mathbf q}{\mathbf q}',\omega}$.
%\vspace{-0.3cm}
\subsection{Commensurate antiferromagnetic order in chargon sector}

\label{SectComm}

In the commensurate antiferromagnetic (AFM) case we assume the spatial dependence of
the order parameter ${\mathbf{\phi }}_{x}=m(0,0,\cos (\mathbf{Qx}))$ with $%
\mathbf{Q}=(\pi ,\pi )$ and introduce also staggered field $A_{0}^{a}=ih\phi
_{x}^{a}/m$. The general form of the susceptibility matrix $\chi _{\mathbf{q}{\mathbf{q}'},\omega }$ at $\mathbf{q},\mathbf{q}'=\mathbf{0},\mathbf{Q}$, allowed by Ward identity (\ref{Wnu0}), is presented in Appendix \ref{WardW}. Since the Ward identities do not fix the $\chi _{{\mathbf{Q}},\omega }^{xx,yy}$ components of the susceptibility, we parameterize these components by the temporal spin stiffness $\chi _{\omega }$ (which is in general
frequency dependent) as
\begin{equation}
\chi _{{\mathbf{Q}},\omega }^{xx,yy}=\frac{m^{2}}{hm+\chi _{\omega }\omega
^{2}},  \label{chiQQAFM}
\end{equation}%
and assume $\chi _{{\mathbf{Q}},\omega }^{xy,yx}=0$.
For $\mathbf{q},\mathbf{q^{\prime }}=\mathbf{0},%
\mathbf{Q}$ we obtain with account of Eq. (\ref{chiQQAFM})
\begin{equation}
\chi _{{\mathbf{q}}\mathbf{q}^{\prime },\omega }^{ab}=d_{\omega }\left( %
\begin{NiceArray}{cc:cc} h \chi_\omega & 0 & 0 & \chi_\omega \omega \\ 0 & h
\chi_\omega & -\chi_\omega \omega & 0 \\ \hdottedline 0 & \chi_\omega \omega
& m & 0 \\ -\chi_\omega \omega & 0 & 0 & m \\ \end{NiceArray}\right) ,
\label{chiQ0AFM}
\end{equation}%
where $d_{\omega }=m/(hm+\omega ^{2}\chi _{\omega })$, the blocks correspond to the values $\mathbf{q},\mathbf{q^{\prime }}=%
\mathbf{0},\mathbf{Q}$, and $a,b=x,y$ numerates components within each block.  One can see that in agreement with spin conservation, discussed in Sect. \ref{WardW1} the
uniform transverse susceptibilities (given by the upper left block of Eq. (\ref{chiQ0AFM})) vanish if the limit $h\rightarrow 0$ is
taken prior to $\omega \rightarrow 0$, 
but they acquire a finite value $\chi
_{\omega =0}$ if the limit $\omega \rightarrow 0$ is taken first. While the limit $\mathbf{q}=\mathbf{0}$  {prior to taking the $\omega \to 0$ limit}, as considered in Eq. (\ref{chiQ0AFM}), excludes the intraband contributions
to the uniform static transverse susceptibility (cf. Ref. \onlinecite{BonettiThesis}), treatment of all interband
terms (which is performed in the absence of the external staggered field $h$)
yields zero transverse susceptibility in accordance with spin conservation (\ref%
{Wnu0}). Therefore,
finite external staggered field, switched off in the end of the calculation,
is crucially important to obtain finite uniform static transverse
susceptibility. Presence of small finite staggered field, which is switched off {\it after} taking ${\bf q}=0$, $\omega\rightarrow 0$ limits,
excludes interband contributions, which do not conserve momentum, 
%$\chi^{xy}_\mathbf{Q,0}$ 
making the corresponding chargon uniform susceptibility finite (see the example of mean field calculation in Appendix \ref{SectMF}).

Inverting
Eq. (\ref{chiQ0AFM}), we find 
\begin{equation}
\kappa _{{\mathbf{q}}\mathbf{q}^{\prime },\omega }^{ab}=\frac{1}{m}\left( %
\begin{NiceArray}{cc:cc} {m\chi_\omega^{-1} } & 0 & 0 & -{\omega } \\ 0 &
{m\chi_\omega^{-1} } & {\omega } & 0 \\ \hdottedline 0 & -{\omega } & {h} &
0 \\ {\omega } & 0 & 0 & {h} \\ \end{NiceArray}\right).   \label{GammaQ0AFM}
\end{equation}%
One can see that the diagonal terms $\kappa _{\mathbf{Q}\mathbf{Q}%
}^{xx,yy}=h/m$, described by the lower right $2\times 2$ block of Eq. (\ref%
{GammaQ0AFM}) \textit{do not} acquire any dynamic frequency-dependent part,
in agreement with the Eq. (\ref{Gamma_Ward2}), but in contrast to the result
of Refs. \onlinecite{Bonetti,BonettiThesis}, based on previous form of Ward identities.
%, and Ref. \cite{BonettiMetzner} using expansion in the gauge fields. 
As it is discussed in
Sect. \ref{GammaSect}, the latter approaches 
assume finiteness of the derivative $\partial _{\tau }^{2}M_{00}^{aa}$,
which in fact vanishes. 
%The latter approach assumes violation of the gauge invariance conditions (\ref{eqq}), (\ref{Wnu0}), which otherwise yield vanishing diagonal second order contributions, related to time derivatives. 
It is important to stress, that absence of the dynamic contributions in the
diagonal $\kappa _{\mathbf{Q}\mathbf{Q}}^{xx,yy}$ terms does not contradict
to the form of dynamic susceptibilities (\ref{chiQQAFM}), which acquire
frequency dependence due to the \textit{off-diagonal} terms in Eq. (\ref%
{GammaQ0AFM}).
%{While the uniform inverse susceptibility $\kappa_{{\mathbf 0}{\mathbf 0}}$ does not depend explicitly on the external staggered field $h$, taking the limit $\mathbf{q}= 0$ prior the $\omega \to 0$  limit  corresponds to the exclusion of intra-band contributions for the uniform inverse susceptibility $\kappa _{\mathbf{0}\mathbf{0}}$, in agreement with the discussion of the Eq. (\ref{chiQ0AFM}) above.}
% \comIG{At finite doping direct calculation of the temporal spin stiffness $\chi_{\omega}$ from the uniform inverse susceptibility $\kappa _{\mathbf{0}\mathbf{0}}$ requires taking the limit $\mathbf{q} \to 0$ before the limit $\omega \to 0$ in order \comAK{This is assumed by Eq. (25).} to exclude intra-band contributions to the susceptibility.  This is only important for the static spin stiffness $\chi_{\omega = 0}$ while in our calculations we have used finite frequency data to study the frequency dependence of the temporal spin stiffness, see the results presented in Sect. \ref{DMFTAFM} below.}
% Importantly, the uniform {\it inverse} susceptibility $\kappa _{\mathbf{0}\mathbf{0}}$ does not depend on the order of the limits $\omega\rightarrow 0$, $h\rightarrow 0$ and can be used for evaluation of the static uniform susceptibility even in the zero staggered magnetic field, see explicit evaluation in Sect. \ref{DMFTAFM} below.  

To obtain the momentum dependence of the susceptibilities, we use the Ward
identity (\ref{Gamma_Ward2}), which takes in momentum space the form 
\begin{align}
q_n q_l M_{q;nl}^{yy,xx}&= {m^2}\kappa^{xx,yy}_{\mathbf{q}+\mathbf{Q},{%
\mathbf{q}+\mathbf{Q}},\omega}-{m h}.   \label{Gamma_com}
\end{align}
At ${\mathbf{q}}=0$ this is consistent with the Eq. (\ref{GammaQ0AFM}). The transverse magnetic susceptibilities
near the wave vector $\mathbf{Q}$ are uniquely determined by the
corresponding current correlation functions $M^{xx,yy}$. 
%The inverse susceptibilities at the momenta close to $\pm {\mathbf Q}$ can be directly obtained from the Eq. (\ref{Gamma_com})  %
%\begin{align}
%&\kappa^{xx,yy}_{\mathbf{q}+\mathbf{Q},{\mathbf{q}+\mathbf{Q}},\omega}=\frac{q_n q_l}{m^2} M_{q;nl}^{yy,xx}-\frac{h}{m},
%\\
%&\kappa^{yy}_{\mathbf{q}+\mathbf{Q},\mathbf{q}+\mathbf{Q},\omega}=\frac{ h}{m}+\frac{q_n q_l}{m^2} M_{nl}^{xx}(0),
%\label{Gamma_com1}
%\end{align}
%where ${\mathbf q}$ assumed to be small. 
We note again that the corresponding components of the inverse
susceptibilities appear to be frequency independent. Neglecting the change
of the other components in the Eq. (\ref{GammaQ0AFM}) with $\mathbf{q}$ and inverting susceptibility matrix, we
readily obtain 
\begin{equation}
\chi^{xx,yy}_{{\pm {\mathbf{Q}}+{\mathbf{q}}},{\pm {\mathbf{Q}}}+{\mathbf{q}}%
,\omega}=\frac{m^2}{h m+\chi_\omega \omega^2+\rho q^2},  \label{chiAFq}
\end{equation}
where the spatial spin stiffness $\rho=M^{xx,yy}_{q\rightarrow 0;nn}$ (we
have taken into account the diagonal form of $M^{xx,yy}_{q\rightarrow 0;nl}$
over spatial indices $n,l$).
%\vspace{-0.5cm}
\subsection{Incommensurate spiral order in chargon sector}

\subsubsection{Susceptibilities at ${\mathbf q}={\mathbf{0}},{\mathbf{Q}}$}

In the case of spiral incommensurate order we
consider the order parameter $\phi _{x}=m(\sin (\mathbf{Qx}),0,\cos (\mathbf{%
Qx}))$ and introduce external staggered field $A_{0}^{a}=ih\phi _{x}^{a}/m$. For calculations we pass to the local reference frame, see Appendix \ref{AppLocal}. The
susceptibilities in the local coordinate frame are diagonal with respect to
momenta. The general form of the susceptibility matrix in the local reference frame, allowed by Ward identities, is presented in Appendix \ref{AppSusc1}. We parameterize the $\bar{\chi}_{{\mathbf{0}},\omega }^{xx}$ and  $\bar{\chi}_{{\pm {\mathbf{Q}}},\omega }^{yy}$ susceptibilities in the local reference frame (denoted by bars here and below), which are not fixed by Ward identities, by (in
general frequency dependent) in-plane and out-of-plane components of temporal spin
stiffnesses $\chi _{2,\omega }$ and $\chi _{1,\omega }$ according to
\begin{equation}
\bar{\chi}_{{\mathbf{0}},\omega }^{xx}=\frac{m^{2}}{hm+\chi _{2,\omega
}\omega ^{2}},\,\,\,\bar{\chi}_{{\pm {\mathbf{Q}}},\omega }^{yy}=\frac{m^{2}%
}{hm+\chi _{1,\omega }\omega ^{2}},\label{chiQQQ}
\end{equation}%
and assume $\bar{\chi}_{{\mathbf{0}},\omega }=\bar{\chi}_{{\mathbf{0}}%
,\omega }^{zx}=0$. From this we obtain the frequency dependence of the
uniform susceptibilities in the local reference frame 
\begin{align}
\bar{\chi}_{{\mathbf{0}},\omega }^{ab}&=\left( 
\begin{array}{ccc}
md_{2,\omega } & \omega \tilde{d}_{2,\omega } & 0 \\ 
-\omega \tilde{d}_{2,\omega } & h\tilde{d}_{2,\omega } & 0 \\ 
0 & 0 & \bar{\chi}_{\mathbf{0},\omega }^{{zz}} 
\end{array}%
\right) _{x,y,z},
\label{chiIncomm0}
\end{align}
where $d_{i,\omega }=m/(hm+\omega ^{2}\chi _{i,\omega })$, $\tilde{d}_{i,\omega }=d_{i,\omega }\chi _{i,\omega }$, and the index $x,y,z$
refers to the respective spin $S^x, S^y, S^z$ reference frame. The blocks of the local susceptibility at the momenta $%
\mathbf{q}=\pm {\mathbf{Q}}$ are conveniently written in the basis $S_{%
\mathbf{q}}^{+},S_{\mathbf{q}}^{y},S_{\mathbf{q}}^{-}$ ($S_{\mathbf{q}}^{\pm
}=S_{\mathbf{q}}^{z}\pm iS_{\mathbf{q}}^{x}$ are the circular in plane spin components) as
\begin{align}
%%%
\bar{\chi}_{-{\mathbf{Q}},\omega }^{ab}&=\left( 
\begin{array}{ccc}
{ \frac{\I h}{\omega} \bar{\chi}^{+y}_{-\mathbf{Q},\omega}} & \bar{\chi}^{+y}_{-\mathbf{Q},\omega}  &  \bar{\chi}^{+-}_{-\mathbf{Q},\omega} \\ 
-\I\omega \tilde{d}_{1,\omega } & m d_{1,\omega } &  \bar{\chi}^{y-}_{-\mathbf{Q},\omega}  \\ 
h \tilde{d}_{1,\omega } & -\I\omega \tilde{d}_{1,\omega } & {\frac{\I h}{\omega} \bar{\chi}^{y-}_{-\mathbf{Q},\omega}}
\end{array}%
\right) _{+,y,-},\,  \label{chiINCOMlocal}
\end{align}%
the susceptibility $\bar{\chi}_{{\mathbf{q}}={\mathbf{Q}},\omega } = (\bar{\chi}_{{\mathbf{q}}=-{\mathbf{Q}},\omega })^+$ can be obtained via hermitian conjugate. In the basis ($S_{-\mathbf{Q}%
}^{+,y,-},S_{\mathbf{0}}^{+,y,-},S_{\mathbf{Q}}^{+,y,-}$) we find then in the global reference frame
\begin{widetext}
\begin{equation}
\chi _{{\mathbf{q}}\mathbf{q}^{\prime },\omega }^{ab}=\left( %
\begin{NiceArray}{ccc:ccc:ccc} 0 & 0 & \bar{\chi}^{zz}_{{\mathbf
0},\omega}+m d_{2,\omega} & 0 & i \omega \tilde{d}_{2,\omega} & 0 &
\bar{\chi}^{zz}_{{\mathbf 0},\omega}-m d_{2,\omega} & 0 & 0 \\ 0 &
d_{1,\omega} m & 0 & -i \omega \tilde{d}_{1,\omega} & 0 & 0 & 0 & 0 & 0 \\
\bar{\chi}^{+-} _{{2{\mathbf Q},\omega}} & 0 & 0 & 0 & 0 & 0 & 0 & 0 & 0 \\
\hdottedline 0 & 0 & 0 & 0 & 0 & h \tilde{d}_{1,\omega} & 0 & i
\tilde{d}_{1,\omega} \omega & 0 \\ 0 & 0 & i \omega \tilde{d}_{2,\omega} & 0
& h \tilde{d}_{2,\omega} & 0 & -i \omega \tilde{d}_{2,\omega} & 0 & 0 \\ 0 &
-i \omega \tilde{d}_{1,\omega} & 0 & h \tilde{d}_{1,\omega} & 0 & 0 & 0 & 0
& 0 \\ \hdottedline 0 & 0 & 0 & 0 & 0 & 0 & 0 & 0 & \bar{\chi}^{+-}
_{2{\mathbf Q},\omega} \\ 0 & 0 & 0 & 0 & 0 & i \omega \tilde{d}_{1,\omega}
& 0 & d_{1,\omega} m & 0 \\ 0 & 0 & \bar{\chi}^{zz}_{{\mathbf 0},\omega}-m
d_{2,\omega} & 0 & -i \omega \tilde{d}_{2,\omega} & 0 &
\bar{\chi}^{zz}_{{\mathbf 0},\omega}+m d_{2,\omega} & 0 & 0
\label{global_chi_structure} \end{NiceArray}\right)_{+,y,-}. 
\end{equation}%
\end{widetext}
As one can see from the limit $\omega \rightarrow 0$ followed by $%
h\rightarrow 0$, the uniform static transverse susceptibilities are given by 
$\chi _{\mathbf{q}=0,\omega \rightarrow 0}^{+-}=2\chi _{\mathbf{q}=0,\omega
\rightarrow 0}^{xx}=2\chi _{\mathbf{q}=0,\omega \rightarrow 0}^{zz}=\chi
_{1,0}$ and $\chi _{\mathbf{q}=0,\omega \rightarrow 0}^{yy}=\chi _{2,0}$, while similarly to the commensurate case the uniform susceptibilities vanish for $h\rightarrow 0$ at finite $\omega$.

Performing the inversion of the susceptibilities in $S^x, S^y, S^z$ basis, we obtain 
\begin{align}
\bar{\kappa}^{ab}_{{\mathbf{0}},\omega}&=\frac{1}{m} \left( 
\begin{array}{ccc}
h & -\omega & 0 \\ 
\omega & {m}{\chi^{-1} _{2,\omega }} & 0 \\ 
0 & 0 & {m}(\bar{\chi}^{zz}_{{\mathbf{0}},\omega})^{-1} 
\end{array}
\right)_{x,y,z}, \label{kappaIncomm}\\
%\end{align}
%\vspace{-0.5cm}
%\begin{align}
{\bar{\kappa}}_{-{\mathbf{Q}},\omega }^{ab}&=%
% \left( 
% \begin{array}{ccc}
% p^+_{\omega} \bar{\kappa}_{-\mathbf{Q},\omega}^{+-}  & 0  & \bar{\kappa}_{-\mathbf{Q},\omega}^{+-}  \\ 
% \frac{2i\omega }{m} & \frac{h}{m} & 0  \\ 
% {4}{\chi _{1,\omega }^{-1}} + p^+_{\omega} p^-_{\omega} \bar{\kappa}_{-\mathbf{Q},\omega}^{+-} & \frac{2i\omega }{m} & % p^-_{\omega} \bar{\kappa}_{-\mathbf{Q},\omega}^{+-} 
% \end{array}%
% \right) _{+,y,-}.
\frac{1}{m}
\left( 
\begin{array}{ccc}
-{\I} {d_\omega ^{+-}\bar{\chi}^{+y}_{-\mathbf{Q},\omega}}  & 0  & \omega d^{+-}_\omega {\tilde{d}_{1,\omega} }  \\ 
2{\I}\omega & {h} & 0  \\ -{\omega}
\bar{\chi}_\omega^2 d^{+-}_\omega & 2{\I}\omega & -{\I} {d^{+-}_\omega \bar{\chi}^{y-}_{-\mathbf{Q},\omega}}
\end{array}%
\right) _{+,y,-},
\notag
\end{align}
where $\bar{\chi}^2_\omega=(\bar{\chi}^{+y}_{-\mathbf{Q},\omega} \bar{\chi}^{y-}_{-\mathbf{Q},\omega} - \bar{\chi}^{yy}_{-\mathbf{Q},\omega} \bar{\chi}^{+-}_{-\mathbf{Q},\omega})/m$,
$d^{+-}_\omega=
%4 m \omega /(\bar{\chi}^{+-}_{-\mathbf{Q},\omega} \omega^2 \tilde{d}_{1,\omega} + h \bar{\chi}^{+y}_{-\mathbf{Q},\omega} \bar{\chi}^{y-}_{-\mathbf{Q},\omega})=
{
4\omega/(\bar{\chi}^{+-}_{-\mathbf{Q},\omega} + h \bar{\chi}^2_\omega)
}.$

{The entire discussion of Sect. \ref{SectComm} concerning the frequency
dependence of the off diagonal components in the commensurate case is
applicable to the incommensurate case, except that the $\bar{\kappa}_{\mathbf{0},\omega}^{zz}$ and $\bar{\kappa}^{\pm\pm}_{\mathbf{Q},\omega}$
components acquire their own dynamics in the latter case.}
%\vspace{-1cm}
\subsubsection{Momentum dependence of susceptibilities near $q={\mathbf{0}},{%
\mathbf{Q}}$}

To obtain susceptibilities at the momenta close to ${\mathbf q}={\mathbf{0}}$ we consider leading contributions to the momentum dependence of the inverse susceptibility matrix (including charge components), allowed by symmetry \cite{Bonetti}, 
%For small momenta ${\mathbf q}$ we obtain
\begin{align}
\bar{\bar{\kappa}}&_{{\mathbf{q}},\omega }^{ab}=m^{-1}\label{kq0}\\
&\times\left( 
\begin{array}{cccc}
{h}+A_{nl}q_{n}q_{l} & \omega  & C_{n}q_{n} & D_{n}q_{n} \\ 
-\omega  & {m}{\chi _{2,\omega }^{-1}} & 0 & 0 \\ 
-C_{n}q_{n} & 0 & {d^{0z}}{\bar{\chi}_{0}^{00}} & -{d^{0z}}{\bar{\chi}%
_{0}^{0z}} \\ 
-D_{n}q_{n} & 0 & -{d^{0z}}{\bar{\chi}_{0}^{0z}} & {d^{0z}}{\bar{\chi}_{0}^{{%
zz}}} 
\end{array}%
\right) _{x,y,z,0},\notag
\end{align}%
where $d^{0z}=m/(\bar{\chi}_{0}^{00}\bar{\chi}_{0}^{{zz}}-(\bar{\chi}%
_{0}^{0z})^{2})$, such that the inversion of Eq. (\ref{kq0}) at ${\mathbf{q}}=0$
yields in the spin sector the result (\ref{chiIncomm0}), double overline
stands for the quantities in the local reference frame which include charge
component. For momenta near $\pm{\mathbf Q}$ we neglect the coupling of charge and spin components, which appears to be small numerically, and obtain
\begin{align}
{\bar{\kappa}}&_{-{\mathbf{Q}+\mathbf{q}},\omega }^{ab}=m^{-1}\label{kq1}\\
&\times\left( 
\begin{array}{ccc}
-{\I} {d_\omega ^{+-}\bar{\chi}^{+y}_{-\mathbf{Q},\omega}}  & 0  & \omega d^{+-}_\omega {\tilde{d}_{1,\omega} }  \\ 
{2\I\omega } & {h+B_{nl}q_{n}q_{l}} & 0  \\ 
-{\omega}
\chi_\omega^2 d^{+-}_\omega & 2{\I}\omega & -{\I} {d^{+-}_\omega \bar{\chi}^{y-}_{-\mathbf{Q},\omega}} 
\end{array}%
\right) _{+,y,-}.
\notag
\end{align}

We further apply Ward identities (\ref{Wnu0}) and (\ref{Gamma_Ward2}) to determine the coeffiecients $A_{nl},B_{nl},C_n,D_n$ (see Appendix
\ref{AppSusc1}). 
%The double bar in Eq. (\ref{kq0}) stands for the matrix of inverse susceptibility, which includes the charge component, 
The resulting susceptibilities 
written in the global reference frame read 
\begin{align}
&\chi _{{\mathbf{q}}{\pm {\mathbf{Q}}},{{\mathbf{q}}\pm {\mathbf{Q}}},\omega
}^{xx,zz}=\frac{1}{4}\left( \bar{\chi}^{xx}_{\mathbf{q},\omega}+\bar{\chi}_{{\mathbf{0}},\omega }^{zz}+\bar{\chi}%
_{2{\mathbf{Q}},\omega }^{+-}\right).\label{chiQQQinc}
\end{align}
The respective susceptibilities in the local reference frame
\begin{align}
\bar{\chi}^{xx}_{\mathbf{q},\omega}&=\frac{m^{2}}{hm+\chi _{2,\omega }\omega
^{2}+\rho _{2,m}q_{m}^{2}},\label{chiQQQinc1}\\
\chi _{{\pm {\mathbf{Q}}},{\pm {%
\mathbf{Q}}},\omega }^{yy}&=\bar{\chi} _{{\pm {\mathbf{Q}}},{\pm {%
\mathbf{Q}}},\omega }^{yy}=\frac{m^{2}}{hm+\chi _{1,\omega }\omega ^{2}+\rho
_{1,m}q_{m}^{2}},\label{chiQQQinc2}  
\end{align}%
where $\rho _{1,n}=2M_{q\rightarrow 0;nn}^{xx,zz}$ and $\rho
_{2,n}=M_{q\rightarrow 0;nn}^{yy}+\Delta \rho _{2,n}$ are the
out-of-plane and in-plane spatial spin stiffnesses (we consider the $q_{x,y}$
coordinates which diagonalize the current correlators $M_{nl}^{0;aa}$ over
the spatial indexes $n,l$), $\Delta \rho_{2,n}$ is the contribution to the in-plane spatial stiffness, 
\begin{align}
\Delta \rho _{2,n}&=\frac{m^{2}}{2}\lim_{q\rightarrow 0}\partial
_{q_{n}}^{2}\left( \frac{1}{\bar{\chi}_{\mathbf{q},0}^{xx}}-\bar{\kappa}_{%
\mathbf{q},0}^{xx}\right)\notag\\
&=\frac{1}{\bar{\chi}_{0}^{zz}}{(D_{n}\bar{\chi}_{0}^{0z}+C_{n}\bar{\chi}_{0}^{zz})^{2}},
\label{Drho}
\end{align}%
which explicit form is given by the Eq. (\ref{Drho1}) below and which originates from the coupling of $x$ component of the susceptibility with $0,z$ components at finite momenta.  The factor of $1/4$ in Eq. (\ref{chiQQQinc})
occurs due to passing to global reference frame and expresses the fact that
one Goldstone mode of $\bar{\chi}_{{\mathbf{q}}=0,0}^{xx}$ in the local
reference frame is split into four modes of $\chi _{\pm {\mathbf{Q}}%
,0}^{xx,zz}$ in the global reference frame. %\begin{equation}
%\Delta \rho_{2,n}=\frac{\left\{  {\phi_{0,n0}^{yz}} \left[({\bar{\chi}_0^{zz}})^{-1}-{\kappa_0^ {z0}}
%   {\bar{\chi}_0^{00}}{\kappa_0^ {0z}}\right]+U \left({\phi_{0,n0}^{yz}}+{\phi_{0,n0}^{y0}} {\bar{\chi}_0^{00}}{\kappa_0^{0z}}\right)\right\}^2}{ ({\bar{\chi}_0^{zz}})^{-1}-\kappa_0^{z0} {\bar{\chi}_0^{00}}\kappa_0^{0z}}
%   \label{rxz}
%\end{equation}

Although the results (\ref{chiQQQ}), 
(\ref{chiQQQinc1}), and (\ref{chiQQQinc2}) formally coincide with those of Refs. \onlinecite{Bonetti,BonettiMetzner,BonettiThesis}, there are
two important differences: the temporal stiffnesses $\chi_{1,2,\omega}$ are
in general frequency dependent in our approach, since their frequency
dependence is not fixed by Ward identities, and the spatial stiffnesses $%
\rho_{1,2}$ are expressed via the respective gauge kernels, including the
correction term in Eq. (\ref{most_important}) and the contribution of the $z$ mode $\Delta \rho_{2,n}$%
.

On approaching the spin symmetric phase of the chargon sector the spin symmetry in
this sector tends to be restored, and the susceptibility $\bar{\chi}_{{\mathbf{0}}%
,\omega}^{zz}=m d_{2,\omega}$ is also divergent. Apart from
that, near the paramagnetic phase the Goldstone mode at ${\mathbf{q}}=0$ in the local reference frame is
mirrored to the wave vector $2{\mathbf{Q}}$, such that $\bar{\chi}^{+-} _{2\mathbf{Q}}=2 m d_{2,\omega}$. Adding the respective momentum
dependencies we find in this limit the in-plane susceptibilities in the global reference frame
\begin{equation}
\chi^{xx,zz}_{{\mathbf{q}}{\pm {\mathbf{Q}}},{{\mathbf{q}}\pm {\mathbf{Q}}}%
,\omega}=\frac{m^2}{h m+\chi_{2,\omega} \omega^2 +\rho_{2,m} q_m^2},
\label{chiQQQpara}
\end{equation}
which have the same form as the out-of-plane susceptibilities (\ref{chiQQQinc2}); the number of Goldstone modes in local and global reference frame
coincides in this case.

\subsection{Explicit expressions for spin stiffnesses}

\label{Explicit}

%--Start
In the following we represent the modified gauge kernel as $M_{q;\mu\rho}={K}%
_{q;\mu\rho}-{K}^{C}_{q;\mu\rho}$, where the two contributions correspond to
the first and second terms in the Eq. (\ref{most_important}). Considering again the local reference frame, we represent the local susceptibility $\bar{\chi}_q$ and the current-spin kernels $\tilde{K}_{q;0\rho}$ and $\tilde{K}_{q;\mu 0}$ (the tilde corresponds to transforming spin, but not the current variables to the local reference frame) via their $U$-irreducible counterparts $\bar{\phi}_q$, $\tilde{\phi}_{q,\mu},\tilde{\phi}^{\mathrm{t}}_{q,\mu}$ by the relations
\begin{align}
\bar{\chi}_q &= ( 1 - \bar{\phi}_q \hat{U} )^{-1} \bar{\phi}_{q}=\bar{\phi}_{q} ( 1 - \hat{U}
\bar{\phi}_q )^{-1},  \label{chiU} \\
\tilde{K}_{q;0\rho} &= ( 1 - \bar{\phi}_q \hat{U} )^{-1} \tilde{\phi}_{q,\rho},\,\,\, 
\tilde{K}%
_{q;\mu 0} = \tilde{\phi}^{\mathrm{t}}_{q,\mu} ( 1 - \hat{U} \bar{\phi}_q )^{-1},\notag
%\label{K0r}
\end{align}
which are similar to those used previously for the dynamic spin susceptibility \cite%
{EdwHertz,AKLambda,My_BS,MyEDMFT,OurFirst}, $\hat{U}= 2U\, \mathrm{diag}(1, 1, 1, -1)$. Accordingly, the
kernel $K_q$ can be split into the $U$-irreducible paramagnetic part $K^{\rm irr}_q$%
, the diamagnetic part $K^{d}_q$, and the $U$-reducible ladder $K^L_q$ part,
\begin{equation}
K_{q;\mu\rho}=K_{q;\mu\rho}^{\rm irr}+K_{q;\mu\rho}^{d} +{K}^{L}_{q;\mu\rho},
%- {K}%
%^{C}_{q;\mu\rho}.
\label{Ksplit}
\end{equation}
where 
\begin{align}
{K}^{L}_{q,\mu\rho} &=\tilde{\phi}^{\mathrm{t}}_{ q,\mu} \left( 1 - \hat{U}
\bar{\phi}_q \right)^{-1} \hat{U} \tilde{\phi}_{ q,\rho}. \label{ladder_current_current}
\end{align}
The explicit form of the $U$-irreducible parts $K^{\rm irr}$ and $K^d$ is
specified below, in Sect. \ref{Numer}. This splitting is itself exact and does not rely on some approximation. The correction term is represented as 
\begin{equation}
{K}^{C}_{q;\mu\rho}=\tilde{\phi}^{\mathrm{t}}_{q,\mu} \left( 1 - \hat{U} \bar{\phi}_q
\right)^{-1} \bar{\kappa}_q \left( 1 - \bar{\phi}_q \hat{U} \right)^{-1}
\tilde{\phi}_{q,\rho},  \label{KC}
\end{equation}
$\bar{\kappa}_q$ is considered as $4\times4$ matrix with zero charge
components. 
%Picking out contributions, which are
%reducible with respect to the bare interaction $U$ according to Eqs. (\ref%
%{chiU}), (\ref{K0r}), we obtain explicit (formally exact) expressions for
%the ladder and correction contributions (all quantities are matrices here
%with respect to spin indices) 

 % Diamagnetic contribution is defined as
%\begin{equation}
%    K^{d}_{mn} = \frac{1}{2}(1-\delta_{\mu,0})(1-\delta_{\rho,0})\sum_{{k}} {\rm Tr}\left\{ \left[(t^{mn}_{\mathbf{k}+{\mathbf Q}/2}+t^{mn}_{\mathbf{k}-{\mathbf Q}/2})\sigma^{0}+(t^{mn}_{\mathbf{k}+{\mathbf Q}/2}-t^{mn}_{\mathbf{k}-{\mathbf Q}/2})\sigma^{y}\right]G_{\mathbf{k}%
%}\right\}  \comAK{??}, 
%\end{equation}
%where $t^{mn}_{\mathbf{k}} = {\partial^2 \epsilon_{\mathbf{k}}}/({\partial k_{m} \partial k_{n}})$, the trace is taken with respect to the spin indexes.
%,and presents only for spacial components $\rho \neq 0$ and $\mu \neq 0$. 

%--End

%Using Eq. (\ref{K0r}) we obtain the ladder and the correction terms in the local reference frame
%As it is discussed in Sect. \ref{GammaSect}, this term appears because of the dependence of the external source $J[A]$ at fixed Legendre parameter $\phi$. Note that within the derivation of Ward identities the spin components of the susceptibility are inverted separately from the charge part, although they are coupled in ladder summation. 
%It can be easily shown that
%\begin{equation}
%\bar{K}^{L}_{q;\mu\rho}
%- \bar{K}^{C}_{q;\mu\rho}
%    =   - \phi^{\rm t}_{q,\mu} %\phi^{-1}_q \phi_{q,\rho}
%    \label{Kcorr}
%\end{equation}

Using explicit form of the inverse susceptibilities (\ref{kq0}) and (\ref{kq1}), evaluating the respective irreducible susceptibilities $\overline{\phi}_q$, one can find that the correction term (\ref{KC}) removes all singular terms,
originating from Goldstone modes of the chargon susceptibility, contained in 
${K}^{L}_{q,\mu\rho}$ (which occur from the inverse matrix in Eq. (\ref%
{ladder_current_current})). These contributions were previously removed in Refs. \onlinecite%
{Bonetti,BonettiMetzner,BonettiThesis} on the basis of their vanishing in
the limit ${\mathbf q}\rightarrow 0$ at $\omega=0$. This is however not fully correct since in the
absence of external staggered field smallness of the kernels $\tilde{\phi}^{\mathrm{t}}_{ q,\mu}$ and $\tilde{\phi}_{ q,\rho}$, proportional
to ${\mathbf q}$ in this limit, is in fact \textit{compensated} by smallness of the
denominator in Eq. (\ref{ladder_current_current}), originating from the presence of Goldstone modes and yields the
contribution, which is in general \textit{finite}. Omitting these
contributions implies either implicit introduction of infinitesimally small
staggered field or fixing the Coulomb gauge $q_m A_{m}=0$ in the nonlinear
sigma model approach. Because of the non-abelian form of the gauge field, the latter gauge fixing is expected however to produce non-trivial ghost fields contributions. 

In our approach this potentially singular contribution is compensated by the $K_{q,nn}^C$ term. 
%For the in-plane mode it is convenient to combine this singular contribution with finite contribution $\Delta \rho_{2,n}$. 
Using explicit form of the irreducible bubble matrices $\phi_q$, determined from the Eq. (\ref{kq0}), and the explicit form of the correction
\begin{equation}
\Delta \rho _{2,n}=\frac{1}{\bar{\chi}_{0}^{zz}}\left[ {%
\tilde{\phi} _{0,n0}^{yz}}+2U({\tilde{\phi} _{0,n0}^{yz}}\bar{\chi}_{0}^{zz}-{\tilde{\phi}
_{0,n0}^{y0}}\bar{\chi}_{0}^{0z})\right] ^{2},
\label{Drho1}
\end{equation}%
%in Eq. (\ref{Drho}), 
%we find $\Delta \rho_{2,n}-K^C_{q,nn} \propto {\mathbf q}^2$ at $\omega=0$ and
%compensates the respective contribution in $K^L$. 
we find that the sum $K^L-K^C+\Delta \rho$ is finite and does not depend explicitly on momentum and staggered magnetic field $h$ in the limit $%
{\mathbf q}\rightarrow 0$, due to cancellation of the terms $O({\mathbf q}^2)$ and $O(h)$. Therefore, the
numerical analysis below can be performed in zero external magnetic field. Keeping
only ladder terms produces incorrect results in this case. Summing all
contributions we finally obtain the respective spatial spin stiffnesses 
\begin{align}
\rho_{1,n}&=2\rho_n=2\left({K}_{0,nn}^{{\rm irr},xx}+{K}_{0,nn}^{d,xx}\right), \\
\rho_{2,n}&={K}_{0,nn}^{{\rm irr},yy}+{K}_{0,nn}^{d,yy}+2U \left[{\tilde{\phi}_{0,0n}^{%
\mathrm{t},y0} (2 U {\bar{\chi}^{00}_0}-1) \tilde{\phi}_{0,n0}^{0y}}\right.\notag\\
&\left.-4 U {%
\tilde{\phi}_{0,0n}^{\mathrm{t},y0}} {\bar{\chi}_0^ {0z}}{\tilde{\phi}_{0,n0}^{zy}}+{%
\tilde{\phi}_{0,0n}^{\mathrm{t},yz}} (2U {\bar{\chi}_0^{zz}}+1)\tilde{\phi}_{0,n0}^{zy}%
\right].\label{r2n}
\end{align}
This result can be obtained alternatively by differentiating the irreducible
susceptibilities over momenta following Refs. \onlinecite{StiffLand,Bonetti,BonettiThesis}
and using the respective Ward identities for spin-current vertices, which follow from Eq. (\ref{Wnu0})  (see Appendix \ref{AppSusc1}). {In case of static mean field theory the obtained result coincides 
%and formally coincides (with modified bubble contributions, which contain dynamic vertex corrections in our case) 
with that proposed 
%for static mean field theory 
in Refs. \onlinecite{Bonetti,BonettiThesis,BonettiMetzner,StiffLand}}. At the same time, {it generalizes previous consideration to include dynamic effects beyond static mean-field theory}. The
cancellation of momentum $q$ and external staggered field dependent terms also shows independence of the spin stiffnesses on the gauge fixing conditions. 
%compensation is however not complete, and the remaining contribution in the right hand side of Eq. (\ref{Kcorr}) appears to be {\it finite} in the incommensurate case, as we show below.
\vspace{-0.1cm}
\section{Dynamical mean-field theory approach}

\label{Numer}

To apply above derived identities for a particular system with long-range
magnetic order we consider a two-dimensional one-band Hubbard model on a
square lattice (\ref{H}) with hopping $t_{ij}=t$ between nearest neighbors
(which is used as a unit of energy) and $t_{ij}=-t^{\prime }$ for
next-nearest neighbors. 
As it is discussed in Refs. \onlinecite{SDW3,SDW4,SDWOur,SBIncomm}, the
incommensurate magnetically ordered states in the hole doped Hubbard model
are thermodynamically unstable within the mean-field approach, which yields
a phase separation \cite{SDWOur,SBIncomm} of incommensurate magnetic order
into domains with incommensurate and commensurate magnetic states. Therefore for detail analysis of spin stiffnesses in a broad doping range, we consider
below DMFT approach \cite{DMFT}, extended to describe incommensurate spin spiral order in the chargon
sector. 

The respective approach for the doped Hubbard model was developed
previously in Refs. \onlinecite%
{DMFT_Incomm_Licht,DMFT_Incomm,DMFT_Incomm1,OurFirst,BonettiThesis}. The main difference of dynamic from static mean field theory is in
the frequency dependence of the self-energy and interaction vertices, see
the discussion in Refs. \onlinecite{OurFirst,ToschiAF}. Further we use the approach of Ref. \onlinecite{OurFirst}, which considers
passing to local reference frame with the order parameter aligned along the $%
z$ axis. As we showed in our previous work \cite{OurFirst}
for different doping levels $x=1-n$ ($n$ is the concentration of electrons)
both commensurate and incommensurate (spiral) long-range magnetic orders
present in DMFT solution of this model. This feature makes this model
well-suited for the study of low-energy magnetic excitations in
strongly-correlated quasi two-dimensional electronic system. All
calculations in the following study were performed for $t^{\prime }/t=0.15$, 
$U=7.5t$ and the temperature $T= 0.1t$.

\subsection{Gauge kernels in the local reference frame}

The magnetic susceptibilities and gauge kernels are represented in DMFT as sum of the ladder diagrams with  local particle-hole irreducible vertices \cite{DMFT,OurRev,OurFirst}. For practical calculations we evaluate the components of the gauge kernel in the local reference frame. The correspondence between the paramagnetic parts of the
gauge kernels in the global and local reference frame is given by (see Appendix \ref{TrKernel})
\begin{align}
K_{q,\mu\nu}^{yy} &= \bar{K}_{q,\mu \nu,++}^{yy} +\bar{K}_{q,\mu \nu,--}^{00} +%
\bar{K}_{q,\mu \nu,+-}^{y0} +\bar{K}_{q,\mu \nu,-+}^{0y},  \notag \\
K_{q,\mu\nu}^{xx} &= \sum\limits_{s=\pm} \left[ \bar{K}_{q,\mu
\nu,ss}^{xx} -i s (\bar{K}_{q,\mu \nu,ss}^{xz} -\bar{K}_{q,\mu
\nu,ss}^{zx}) \right.\notag\\
&\left.+\bar{K}_{q,\mu \nu,ss}^{zz}\right].   \label{Kxx}
\end{align}
We further split each term in Eqs. (\ref{Kxx}) according to the representation (\ref{Ksplit}).
 The $U$-irreducible part of the kernels in the local coordinate frame can be then represented as the sum of the bare bubble and the $U$-irredicible vertex correction 
\begin{align}
%\bar{K}^{\mathrm{p}} &= \bar{K}^{b,\mathrm{p}}_{q,\mu\rho} + \bar{K}%
%^{L}_{q,\mu\rho},   %\label{current_current_computational_equation} \\
\bar{K}^{\rm irr}_{q,\mu\rho,ss'}&=\sum_\nu \bar{K}^{0}_{q;\mu\rho,ss'}(\nu) + \bar{K}%
^\phi_{q;\mu\rho,ss'}.
\end{align}
The bare current-current bubble $\bar{K}^0$ in the local
reference frame reads 
\begin{align}
\bar{K}^{0;\alpha \beta}_{q;\mu\rho,ss'}(\nu) = &\sum_{\mathbf{k}} T^{\mu,s}_{{%
\mathbf{k}},{\mathbf{q}}} \mathrm{Tr} \left[\sigma^{\alpha} G_{k}
\sigma^{\beta} G_{k+q} \right] T^{\rho,s'}_{{\mathbf{k}},{\mathbf{q}}},
\label{K0}
\end{align}
where the trace is taken with respect to spin variables, $G_k^{\sigma\sigma^{\prime
}}$ is the electron Green's function in the local reference frame,
\begin{align}
T^{\mu \pm}_{{\mathbf k},{\mathbf q}} &= (T^\mu_{{\mathbf k}-{\mathbf Q}/2,{\mathbf q}} \pm T^\mu_{{\mathbf k}+{\mathbf Q}/2,{\mathbf q}})/{2}\,\,\,(\alpha,\beta=0,y),\notag\\
T^{\mu \pm}_{{\mathbf k},{\mathbf q}} &= T^\mu_{{\mathbf k}+s{\mathbf Q}/2,{\mathbf q}}\,\,\,(\alpha,\beta=x,z),
\end{align}
$T^\mu_{{\mathbf{k}},{%
\mathbf{q}}}=({t_{\mathbf{k}}^{\mu} + t^{\mu}_{\mathbf{k} + \mathbf{q}}})/2$%
,  are the current vertices, $t^{m}_{\mathbf{k}} = {\partial}\epsilon_{%
\mathbf{k}}/{\partial k^{m}}$, $t^{0}_k\equiv \I$. The same equations (\ref{Kxx}) with $T^{0}_{{\mathbf{k}},%
{\mathbf{q}}} = T^{0,+}_{{\mathbf{k}},{\mathbf{q}}} = \I$, $%
T^{0,-}_{{\mathbf{k}},{\mathbf{q}}} = 0$ are applicable for spin
susceptibilities. The ladder and correction terms in the local reference frame (including transforming the current variables to this reference frame) are given by the Eqs. (\ref{ladder_current_current}) and (\ref{KC}) with the replacement $\tilde{\phi}\rightarrow \bar{\phi}$, where we use the particle-hole $U$-irreducible bubbles in the local reference frame
\begin{align}
\bar{\phi}_{q,\rho,s} &= \sum\limits_{\nu} \gamma_{q} (\nu) \bar{K}^0_{q,0\rho,0s}
(\nu),\\ \bar{\phi}^{\mathrm{t}}_{q,\mu,s} &= \sum\limits_{\nu} \bar{K}%
^0_{q,\mu0,s0} (\nu) \gamma^{\mathrm{t}}_{q} (\nu),
\end{align}
$T^{0,0}_{{\mathbf{k}},{%
\mathbf{q}}}\equiv {\I}$, and matrix multiplications are assumed here and in the
following expressions and triangular vertices are defined by 
\begin{align}
\gamma_{q}(\nu) &= \sum\limits_{\nu^{\prime }} \left[ \hat{I} - \bar{K}%
^{0}_{q;00,++} {\tilde{\Phi}}_{\omega} \right]_{\nu^{\prime }\nu} ^{-1},\label{gamma10}\\
\gamma^{\mathrm{t}}_{q}(\nu) &= \sum\limits_{\nu^{\prime }} \left[ \hat{I} - {%
\tilde{\Phi}}_{\omega} \bar{K}^{0}_{q;00,++} \right]_{\nu \nu^{\prime }} ^{-1}.
\label{gamma1} 
\end{align}
In Eqs. (\ref{gamma10}) and (\ref{gamma1}) the inversion is performed with
respect to the frequency and spin indexes (considered as multi index), $\bar{%
K}^{0}_{q;00}$ without frequency argument is considered as the diagonal
matrix with respect to frequencies, ${\tilde{\Phi}}_{\omega}={\hat \Phi}%
_{\omega} - {\hat{U}}$, ${\hat \Phi}_{\omega}$ is the particle-hole
irreducible local vertex, expressed in the spin basis.

The $U$-irreducible vertex correction to the kernel can be represented as 
\begin{equation}
\bar{K}^{\phi}_{q;\mu\rho,ss'}= \sum\limits_{\nu} \bar{K}^0_{q,\mu0,s0} (\nu) 
\widetilde{\gamma}_{q;\rho,s'} (\nu).  \label{Kfi}
\end{equation}
where 
\begin{equation}
\widetilde{\gamma}_{q;\rho,s'} (\nu) = \sum\limits_{\nu^{\prime
},\nu^{\prime \prime }} \left[ \hat{I} - {\tilde{\Phi}}_{\omega} \bar{K}%
^{0}_{q;00} \right]_{\nu \nu^{\prime \prime }} ^{-1} \left[{\tilde{\Phi}}%
_{\omega} \bar{K}^0_{q,0\rho,0s'}\right]_{\nu^{\prime \prime }\nu^{\prime }}.
\label{gamma2}
\end{equation}

The diamagnetic part is expressed through its counterpart in the local
reference frame as 
\begin{equation}
K_{\mu\nu} ^{d} =\bar{K}_{ \mu\nu,+}^{d, 00} +\bar{K}_{ \mu\nu,-}^{d, yy}, 
\label{Kdg}
\end{equation}
where 
\begin{align}
\bar{K}_{ \mu\nu,s}^{d,ab} = -\delta_{ab}(1-\delta_{\mu 0})(1-\delta_{\nu 0})
&\sum_k T^{\mu\nu,s}_{\mathbf{k}} \mathrm{Tr}\left[ \sigma^a G_{k} \right],
\label{Kd}
\end{align}
$T^{\mu \nu \pm}_{{\mathbf{k}}} = (t^{\mu\nu}_{{\mathbf{k}}-{\mathbf{Q}%
}/2} \pm t^{\mu\nu}_{{\mathbf{k}}+{\mathbf{Q}}/2})/{2}$, $t^{mn}_k = {%
\partial}^2\epsilon_{\mathbf{k}}/({\partial k^{m}\partial k^{n}})$.

% This formula was also used before simplification was discovered \comAK{Something is wrong here, since in $\phi_{R q}^{\rho}$ the quantity $\chi^{(0) \rho}_{R q} (\nu)$ stands {\it at the right}, not left. Is that $\phi_{L q}^{\rho}$?}
% \begin{equation}
% \phi_{R q}^{\rho} = \sum\limits_\nu \chi^{(0) \rho}_{R q} (\nu) + % \sum\limits_\rho \chi^{(0)}_{q} (\nu) % \widetilde{\gamma}_{R q}^\rho (\nu)
% \end{equation}
%Obtained correlation function can be transformed to the global coordinate frame according to (\ref{}) to obtain physically observed quantities such that space spin stiffness.

Below we consider the results of the numerical calculations, performed within CT-QMC method, realized in
iQIST package \cite{iQIST}, keeping 160 (positive and negative) fermionic Matsubara frequencies. We
also use corrections on the finiteness of frequency box, see Refs. \onlinecite%
{My_BS,MyEDMFT}. We have verified that for static self-energies and
vertices the mean field results for the susceptibilities are reproduced (see also Appendix \ref%
{SectMF}). 
%and Fig. \ref{afm_stif})}. 
In the DMFT calculations below for simplicity we
consider zero external staggered magnetic field $h = 0$. {The wave vector ${\mathbf Q}$ of magnetic instability in the chargon sector is obtained from the minimum of smallest eigenvalue of $1 - U \phi_{\mathbf{Q},0}$, see Appendix 
\ref{Det}}. %was absent as we 
%susceptibility-like terms with $\left[ 1 - \phi_q U \right]^{-1}$ and $\left[ 1 - U \phi_q \right]^{-1}$. It is natural to trace this cancellation effect back to the two terms: part of the true gauge kernel of the form
%%
%\begin{equation}
%    \phi_{L q}^{\mu} \left[ 1 - U \phi_q \right]^{-1} U \phi_{R q}^{\rho}
%\end{equation}
%
%and its counterpart from the correction
%
%\begin{equation}
%    \phi_{L q}^{\mu} \phi^{-1}_q \left[ 1 - \phi_q U \right]^{-1} \phi_{R q}^{\rho}
%\end{equation}

\subsection{DMFT approach in the antiferromagnetic phase}
\label{DMFTAFM}

\begin{figure}[t]
{ \centering \includegraphics[scale=1.04]{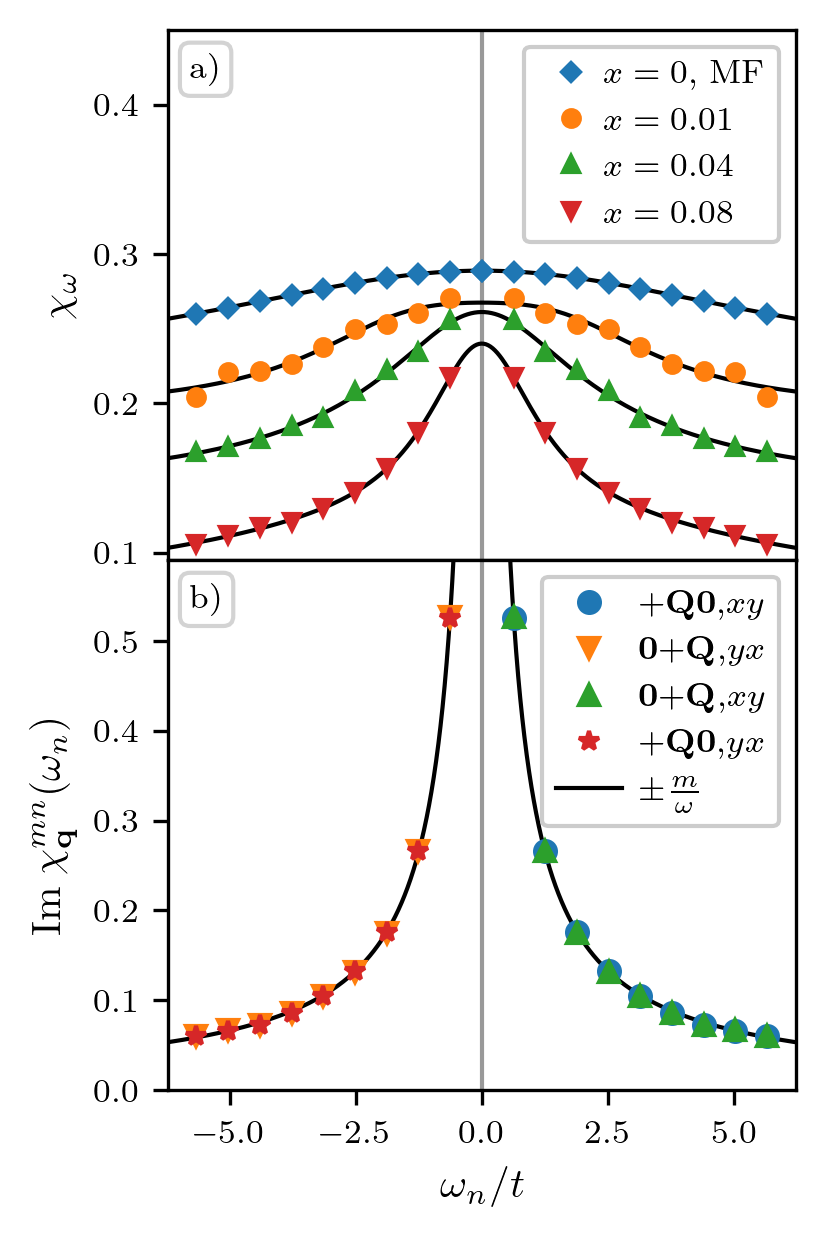} }
\caption{(Color online). a) Frequency dependence of the temporal stiffness  $%
\chi_\omega$ for the AFM case for various fillings in DMFT approach, the result of mean field (MF) approach at half filling is presented for comparison. b)
Susceptibility components for AFM case at $x = 0.05$ compared to the Ward
identities result (\ref{chiQ0AFM}).}
\label{afm_stif}
\end{figure}

\begin{figure}[t]
{\centering \includegraphics[scale=1.0]{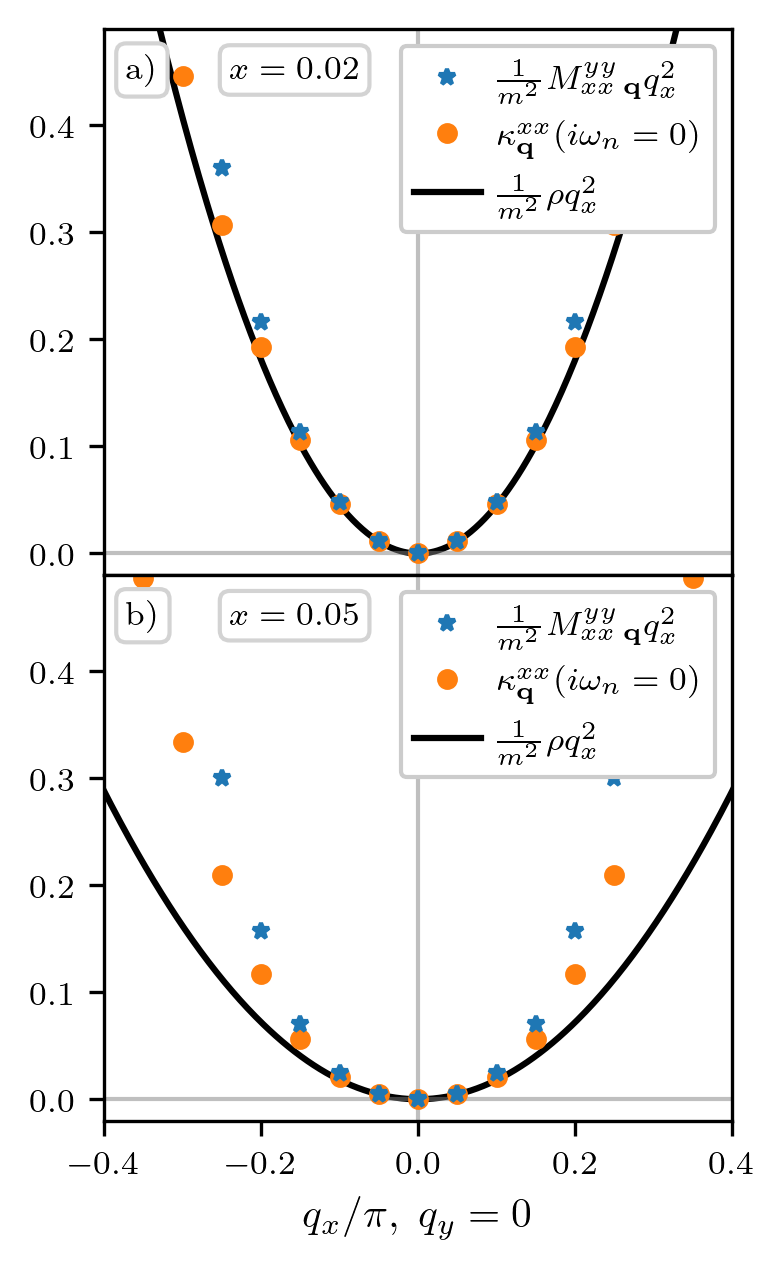}}
\caption{(Color online). Wave vector dependence of the inverse
susceptibility $\protect\kappa$ in the AFM case for hole doping levels $x =
0.02$ (a) and $x = 0.05$ (b). Orange circles correspond to the result of the DMFT calculation, solid lines correspond to the long-wavelength limit $\mathbf{q}
\to 0$, obtained from Ward identities, stars is the respective
dependence, obtained with account of the momentum dependence of the gauge
kernels.}
\label{afm_ward_q_fig}
\end{figure}

We consider first the antiferromagnetic order in the chargon sector, which
occurs at half filling or small doping. We have verified that the dynamic
transverse uniform susceptibility of chargons vanishes at zero staggered
field, as it is required by Ward identities. To determine the (frequency
dependent) temporal stiffness, we use the relation (\ref{chiQQAFM}), $%
\chi_\omega=m^2/(\chi_{\mathbf{Q}}^{yy}\omega^2)$. Since the susceptibility $\chi_{%
\mathbf{Q}}^{yy}$ at finite frequency changes continuously with the staggered field, the
stiffness $\chi_\omega$, obtained in the zero-field calculation, can be also considered as a limit $h\rightarrow 0$ of the
respective uniform susceptibility. 
%{We furthermore determine the uniform transverse susceptibilities $\chi_0$ from the irreducible susceptibility matrix elements $\bar{\phi}_{0}^{xx}=\bar{\phi}_{0}^{yy}=\chi_0/(1+2U\chi_0)$ }. 

If no dynamical effects were present, one would
obtain almost frequency-independent temporal stiffness $\chi$ in the low frequency limit. In DMFT approach the
temporal stiffness $\chi_{\omega}$ becomes essentially dynamic and does not necessarily reduce to a
constant. The calculated frequency dependencies of temporal stiffness $\chi_{\omega}$ for various hole
dopings $x$ are presented in Fig. \ref{afm_stif}(a). Near half-filling $x
\to 0$ the frequency dependence of temporal stiffness becomes less significant and its static limit is
almost equal to the mean-field results for half-filling. With increase of
the doping a peak develops at $\omega = 0$. {The change of static temporal stiffness limit $\chi_{0} $ with hole doping is opposite to its high-frequency limit at ${\omega_n \gg t}$: the latter susceptibility decreases with doping, while the former increases.} As a result, the frequency dependence becomes more pronounced with doping. In Fig. \ref{afm_stif}(b) we present various susceptibility components of chargon sector, computed in the AFM state  at finite hole doping 
$x = 0.05$. One can see that the considered components fulfill the Ward
identities. It is important to emphasize the computed non-diagonal
components are parameter-free and present solely due to the presence of the
long-range magnetic order in the chargon sector.

The momentum dependence of inverse susceptibility, obtained from the
current-current correlation function, is compared to the full momentum
dependence obtained in DMFT in Fig. \ref{afm_ward_q_fig} for two doping
levels. We see that the results match each other up to quadratic terms $%
O(q^2)$ as it follows from the Ward identities. Note that although the
obtained relations are not meant to be correct apart from the quadratic
order $\mathbf{q}^2$, the $\mathbf{q}$-dependent current-current correlation
function gives better agreement with the inverse susceptibility than
constant value of spin stiffness $\rho$ away from half-filling, where
inverse susceptibility is well described by constant stiffness in a wide
range of wave vectors.

In the commensurate case the contributions, containing vertex corrections,
namely the $U$-irreducible term $K^{\phi}$, the ladder contribution $K^L$, the correction term $K^C$ (\ref{KC}),
and the contribution $\Delta \rho_{2,n}$ vanish at $q=0$,
since the bubbles $\bar{K}^{0}_{0;0\rho}(\nu)$ and $\bar{K}^{0}_{0;\mu0}(\nu)
$, together with the respective bubbles $\phi_{0,\rho}(\nu)$ and $\phi^{%
\mathrm{t}}_{0,\mu}(\nu)$ vanish because of the oddness of the function,
which is summed in the Eq. (\ref{K0}) with respect to ${\mathbf{k}}%
\rightarrow {\mathbf{k}+{(\pi,\pi)}}$ (cf. Refs. \onlinecite{Bonetti,BonettiThesis}%
). Therefore, these terms do not contribute to the spin stiffnesses in the
commensurate case. The situation in this respect is the same as for the
optical conductivity in DMFT, where it was argued that the vertex
corrections vanish \cite{Khurana,DMFT}.
%\vspace{-0.5cm}
\subsection{DMFT approach in the incommensurate phase}
\label{DMFTIncomm}

\begin{figure}[t]
{\centering \includegraphics[scale=0.9]{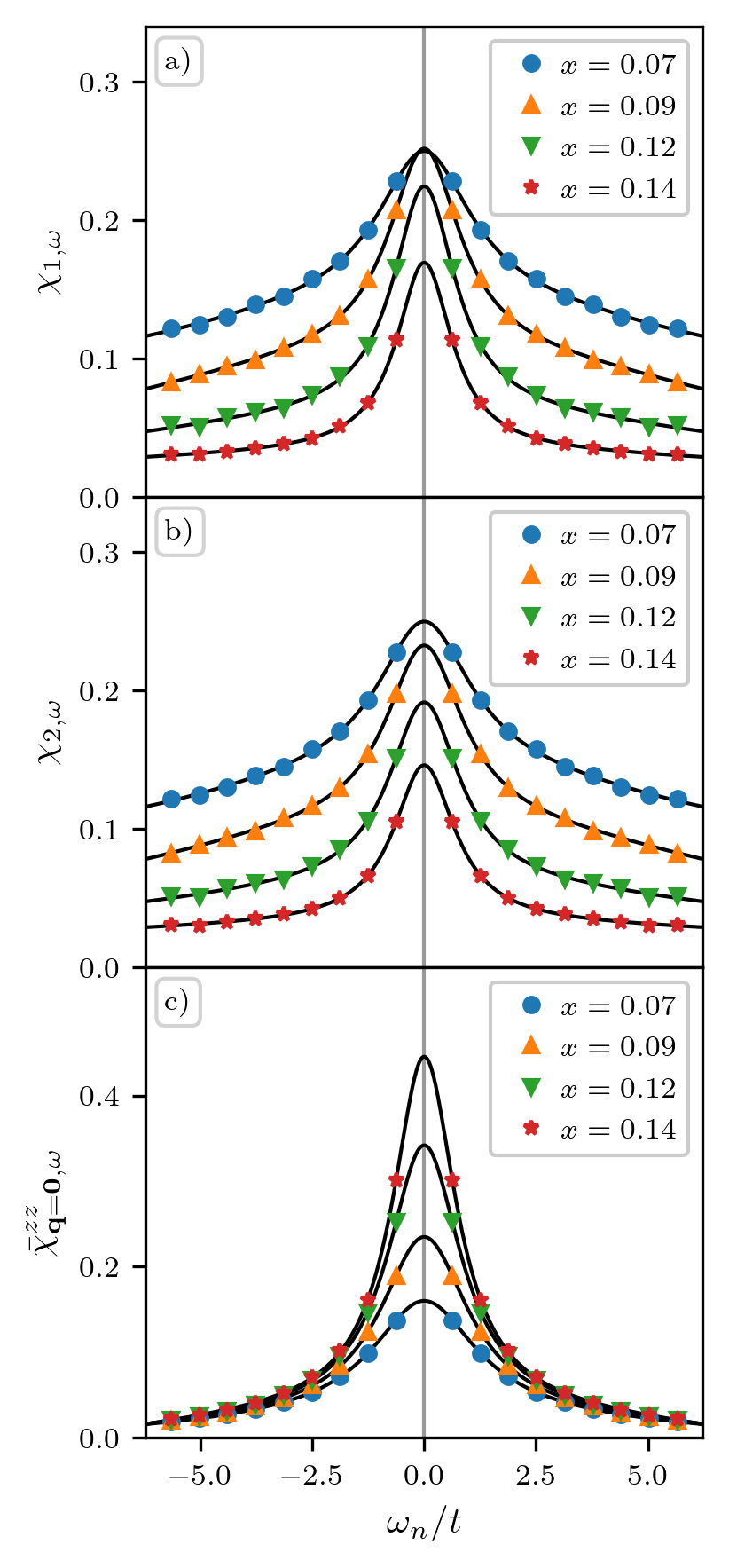} }
\caption{(Color online). The frequency dependencies of the out-of-plane $\protect\chi_{1,\protect%
\omega}$ (a) and in-plane $\protect\chi_{2,\protect\omega}$ (b) temporal stiffnesses in the incommensurate
case, as well as the component of the susceptibility $\bar{\chi}^{zz}_{{\mathbf{0}},\omega}$ (c) for various fillings.}
\label{FigChiQ}
\end{figure}

\begin{figure}[t]
{\centering \includegraphics[scale=1.0]{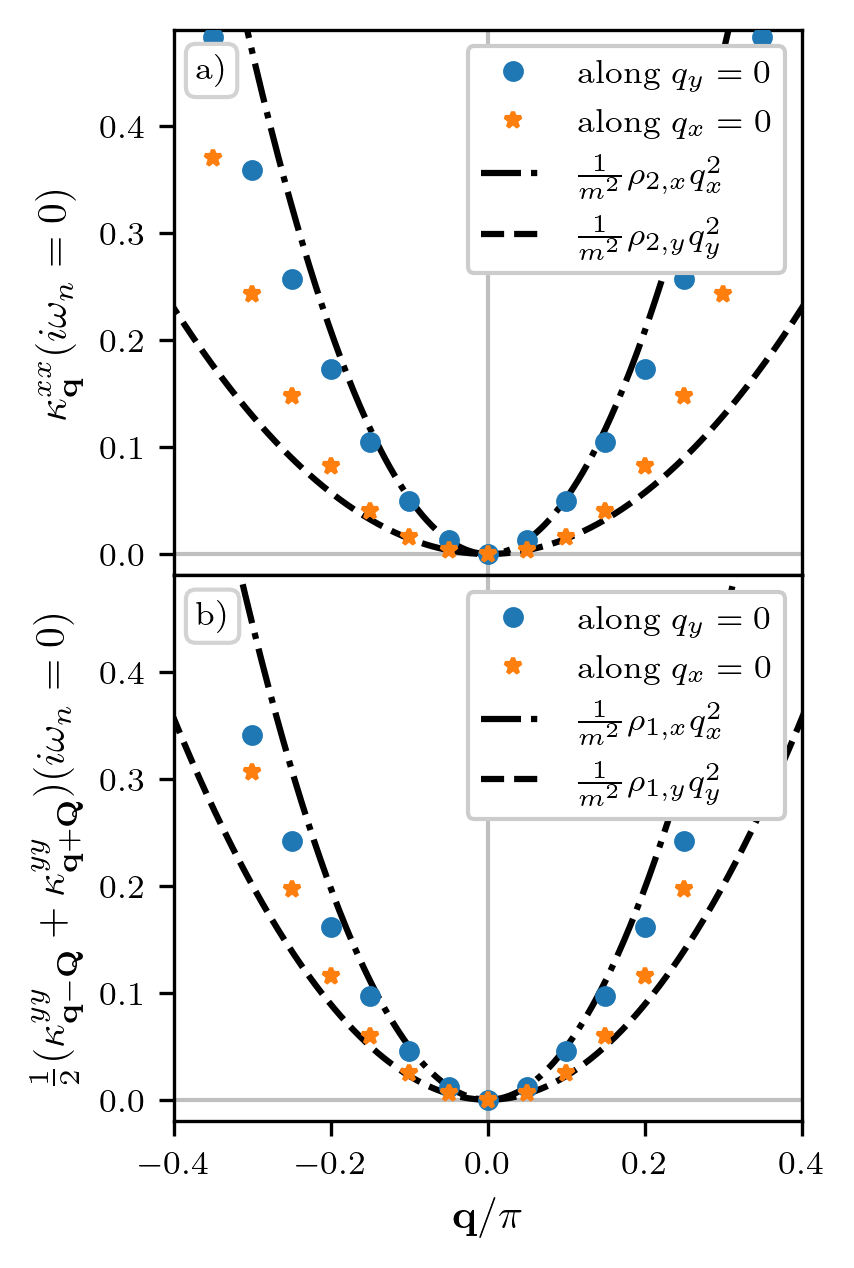}}
\caption{(Color online). Wave vector dependence of the in-plane $\protect%
\bar{\kappa}^{xx}$ (a) and out-of-plane $\bar{\kappa}^{yy}$ (b)
components of the static inverse susceptibility in the
local reference frame in the incommensurate case for hole doping levels $x =
0.13$. Blue circles and orange stars correspond to the DMFT result in
different directions, dash-dotted and dotted lines show the long-wavelength
limit $\mathbf{q} \to 0$, following from the Ward identities.}
\label{chiq_Incomm}
\end{figure}

According to the Eq. (\ref{global_chi_structure}) in the incommensurate case
the frequency dependence of the susceptibilities is characterized by two
temporal stiffnesses $\chi_{1,2,\omega}$ and the uniform dynamic
susceptibility $\bar{\chi}^{zz}_{{\mathbf{0}},\omega}$ in the local
coordinate frame. Using our definition of Eq. (\ref{chiQQQ}) we
extract temporal spin stiffnesses as $\chi_{1,\omega}={m^2}/(\chi_{\mathbf{Q}%
,\omega_n}^{yy}\omega_n^2)$ and $\chi_{2,\omega_n}={2 m^2}/(\chi_{\mathbf{0}%
,\omega_n}^{xx}\omega_n^2)$. 
%{We also determine the uniform static transverse susceptibilities determined from the $U$-irreducible susceptibility matrix elements $\bar{\phi}_0^{yy}=\chi_{2,0}/(1+2U \chi_{2,0})$ and $\bar{\phi}_{\mathbf Q}^{+-}=\chi_{1,0}/(1+U \chi_{1,0})$.} 
%\comIG{ via $\chi_{2,\omega}=1/\bar{\kappa}_{{\mathbf 0},\omega}^{yy}$ and  $\chi_{2,\omega}= \frac{\bar{\kappa}_{-{\mathbf Q},\omega}^{+-}}{\bar{\kappa}_{-{\mathbf Q},\omega}^{+-} \bar{\kappa}_{-{\mathbf Q},\omega}^{-+} - \bar{\kappa}_{-{\mathbf Q},\omega}^{++} \bar{\kappa}_{-{\mathbf Q},\omega}^{--}}$}  
The results of the numerical calculations are
shown in Fig. \ref{FigChiQ}. We find that in the incommensurate phase away
from half filling the frequency dependence of temporal stiffnesses becomes
even more essential than in the AFM phase. With the increase of
doping the temporal stiffnesses $\chi_{1,2}(\omega_n)$ develop a sharp peak at $%
\omega_n = 0$, with different height for longitudinal and transverse
channels. The susceptibility $\bar{\chi}^{zz}_{{\mathbf{0}},\omega}$
(together with $\bar{\chi}^{+-}_{2{\mathbf{Q}},\omega}$ component,
not shown) continuously increases with doping, and its zero frequency
component diverges at the incommensurate magnetic to paramagnetic
transition, which provides a possibility of passing from Eq. (\ref{chiQQQinc}%
) to the Eq. (\ref{chiQQQpara}) and formation of a single soft mode on the
paramagnetic side. 
%However, numerical results show that the limit $\omega_n \to \infty$ is universal for both channels. 
%\begin{figure}[t]
%\includegraphics[scale=0.8]{chi_zz}
%\caption{(Color online). aaa }
%\label{FigChiQ}
%\end{figure}
%, to compute singular behaviour of the susceptibility in the global reference frame in the vicinity of the goldstone poles. 
%These functions specify the structure of the susceptibility in the local reference frame (\ref{global_chi_structure}). 
We have verified that the components of the susceptibility $\bar{\chi}_{\mathbf{0},\omega}$ and $\bar{\chi}_{\pm\mathbf{Q},\omega}$ in the the local reference frame, computed within DMFT, coincide
with the corresponding analytical results (\ref{chiINCOMlocal}), including the off-diagonal contributions to the local susceptibility $\bar{\chi}^{\pm y}$, which have parameter-free analytic form $\pm {m}/{%
\omega}$. 
%In Fig.\ref{FigChiQ}e) we show other  as it follows from the Ward identities.

\begin{figure}[b]
{\centering \includegraphics[scale=1.0]{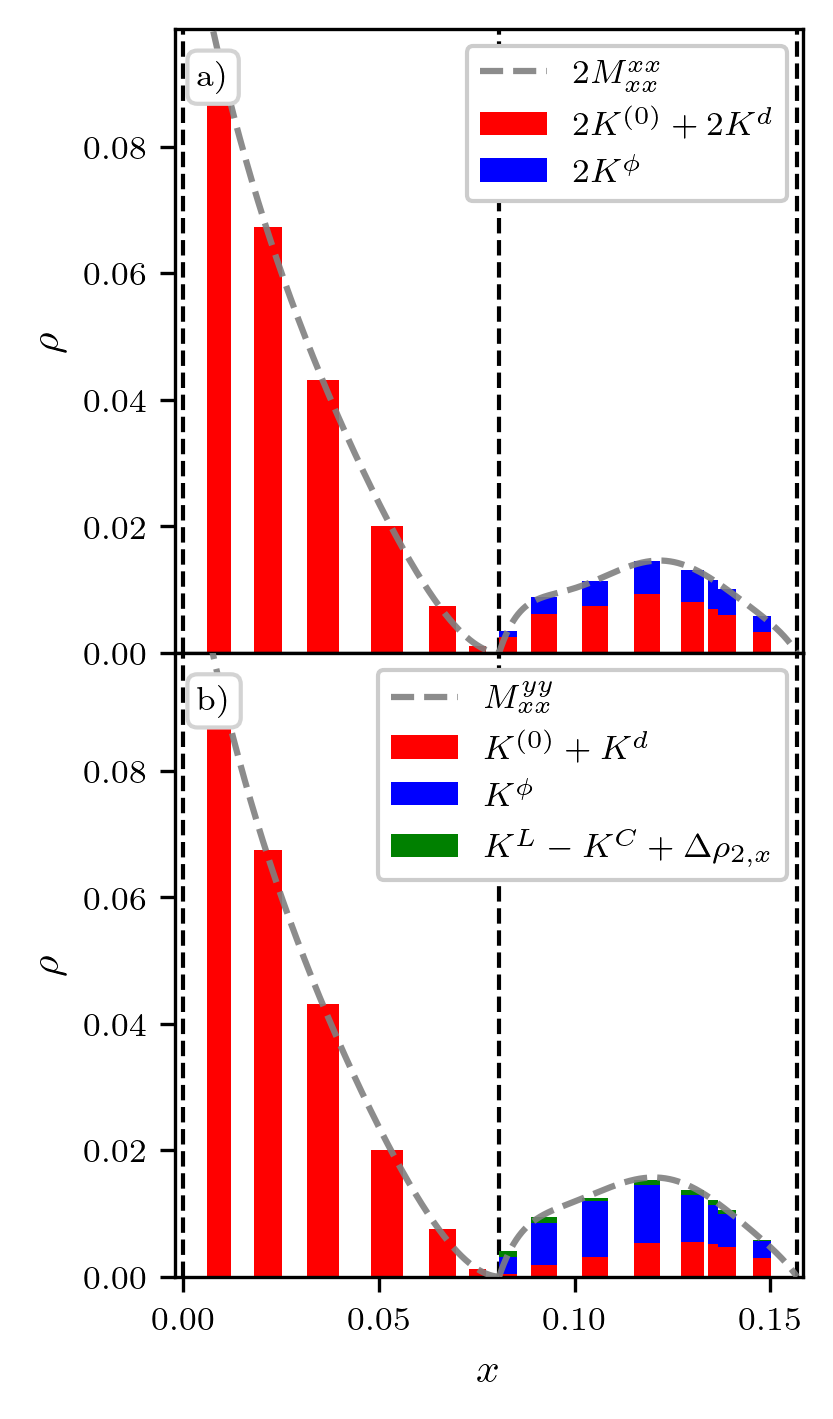}}
% \includegraphics[scale=1.0]{figures/m_xx_xx.png} 
%\includegraphics[scale=0.8]{figures/m_xx_yy.png}
%\caption{(Color online). aaa}
%\end{figure}
%\begin{figure}[h!]
% \includegraphics[scale=1.0]{figures/m_yy_xx.png} 
%\includegraphics[scale=0.8]{figures/m_yy_yy.png}
\caption{(Color online). The bar graph, showing various contributions to the
spatial out-of-plane spin stiffness $\protect\rho_{1,x}$ (a) and the in
plane spin stiffness $\protect\rho_{2,x}$ (b). Red part is the sum of
bare para- and diamagnetic contributions, blue part is the $U$-irreducible vertex correction 
$K^\phi$, the remaining green part, which originates from the $U$-reducible vertex correction $K^L-K^C$, together with the contribution $\Delta\rho_{2,n}$, is negligibly small for the in-plane mode and vanishes for the out of plane mode. Vertical dashed lines mark the commensurate-incommensurate and
incommensurate-paramagnetic transitions.}
\label{Bars}
\end{figure}

%It is important to note that in the case of DMFT where irreducible interaction vertex does not depend of wave vectors we can explicitly show what is the significance of the (\ref{most_important}) correction. We should start from general expression (\ref{current_current_computational_equation}). Explicit expressions after the Fourier transformation are

%\begin{equation}
%    K^{\mu,\nu}_q = K^{(0)\;\mu\nu}_{ \; q} + \widetilde{\phi}^{\mu\nu}_{ \; q} + \phi_{L q}^{\mu} \left[ 1 - U \phi_q \right]^{-1} U \phi_{R q}^{\nu}
%\end{equation}

The momentum dependences of inverse susceptibilities are shown and compared
to the results of Ward identities in Fig. \ref{chiq_Incomm}. It can be seen
that in the incommensurate case two Goldstone modes result in four different
spatial stiffnesses, corresponding to the in-plane and out-of-plane
fluctuations along $q_x$ and $q_y$ directions. The obtained spin stiffnesses
appear to be equal to the long wavelength limit of the current correlation
functions $M_{\mathbf{q} \to 0}$. At finite doping the susceptibility is not
well described by quadratic dependence, apart from the small $|{\mathbf{q}}|$
values. Also in the incommensurate case using the momentum dependent
current-current correlation functions $M_{\mathbf{q}}$ instead of the
respective spin stiffnesses does not significantly improve the description
of the momentum dependence, in contrast to the commensurate case. This
feature might be a consequence of the suppression of bubble contributions
and non-vanishing vertex corrections in the incommensurate case (see below).

\begin{figure}[t]
\includegraphics[scale=1.0]{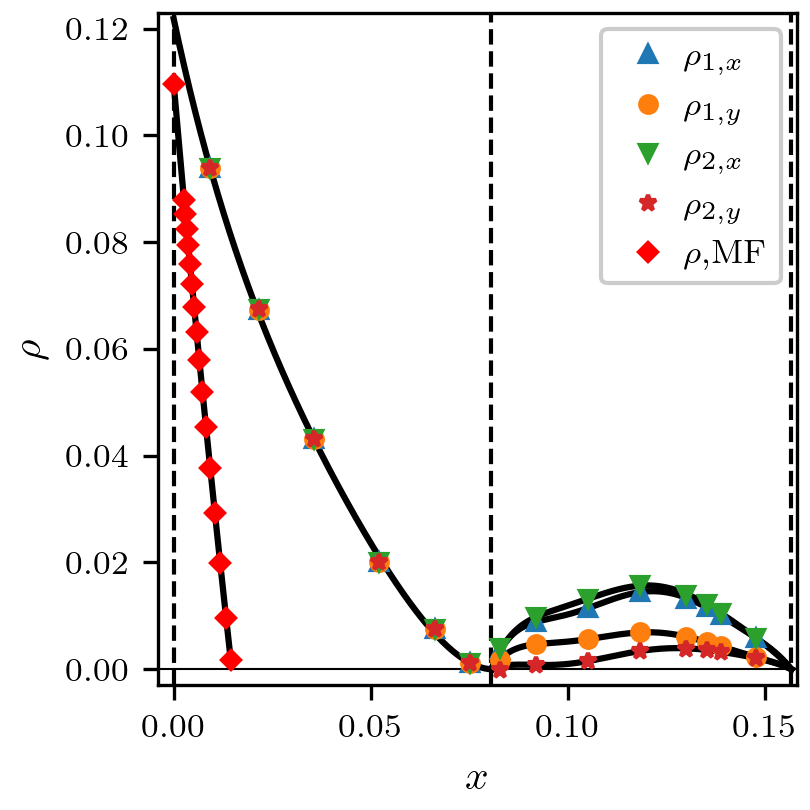}
\caption{(Color online). Spatial stiffness as a function of doping level $x$ in DMFT (triangles, circles, stars) and mean field (rhombs) approaches. Vertical dashed lines mark the commensurate-incommensurate and
incommensurate-paramagnetic transitions.}.
\label{rhorho}
\end{figure}

As well as in the commensurate case, the vertex corrections $K^{\phi,L,C}_{{%
\mathbf{0}},0,yy}$ and $\Delta\rho_{2,y}$ to the respective spin stiffnesses $\rho_{n,y}$, determined by the components of the modified kernel $M_{yy}
$, vanish by symmetry, since they are related to the components of the
current along the direction of ${Q_y=\pi}$ component. We also find vanishing of vertex corrections $K^{L,xx}_{{%
\mathbf{0}},0,xx}-K^{C,xx}_{{%
\mathbf{0}},0,xx}$ to the respective spin stiffnesses $\rho_{1,x}$. The respective
contributions to the $M_{xx}$ components of the modified kernel at $\mathrm{i%
} \omega_n = 0$, which determine $\rho_{1,x}$ and $\rho_{2,x}$ stiffnesses,
are plotted as a functions of {$x$} in Fig. \ref{Bars}. We find the $U$-reducible contribution $K^L-K^C+\Delta\rho_{2,n}$ (containing also the contribution of the $z$-component $\Delta \rho_{2,n}$) is negligibly small.  Despite
partial cancellation of bubble part of paramagnetic response $K^{0}_{q}$
and its diamagnetic part $K^{d}$, for $M_{xx}^{xx}$ component
these bare contributions constitute substantial part of the spatial spin
stiffness. Yet, the $U$-irreducible vertex correction contribution $K^{\phi}$, {which occurs due to dynamic effects}, is also
non-zero in this case, and provides approximately $1/3$ to $1/4$ amount of
the respective spin stiffness. On the other hand, for $M_{xx}^{yy}$ component 
%in the AFM case or when wave vector $\mathbf{q}$ lies along commensurate direction the two terms exactly cancel each other. However it is not the case 
in the incommensurate case 
%when the wavevector $\mathbf{q}$ lies along incommensurate direction. In this case 
the $U$-irreducible vertex part provides the major part of the spin stiffness. Therefore, neglecting {dynamic} vertex corrections in that case would yield dramatic unerestimate of spin stiffness.

Finally, we provide the doping dependence of the respective spatial spin
stiffnesses in Fig. \ref{rhorho}. One can see that in DMFT approach with the deviation from
half filling the spin stiffnesses first decrease and vanish at the
commensurate-incommensurate transition. This vanishing of spin stiffness can
be easily understood considering the incommensurate case in the vicinity of
such a transition, where Goldstone modes are present at both wave vectors ${%
\mathbf{q}}=\pm {\mathbf{Q}}$ in the global reference frame (see Eq. (\ref%
{chiQQQinc})). Therefore, approaching the commensurate phase necessarily
yields $q^4$ behavior of the inverse susceptibility, which implies vanishing
of the spin stiffness {(see Appendix 
\ref{Det})}. With further increase of doping the spin siffnesses
first increase, and then decrease again to zero at the incommensurate
magnetic-paramagnetic transition, where their smallness compensate the
smallness of the numerator in Eq. (\ref{chiQQQpara}). 

For comparison, we show the results of the mean field approach in the antiferromagnetic phase, where stable magnetic excitations exist in this approach. One can see, that the mean field spin stiffness quickly decreses with doping, and vanishes at the few percents of doping. At larger dopings, we were not able to find stable incommensurate magnetic solution, in agreement with earlier results \cite{SDW3,SDW4,SDWOur,SBIncomm}. Quick disappearence of static long-range order in mean-field theory is reminiscent of the experimental data on high-$T_c$ compounds. At the same time, the persistence of stable magnetic excitations in dynamical mean field theory  shows that dynamic nature of electronic correlations allows for preserving incommensurate magnetic order in the chargon sector in a broad doping range. With account of spinon fluctuations (see, e.g., Ref. \onlinecite{BonettiMetzner}), this corresponds to the {\it short-range} magnetic order of the entire system. We note that strong suppression of correlation length near the commensurate-incommensurate transition was observed for La$_{2-x}$Sr$_x$CuO$_4$ in Ref. \onlinecite{Neutron}, which agrees with the obtained suppression of the spatial spin stiffnesses near this transition.  Therefore, the dynamical mean field theory appears more appropriate for description of short range magnetic order.

\section{Conclusions}

In summary, we have derived corrected Ward identities for the frequency and momentum dependence of the susceptibilities in the commensurate- and incommensurate magnetic phases near it’s singularity. It was shown that Ward identities relate the temporal spin stiffnesses, which are in general dynamic, and spatial stiffnesses to the respective quantities, which can be extracted from the microscopic analysis. The obtained results for the spatial spin stiffnesses contain contributions of the bare para- and diamagnetic terms, as well as the two types ($U$-irreducible $K^\phi$ and $U$-reducible $K^L$) of vertex corrections, together with two additional correction terms $K^C$ and $\Delta \rho$. 

%We have also developed an explicit formalism which allows to decompose initial electronic degrees of freedom to chargons and spinons for a magnetically ordered state. We have showed that this decomposition is directly related to the breaking of gauge symmetry. Dynamics of both particles also depends on the gauge.

We have verified obtained identities numerically in the framework of dynamic mean field theory for the two-dimensional Hubbard model with nearest- and next-nearest-neighbor hopping in the antiferromagnetic and incommensurate cases. In particular, we have obtained temporal and spatial spin stiffnesses for various hole dopings. Several important results were obtained. First, with increasing of doping temporal stiffnesses acquire strong frequency dependence, their static limit differs significantly from the high-frequency asymptotics. Second, spatial stiffness was found to vanish at the transition from antiferromagnetic to the state with spiral magnetic order. Third, our calculations show that although there are no vertex corrections to the spatial spin stiffnesses in the antiferromagnetic case, these corrections are finite in the incommensurate state. Their effect is different for the different excitation modes and different spatial directions.
% Vertex corrections are still absent along commensurate direction, while the $U$-irreducible vertex corrections become significant for the incommensurate direction. The effect of the $U$-reducible vertex corrections was found to be numerically small.
Vertex corrections are  absent along commensurate direction, while they are finite along incommensurate direction. $U$-irreducible vertex corrections contribute significantly to the spatial spin stiffness, while the effect of the $U$-reducible vertex corrections was found to be numerically small.
The obtained vertex corrections to the spatial spin stiffness, which are absent in the static mean-field theory, seem to be the reason of stabilization of incommensurate long range magnetic order in the chargon sector, obtained previously in Ref. \onlinecite{OurFirst}. 
Therefore, strong dynamic effects are essential for stabilization of magnetic order, especially incommensurate, at significant doping levels.

The incommensurate long range magnetic order in the chargon sector, obtained in the present study, is observed as the {\it short range} magnetic order in cuprate high-$T_c$ compounds. With obtained spin stiffnesses this short range order can be straightforwardly described, considering fluctuations in the spinon sector according to the $1/N$ expansion for $CP^{N-1}$ model \cite{Auerbach,CSSen,Azaria,Campostrini,BonettiMetzner} or $O(N)/(O(N-2)\times O(2))$ nonlinear sigma model \cite{KII} for magnetic degrees of freedom. This represents the topic for the forthcoming study. The performed study shows that the {\it dynamic} effects in the chargon sector, as well as frequency dependence of the resulting temporal spin stiffness are crucial for this study. The generalization and application of the obtained results to the frustrated magnetic systems is also of certain interest. 
The derivation of corrected Ward identities for systems with broken symmetry can also be applied for the superconductivity case in the future studies. 

{\it Note added in Proof.} Recently, we were informed about the Erratum \cite{Errat} to Ref. \onlinecite{Bonetti}, which also suggests introducing small external, generally nonuniform, magnetic field for derivation of the corrected form of Ward identities for both, commensurate and incommensurate types of magnetic order.

%Another application this formalism is the study of magnetically disordered phases of matter with strong short range magnetic order such a spin liquids and the effect of strong frustration on the dynamical effects.

% In summary, we have derived corrected Ward identities for the frequency and
% momentum dependence of the susceptibilities in the commensurate- and
% incommensurate magnetic phases. We relate the temporal spin stiffnesses, which are in general dynamic, to the respective quantities, which can be extracted from the microscopic analysis. The obtained results for the spatial spin
% stiffnesses contain contributions of the bare para- and diamagnetic terms,
% as well as the two types ($U$-irreducible $K^\phi$ and $U-reducible$ $K^L$) vertex corrections, together with two additional contributions ( correction terms) $K^C$ and $\Delta_\rho$. While we find the sum of the contributions $K^L-K^C+\Delta \rho$ vanishing in the antiferromagnetic phase and for spin stiffnesses $1$ numerical small

\section*{Acknowledgements}

This study is supported by the Russian Science Foundation (Grant No. 24-12-00186). 

\begin{widetext}
\appendix
\renewcommand\thefigure{C\arabic{figure}}
\setcounter{figure}{0}

\section{General definitions and derivation of the current, spin
susceptibilities, and gauge kernel}
\label{AppGeneral} 

We use the definition of spin operators 
\begin{equation}
S_i^{\alpha} = \frac{1}{2} c^+_i \sigma^{\alpha} c_i.
\end{equation}
After the Fourier transformations 
%(up to the N factor, see The Theory of Magnetism by Daniel C. Mattis , p. 247 of Russian edition)
\begin{align}
c_k &= \sum_i e^{-\mathrm{i} k r_i} c_i,\, c_k^+ = \sum_i e^{\mathrm{i} k
r_i} c_i^+ 
%\\
%c_i &= \sum_k e^{\mathrm{i} k r_i} c_k ,\, c_i^+ = \sum_k e^{-\mathrm{i} k
%r_i} c_k^+
\end{align}
this results in the following expression for Fourier components 
\begin{equation}
S_q^{\alpha} = \sum_i e^{-\mathrm{i} q r_i}S_i^{\alpha} = \sum_k c^{+}_{k} 
\frac{\sigma^{\alpha}}{2} c_{k+q},
\end{equation}
and $S^{\alpha +}_{q}=S^{\alpha}_{-q}$. 
%creates "spin wave" with momentum $-q$ ($S^{\alpha}_i$ is a hermitian)
%\begin{equation}
%    S_q^{\alpha +} = \sum_i e^{+\I q r_i} S_i^{\alpha} = \sum_k c^{+}_{k+q} \frac{\sigma^{\alpha}}{2} c_{k} = S^{\alpha}_{-q}
%\end{equation}
The spin susceptibility 
\begin{equation}
\chi^{\alpha,\beta}_{q,q^{\prime }} = {-} \langle
\langle S^{\alpha}_{q} | S^{\beta}_{-q^{\prime }} \rangle \rangle = \langle
S^{\alpha}_{q} S^{\beta}_{-q^{\prime }} \rangle - \delta_{q,0}
\delta_{q^{\prime },0} \langle S^{\alpha}_{q=0} \rangle \langle
S^{\beta}_{q^{\prime }=0} \rangle  \label{spin_spin_correlation_function}
\end{equation}
reads 
\begin{align}
\chi^{\alpha,\beta}_{q,q^{\prime }} &= {-}
\sum_{k,k^{\prime }} \langle \langle c^+_{k} \frac{\sigma^\alpha}{2} c_{k+q}
| c^+_{k+q^{\prime }} \frac{\sigma^\beta}{2} c_{k} \rangle \rangle =
\label{spin_spin_cor_fun_c} \\
&= \sum_{k,k^{\prime }} \Bigg[ \langle c^+_{k,\sigma} \frac{%
\sigma^\alpha_{\sigma, \sigma^{\prime }}}{2} c_{k+q, \sigma^{\prime }}
c^+_{k^{\prime }+q^{\prime },\sigma^{\prime \prime \prime }} \frac{%
\sigma^\beta_{\sigma^{\prime \prime \prime },\sigma^{\prime \prime }}}{2}
c_{k^{\prime },\sigma^{\prime \prime }} \rangle - \delta_{q,0}
\delta_{q^{\prime },0} \delta_{\sigma, \sigma^{\prime }}
\delta_{\sigma^{\prime \prime \prime }, \sigma^{\prime \prime }} \langle
c^+_{k,\sigma} \frac{\sigma^\alpha_{\sigma, \sigma^{\prime }}}{2} c_{k,
\sigma^{\prime }} \rangle \langle c^+_{k^{\prime },\sigma^{\prime \prime
\prime }} \frac{\sigma^\beta_{\sigma^{\prime \prime \prime },\sigma^{\prime
\prime }}}{2} c_{k^{\prime }, \sigma^{\prime \prime }} \rangle \Bigg].  \notag
\end{align}
%and can be expressed in terms of the generalized susceptibility 
%(up to a temperature factor) which can also be nonlocal due to the presence of magnetic order 
%($\mathfrak{m} = (\sigma,\sigma^{\prime }),\mathfrak{m^{\prime }} =
%(\sigma^{\prime \prime }, \sigma^{\prime \prime \prime })$) 
%\begin{align}
%\mathcal{\chi}_{q,q^{\prime }}^{\mathfrak{m}\mathfrak{m^{\prime }}} &= %
%\textcolor{OliveGreen}{-} \sum_{k,k^{\prime }}\langle\langle{c^+_{k,\sigma}
%c_{k+q,\sigma^{\prime }}| c^+_{k^{\prime }+q^{\prime },\sigma^{\prime \prime
%\prime }} c_{k^{\prime },\sigma^{\prime \prime }}}\rangle\rangle =  \notag \\
%&= \sum_{k,k^{\prime }} \left[ \langle c^+_{k,\sigma} c_{k+q, \sigma^{\prime
%}} c^+_{k^{\prime }+q^{\prime },\sigma^{\prime \prime \prime }} c_{k^{\prime
%},\sigma^{\prime \prime }} \rangle - \delta_{q,0} \delta_{q^{\prime },0}
%\delta_{\sigma, \sigma^{\prime }} \delta_{\sigma^{\prime \prime \prime },
%\sigma^{\prime \prime }} \langle c^+_{k,\sigma} c_{k, \sigma} \rangle
%\langle c^+_{k^{\prime },\sigma^{\prime \prime }} c_{k^{\prime },
%\sigma^{\prime \prime }} \rangle \right]  \label{chi_definition}
%\end{align}

The current operators are defined similarly to the $U(1)$ case (see, e.g.,
Ref. \onlinecite{KatsLicht}). %in the following form
%\begin{equation}
%    J^{\alpha}_{i,\mu} = \sum_{j} t_{ij} r^{\mu}_{ij} c^+_i \frac{\sigma^{\alpha}}{2} c_j
%    \label{current_operator_definition}
%\end{equation}
%Than the Fourier component takes the following form
%\begin{equation}
%    J_{q,\mu}^{\alpha} = \sum_i e^{-\I q r_i} J_{i,\mu}^{\alpha} = \sum_k \bar{t}^{\mu}_k c^+_k c_{k+q},
%\end{equation}
%where $\bar{t}^{\mu}_k$ is defined as
%\begin{equation}
%    \bar{t}^{\mu}_k = \sum_{r_{ij}} e^{-\I k r_{ij}} t_{ij} r^{\mu}_{ij}
%    \label{current_vertex_def}
%\end{equation}
%Note that according to (\ref{current_operator_definition}) $J^{\alpha}_{i,\mu}$ is not hermitian. However one can arrive at another - hermitian - definition of the current operator. 
We start with the respective part of the action 
\begin{equation}
{\mathcal{S}}_0=-\sum_{i j} c_i^{+} t_{i j} \mathcal{R}^+_i \mathcal{R}_j c_j
\end{equation}
and rewrite it in the form where hermicity is explicit and $j = i + \Delta_m$%
, where ${m}$ indicated direction ${m} \in x,y$, assuming $t_{\Delta_m} =
t_{-\Delta_m}$ 
\begin{equation}
{\mathcal{S}}_0=-\frac{1}{2} \sum_{i} \sum_{\Delta_m} t_{\Delta_m} \left(
c^+_{i + \Delta_m} \mathcal{R}^+_{i + \Delta_m} \mathcal{R}_{i} c_i + c_i^+ 
\mathcal{R}^+_{i} \mathcal{R}_{i+\Delta_m} c_{i + \Delta_m}\right).
\end{equation}
This part of the action can be rewritten in the form 
\begin{equation}
{\mathcal{S}}_0 = - \frac{1}{2} \sum_i \sum_{\Delta} t_{\Delta} c^+_i \left[
e^{\Delta_m \left( \overleftarrow{\partial}_{{m}} + \mathrm{i} A_{i {m}}
\right)} + e^{\Delta_m \left( \overrightarrow{\partial}_{{m}} - \mathrm{i}
A_{i {m}} \right)} \right] c_i,
\end{equation}
where $\Delta$ enumerates all neighbours connected
by finite hopping $t_{\Delta}$, $\overleftarrow{\partial}_{m}
$ and $\overrightarrow{\partial}_{m}$ are generators of left and right
lattice translations in the directions $m = x,y$ respectively. $A_{i m}$ is
defined according to (\ref{AR}).

If we consider only slowly varying fields $A_{i m}$ we can neglect the contributions coming from derivatives of $A_{i
m}$ fields. In this case action simplifies to the usual form of Peierls
substitution, 
\begin{equation}
{\mathcal{S}}_0 \approx - \frac{1}{2} \sum_i \sum_{\Delta} t_{\Delta} \left(
c^+_{i + \Delta} e^{+ \mathrm{i} \Delta_m A_{i {m}} } c_i + c^+_i e^{ - 
\mathrm{i} \Delta_m A_{i {m}} } c^+_{i + \Delta} \right).
\end{equation}
%\begin{align}
%    c^+_{i + \Delta} &= c^+_i \overleftarrow{T}_{\Delta} = c^+_i e^{\Delta^{\alpha} \overleftarrow{\partial}_{\alpha}}, \nonumber \\
%    c_{i + \Delta} &= \overrightarrow{T}_{\Delta} c_i = e^{\Delta^{\alpha} \overrightarrow{\partial}_{\alpha}} c_i \nonumber
%\end{align}
%
%\begin{equation}
%   \left( \overleftarrow{T}_{\Delta} \right)^+ = \overrightarrow{T}_{\Delta} \nonumber
%\end{equation}
% Expanding $\mathcal{R}$ as
% \begin{align}
% \mathcal{R}_{i + \Delta_m} \approx \mathcal{R}_{i} - \I \mathcal{R}_i A_{i {m}} \Delta_m,\\
%  \mathcal{R}^+_{i + \Delta_m} \approx \mathcal{R}_{i}^+ + \I  A_{i {m}} \mathcal{R}^+_i \Delta_m
% \end{align}
% which keeps $\mathcal{R}_i^+ \mathcal{R}_i = 1$ and $A_{i {m}}% is hermitian, we get 
%\begin{multline}
%    {\mathcal S}_0=-\frac{1}{2} \sum_{i,\Delta_m} t_{\Delta_m} \left( c^+_{i + \Delta_m} c_i + c_i^+ c_{i + \Delta_m} \right) - \frac{\I}{2} \sum_{i} \sum_{\Delta_m} t_{\Delta_m} \Delta_m \left( c^+_{i + \Delta_m} A_{i {m}} c_i - c_i^+ A_{i {m}} c_{i + \Delta_m} \right) + \\
%    + \frac{1}{4} \sum_i \sum_{\Delta_m} t_{\Delta_m} \Delta^{{m} \; 2} \left( c^+_{i + \Delta_m} A_{i {m}}^2 c_i + c^+_i A_{i {m}}^2 c_{i+\Delta_m} \right) - \\
%    - \frac{\I}{2} \sum_i \sum_{\Delta_m} t_{\Delta_m} \Delta^{{m} \; 2} \left( c^+_{i + \Delta_m} \partial_{{m}} A_{i {m}} c_i - c^+_i \partial_{{m}} A_{i {m}} c_{i+\Delta_m} \right) + \ldots
%\end{multline}
It can be further expanded in series with respect to the powers of $A_{i m}$
field 
\begin{align}
{\mathcal{S}}_0=&-\frac{1}{2} \sum_{i,\Delta} t_{\Delta} \left( c^+_{i +
\Delta} c_i + c_i^+ c_{i + \Delta} \right) - \frac{\mathrm{i}}{2} \sum_{i}
\sum_{\Delta} t_{\Delta} \Delta_m \left( c^+_{i + \Delta} A_{i {m}} c_i -
c_i^+ A_{i {m}} c_{i + \Delta} \right) + \\
&+ \frac{1}{4} \sum_i \sum_{\Delta} t_{\Delta} \Delta_m \Delta_{n} \left(
c^+_{i + \Delta} A_{i {m}} A_{i,{n}} c_i + c^+_i A_{i {m}} A_{i,{n}}
c_{i+\Delta} \right) +\ldots
\end{align}
Then we get current operator which is explicitly Hermitian,
%\begin{align}
%    J_{i {m}}^{\alpha} &= \frac{i}{2} \sum_{\Delta_m} t_{\Delta_m} \Delta_m \left( c^+_{i + \Delta_m} \frac{\sigma^{\alpha}}{2} c_i - c_i^+ \frac{\sigma^{\alpha}}{2} c_{i + \Delta_m} \right)\notag\\
%    &= \frac{i}{2} \sum_{j} t_{ji} r^{{m}}_{ji} \left( c^+_{j} \frac{\sigma^{\alpha}}{2} c_i - c_i^+ \frac{\sigma^{\alpha}}{2} c_{j} \right)
%\end{align}
\begin{align}
J_{i {m}}^{\alpha} &= \frac{\I}{2} \sum_{\Delta} t_{\Delta} \Delta_m \left(
c^+_{i + \Delta} \frac{\sigma^{\alpha}}{2} c_i - c_i^+ \frac{\sigma^{\alpha}%
}{2} c_{i + \Delta} \right),  
%\notag 
%\\
%&= \frac{i}{2} \sum_{j} t_{ji} r^{{m}}_{ji} \left( c^+_{j} \frac{%
%\sigma^{\alpha}}{2} c_i - c_i^+ \frac{\sigma^{\alpha}}{2} c_{j} \right) 
%\notag
\end{align}
with the respective Fourier transform 
\begin{align}
J_{q,{m}}^{\alpha} &=
%\frac{1}{2} \sum_k \left( t^{{m}}_{\mathbf{k}} c_k^+ 
%\frac{\sigma^\alpha}{2} c_{k+q} - t^{{m}}_{-{\mathbf{k}}} c_{k-q}^+ \frac{%
%\sigma^\alpha}{2} c_{k} \right)  \notag \\
\frac{1}{2} \sum_k \left( t_{\mathbf{k}}^{{m}} - t^{{m}}_{-{\mathbf{k}} - 
\mathbf{q}} \right) c_k^+ \frac{\sigma^{\alpha}}{2} c_{k+q},
\end{align}
where $t^{{m}}_{\mathbf{k}} = -\mathrm{i} \sum\limits_{r_{ij}} t_{ij} r^{{m}%
}_{ij} e^{-\mathrm{i} k r_{ij}} = {\partial} \epsilon_{\mathbf{k}}/{\partial
k^{{m}}}$, $\epsilon_{\mathbf{k}} = \sum\limits_{r_{ij}} e^{-\mathrm{i} k
r_{ij}} t_{ij}$. In the presence of inversion symmetry  $\epsilon_{\mathbf{k}} = \epsilon_{-{\mathbf{k}}}$,
$t^{{m}}_{-{\mathbf{k}}} = -t^{{m}}_{\mathbf{k}}$, and
\begin{equation}
J_{q,{m}}^{\alpha} =\sum_k T^{m}_{{\mathbf{k}},{\mathbf{q}}} c_k^+ \frac{%
\sigma^{\alpha}}{2} c_{k+q},  \label{current_operator}
\end{equation}
where $T^{m}_{{\mathbf{k}},{\mathbf{q}}}=({t_{\mathbf{k}}^{{m}} + t^{{m}}_{{%
\mathbf{k}} + {\mathbf{q}}}})/2$. It can be easily verified that 
%\begin{equation}
$J_{q,{m}}^{\alpha+} = J_{-q,{m}}^{\alpha}$, %\end{equation}
as it should be for a Fourier transform of a hermitian operator.

Defining $T_{{\mathbf{k}},{\mathbf{q}}}^{0}=\I$, the gauge kernel (including
its temporal, i.e. spin components) then takes the form 
%Now we can define the current-current correlation functions of the form (no macroscopic current presents) similarly with (\ref{spin_spin_correlation_function})
\begin{equation*}
K_{q,q^{\prime };{\mu },{\nu }}^{\alpha ,\beta }=\sum_{j}\int d\tau \frac{%
\delta {W}}{\delta A_{i,{\mu }}^{\alpha }\delta A_{j,{\nu }}^{\beta }}%
e^{iq(x_{j}-x_{i})}=\langle \langle J_{q,{\mu }}^{\alpha }|J_{-q^{\prime },{%
\nu }}^{\beta }\rangle \rangle {+}K_{{\mu }{\nu }}^{d},
\end{equation*}%
where 
\begin{align}
    \langle \langle J_{q,{\mu }}^{\alpha }|J_{-q^{\prime },{\nu }}^{\beta}\rangle \rangle = -& \sum_{k,k^{\prime }}T_{{\mathbf{k}},{\mathbf{q}}}^{\mu}\langle c_{k,\sigma }^{+}\frac{\sigma _{\sigma ,\sigma ^{\prime }}^{\alpha }%
    }{2}c_{k+q,\sigma ^{\prime }}c_{k^{\prime }+q^{\prime },\sigma ^{\prime
    \prime \prime }}^{+}\frac{\sigma _{\sigma ^{\prime \prime \prime },\sigma
    ^{\prime \prime }}^{\beta }}{2}c_{k^{\prime },\sigma ^{\prime \prime
    }}\rangle T_{{\mathbf{k}}^{\prime },{\mathbf{q}}^{\prime }}^{\nu }, \\
    K_{{\mu }{\nu }}^{d}=-\frac{1}{2} & \sum_{{k}}t_{\mathbf{k}}^{{\mu }{\nu }}
    \langle c_{{k}}^{+}\frac{\sigma ^{0}}{2}c_{{k}}\rangle (1-\delta _{\mu0})(1-\delta _{\nu 0})  \label{current_current_cor_fun_c}
\end{align}%
are the para- and diamagnetic contributions to the current correlation
function, $t_{\mathbf{k}}^{mn}={\partial ^{2}\epsilon _{\mathbf{k}}}/({\partial k_{m}\partial k_{n}})$.

\section{Gauge transformation and Ward identities}
\label{AppWard}
The dependence of the action on the matrix $\mathcal{R}$ has the form 
\begin{equation}
S[c,c^{+},\mathcal{R}]=\sum_{ij}\int\limits_{0}^{\beta }d\tau \;c_{i}^{+}%
\left[ \left( \frac{\partial }{\partial \tau }-\mu +\mathcal{R}_{i}^{+}\frac{%
\partial }{\partial \tau }\mathcal{R}_{i}\right) \delta _{ij}-t_{ij}\mathcal{%
R}_{i}^{+}\mathcal{R}_{j}\right] c_{j}+U\sum_{i}\int_{0}^{\beta }d\tau
\;n_{i\uparrow }n_{i\downarrow }.
\end{equation}%
The gauge transformation parameterized by $SU(2)$ field $\mathcal{V}_{i}(\tau )$ applied to the fields $A_\mu$ read%
\begin{align}
& A_{\mu i}\rightarrow \mathcal{V}_{i}A_{\mu i}\mathcal{V}_{i}^{+}+\mathrm{i}%
\mathcal{V}_{i}\partial _{\mu }\mathcal{V}_{i}^{+}.
\end{align}
For infinitesimal transformations we obtain Eq. (\ref{TransformA}) of the main text.%
%
%Само калибровочное преобразовние, реализуемое через  дейcтвует так:
%Derivation of Ward identities

\label{WardSect}

\subsection{Ward identities for the functional $W$}
\label{WardW}

The Ward identities are derived from the condition $\delta W\left[\mathcal{R}[\mathcal{V}]\right]=0$. The main Ward identity for the functional $W$ was derived in Ref. \onlinecite%
{Bonetti}. 
%При использовании условия калибровочной инвариантности следует учесть, что в формулы преобразования полей $A_{\mu i}(\tau)$ и $J_i(\tau)$ входят не только сами компоненты калибровочного поля $\mathcal{V}^a$, но и его производные $\partial_\mu \mathcal{V}^a$. 
The variation of the functional $W$ takes the form (integration over
imaginary time and summation over lattice indices are implicitly assumed) 
\begin{equation}
\delta W = \frac{\delta W}{\delta A^a_\mu} \frac{\delta A^a_\mu}{\delta 
\mathcal{V}^b} \delta \mathcal{V}^b + \frac{\delta W}{\delta A^a_\mu } \frac{%
\delta A^a_\mu}{\delta ( \partial_\nu \mathcal{V}^b )} \delta ( \partial_\nu 
\mathcal{V}^b ). 
\end{equation}
The variational derivatives of the fields $A_{\mu i}(\tau)$ 
%и $J_i(\tau)$ 
over gauge field $\mathcal{V}^a$ take the form 
\begin{align}
& \frac{\delta A^a_\mu}{\delta ( \partial_\nu \mathcal{V}^b )} = -
\delta_{a,b} \delta_{\mu,\nu}, \,\,\, \frac{\delta A^a_\mu}{\delta \mathcal{V%
}^b } = \varepsilon_{a b c} A^{c}_\mu.
\end{align}
%и воспользуемся тождеством
%\begin{equation}
%        \frac{\delta W}{\delta A^a_\mu } \frac{\delta A^a_\mu}{\delta ( \partial_\nu \mathcal{V}^b )} \delta ( \partial_\nu \mathcal{V}^b ) = - \frac{\delta W}{\delta A^a_\mu } \delta ( \partial_\mu \mathcal{V}^a ) = + \partial_\mu \left( \frac{\delta W}{\delta A^a_\mu } \right) \delta \mathcal{V}^a
%\end{equation}
Then the condition $\delta W = 0$ yields 
\begin{equation}
\partial_\mu \left( \frac{\delta W}{\delta A^a_\mu} \right) - \varepsilon_{a
b c} \frac{\delta W}{\delta A^b_\mu} A^c_\mu = 0.  \label{initial_w_id}
\end{equation}
%верное для любого $a \in {x,y,z}$.
The Eq. (\ref{initial_w_id}) coincides with that derived in Ref. \onlinecite%
{Bonetti}. It describes motion of spins in the external magnetic and current
fields. We note that this equation does not contain explicitly the internal
(mean) fields. %, since they are cancelled between various contributions.
%\comAKK{For $A_{\mu\neq 0}=0$  this is essentially the Landau-Lifshitz equation. But where is the magnetic exchange contribution??}
%\par\noindent\rule{\textwidth}{0.5pt}
%{\color{teal} Derivation without Legendre transformation .}
{By differentiating over $A_{\nu }^{d}$ we obtain the equation 
\begin{equation}
\partial _{\mu ,x}K_{\mu ,x;\nu ,x^{\prime }}^{ad}+{\epsilon _{acb}A_{\mu
,x}^{c}K_{\mu ,x;\nu ,x^{\prime }}^{bd}}+\varepsilon _{abd}j_{\nu
,x}^{b}\delta _{x,x^{\prime }}=0,  \label{WardWGen}
\end{equation}%
where $j_{\nu ,x}^{b}=-{\delta W}/{\delta A_{\nu ,x}^{b}}$ is the spin
current (including the $\nu =0$ spin density component). 
For $\nu =n>0$ the
last term in the left hand side vanishes in the equilibrium and we find} Eq. (\ref{eqq}) of the main text.
Restricting the non-zero component of the gauge field in the equilibrium to $%
A_{0,x}$ only (which is proportional to the external non-uniform magnetic
field), putting $\nu=0$ in Eq. (\ref{WardWGen}), we find Eq. (\ref{Wnu0}) of the main text.

In the commensurate antiferromagnetic case the transverse susceptibilities $\chi ^{ab}$ with $%
a,b=x,y$ are decoupled from $\chi ^{zz}$. However, the diagonal
susceptibilities $\chi ^{xx,yy}$ are coupled to the off-diagonal ones $\chi
^{xy,yx}$ by the dynamic terms. The general form of the transverse
susceptibility at $\mathbf{q},\mathbf{q^{\prime }}=\mathbf{0},\mathbf{Q}$,
allowed by the Ward identity (\ref{Wnu0}), together with the conjugated one,
obtained by the interchange $x\leftrightarrow x^{\prime }$, in the basis $%
a,b=x,y$ takes the form 
\begin{equation}
\chi _{{\mathbf{q}}\mathbf{q}^{\prime },\omega }^{ab}=\frac{1}{\omega ^{2}}%
\left( \begin{NiceArray}{cc:cc} 
{h r^y_{\mathbf{Q},\omega}} & {h^2 \chi _{\mathbf{Q},\omega}^{{yx}}} & -\omega{h \chi _{\mathbf{Q},\omega}^{{yx}}} & \omega r^y_{\mathbf{Q},\omega} \\ 
{h^2 \chi _{\mathbf{Q},\omega}^{{xy}}} & {h r^x_{\mathbf{Q},\omega}} & - \omega r^x_{\mathbf{Q},\omega} & \omega{h \chi _{\mathbf{Q},\omega}^{{xy}}} \\ \hdottedline 
\omega{h \chi _{\mathbf{Q},\omega}^{{xy}}} & \omega r^x_{\mathbf{Q},\omega} & \omega^2 \chi
_{\mathbf{Q},\omega}^{{xx}} & \omega^2\chi _{\mathbf{Q},\omega}^{{xy}} \\
-\omega r^y_{\mathbf{Q},\omega} & -\omega{h \chi
_{\mathbf{Q},\omega}^{{yx}}} & \omega^2\chi _{\mathbf{Q},\omega}^{{yx}} &
\omega^2\chi _{\mathbf{Q},\omega}^{{yy}} \\ \end{NiceArray}\right), 
\end{equation}%
where $r^a_{\mathbf{q},\omega}=m-h \chi _{\mathbf{q},\omega}^{{aa}}$, $\chi _{\mathbf{Q},\omega }^{ab}\equiv \chi _{\mathbf{QQ},\omega }^{ab}
$ and the blocks correspond to the values $\mathbf{q},\mathbf{q^{\prime }}=%
\mathbf{0},\mathbf{Q}$. The staggered components $\chi _{\mathbf{Q},\omega}$ are not fixed by Ward identities, and have to be determined from the microscopic theory. In the main text we express them in Eq. (\ref{chiQQAFM}) through the frequency dependence of $\chi_\omega$, which appears to be equal to the
uniform transverse suscetibility according to the Eq. (\ref{chiQ0AFM}).

\subsection{ 
%relation of the derivatives of $\Gamma$ and $W$, and
Ward identities for the Legendre transformed functional $%
\Gamma $}
\label{WardGammaSect}

{Using Eq. (\ref{DG1}), we obtain the Ward identity %\begin{equation}
%        \partial_\mu \left( \frac{\delta \Gamma}{\delta A^a_\mu} \right) - \varepsilon_{a b c} \left( \frac{\delta \Gamma}{\delta A^b_\mu} A^c_\mu - \phi^b \frac{\delta \Gamma}{\delta \phi^c} \right) = 0
%\end{equation}
\begin{equation}
\partial _{\mu }\left( \frac{\delta \Gamma }{\delta A_{\mu ,x}^{a}}\right)
-\varepsilon _{abc}\left( \frac{\delta \Gamma }{\delta A_{m,x}^{b}}%
A_{m,x}^{c}-\phi _{x}^{b}\left( iA_{0}^{c}+\frac{\delta \Gamma }{\delta \phi
_{x}^{c}}\right) \right) =0,  \label{WardGamma}
\end{equation}%
which is equivalent to that used in Refs. \onlinecite%
{Bonetti,BonettiMetzner,BonettiThesis} 
\begin{equation}
\partial _{\mu }\left( \frac{\delta \Gamma }{\delta A_{\mu ,x}^{a}}\right)
-\varepsilon _{abc}\left( \frac{\delta \Gamma }{\delta A_{\mu ,x}^{b}}A_{\mu
,x}^{c}+\frac{\delta \Gamma }{\delta \phi _{x}^{b}}\phi _{x}^{c}\right) =0.
\end{equation}%
%which coincides with that of Refs. \cite{Bonetti,BonettiThesis,BonettiMetzner}. %. However, the Eq. (\ref{Gamma_vanishpart}) implies that the Eq. (\ref{WardGamma}) can be rewritten as
%\begin{equation}
%         i \partial_\tau \phi^a+\partial_m \widetilde{j}_m^a=\varepsilon_{a b c} \left[ \phi^b \left(i A^c_0 +\frac{\delta \Gamma}{\delta \phi^c}\right) -\widetilde{j}^b_m A^c_m
%           \right],
%\end{equation}
%where $\widetilde{j}_m^a=-\delta \Gamma/\delta A_m^a$. 
%\comAK{The second term in () brackets seems to represent the mean field. Notably, it has zero equilibrium value (!), cf. antiferromagnetism without anomalous averages. 
Differentiating over $\phi _{x^{\prime }}^{d}$ and $A_{n,x^{\prime }}^{d}$
we obtain 
\begin{align}
\partial _{m}\frac{\delta ^{2}\Gamma }{\delta A_{m,x}^{a}\delta \phi
_{x^{\prime }}^{d}}+\varepsilon _{abc}\phi _{x}^{b}\frac{\delta ^{2}\Gamma }{%
\delta \phi _{x}^{c}\delta \phi _{x^{\prime }}^{d}}& =i\left( \partial
_{\tau }\delta _{ad}-{i\varepsilon _{adc}A_{0,x}^{c}}\right) \delta
_{xx^{\prime }},  \label{gamma_eqv_1} \\
\partial _{m}\frac{\delta ^{2}\Gamma }{\delta A_{m,x}^{a}\delta
A_{n,x^{\prime }}^{d}}+\varepsilon _{abc}\phi _{x}^{b}\frac{\delta
^{2}\Gamma }{\delta \phi _{x}^{c}\delta A_{n,x^{\prime }}^{d}}& =0.
\label{gamma_eqv_2}
\end{align}%
} %Such that 
%\begin{align}
%&{\Gamma^{yx}_{\pm {\mathbf Q},0}}=\omega/m,\,\Gamma^{xy}_{\pm {\mathbf Q},0}=-\omega/m\\  
%&{\Gamma^{yy}_{\pm {\mathbf Q},\pm {\mathbf Q}}} ={\Gamma^{xx}_{\pm {\mathbf Q},\pm {\mathbf Q}}}={\Gamma^{zz}_{\pm {\mathbf Q},\pm {\mathbf Q}}}=-h_{Q}/m\\
%&{\Gamma^{yz}_{\pm {\mathbf Q},0}}=\mp i \omega/m;{\Gamma^{zy}_{\pm {\mathbf Q},0}}=\mp i \omega/m
%\end{align}
%\begin{equation}
%           \partial_{m,x} \partial_{n,x'} \frac{\delta^2 \Gamma}{ \delta A_{m,x}^a \delta A^d_{n,x'}} + \varepsilon_{d \widetilde{b} \widetilde{c}}  \phi_{x'}^{\widetilde{b}} \partial_{m,x}\left( \frac{\delta^2 \Gamma}{\delta A_{m,x}^a \delta \phi_{x'}^{\widetilde{c}} }\right)
%            = 0
%           \label{gamma_eqv_2}
%\end{equation}
%\begin{equation}
%           \partial_{m,x} \partial_{n,x'} \frac{\delta^2 \Gamma}{ \delta A_{m,x}^a \delta A^d_{n,x'}} = \varepsilon_{d \widetilde{b} \widetilde{c}}  \phi_{x'}^{\widetilde{b}} \left( -i \partial_\tau \delta_{a \widetilde{c}} \delta_{x x'}  + \varepsilon_{a b c}\phi^b_x   \frac{\delta^2 \Gamma}{\delta \phi^{c}_x \delta \phi^{\widetilde{c}}_{x'}}\comAKK{+i\varepsilon_{a \widetilde{c} c} A_{0,x}^c\delta_{xx'}}\right)
%           \label{gamma_eqv_2}
%\end{equation}
%\begin{equation}
%            i q_m \frac{\delta^2 \Gamma}{\delta A^a_{m,q} \delta A_{n,q'}^d} + \varepsilon_{a b c} \phi^b \left( \frac{\delta^2 \Gamma}{\delta \phi_{q\pm Q}^c A_{n,q'}^d}\right)
%            = 0
%\end{equation}
{From equations (\ref{gamma_eqv_1}), 
 (\ref{gamma_eqv_2}) we obtain the Ward identity (\ref{Gamma_Ward1}) for the second derivatives. 

\section{Mean field theory at half filling}

\label{SectMF}

\begin{figure}[b]
\includegraphics[scale=1.0]{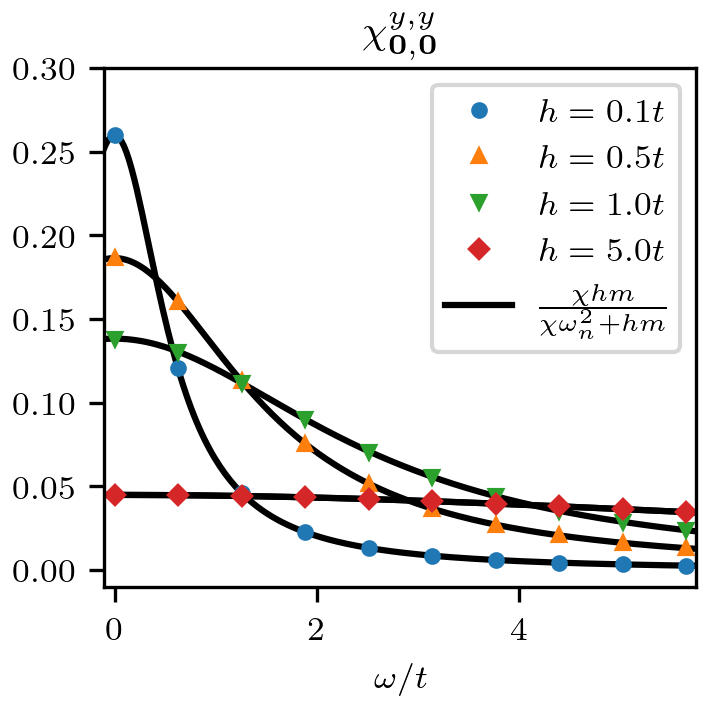} %
\includegraphics[scale=1.0]{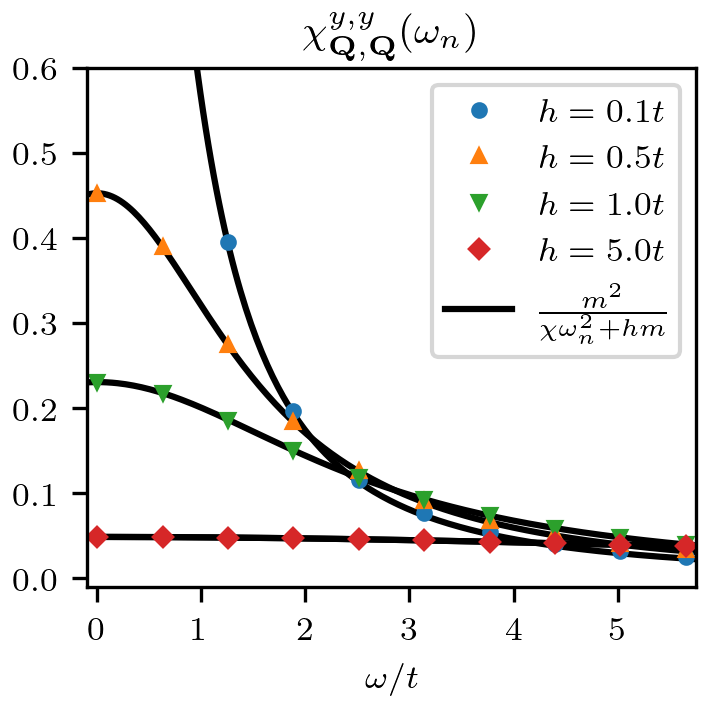} 
\caption{(Color online). The frequency dependence of the uniform $\protect%
\chi^{yy}_{00}$ and $\protect\chi^{yy}_{\mathbf{Q}{\mathbf{Q}}}$ transverse
susceptibilities for the half filled Hubbard model in the mean field
approximation.}
\label{MF_figure}
\end{figure}

As an example demonstrating importance of the external staggered field for obtaining correct uniform susceptibility we consider the mean-field approximation for considering
the chargon sector at half-filling. The mean-field equations for were
derived in Refs. \onlinecite{SDW1,SDW2}; in Ref. \onlinecite{OurFirst} it was shown how
they can be obtained in the local reference frame; they read 
\begin{align}
m & =\frac{1}{2}\sum_{\mathbf{k}}\frac{\Delta}{E_{\mathbf{k}}}\left[ f(E_{%
\mathbf{k}}^{v})-f(E_{\mathbf{k}}^{c})\right] ,n=\sum_{\mathbf{k}}\left[
f(E_{\mathbf{k}}^{v})+f(E_{\mathbf{k}}^{c})\right]=1,
\end{align}
where $\Delta=Um+h$, $E_{\mathbf{k}}=\sqrt{(\varepsilon_{\mathbf{k}%
}^{-})^{2}+\Delta^{2}}$, $\varepsilon_{\mathbf{k}}^{\pm}=(\varepsilon_{%
\mathbf{k}}\pm \varepsilon_{\mathbf{k+Q}})/2$, $E_{k}^{c,v} =\varepsilon_{%
\mathbf{k}}^{+}\pm E_{\mathbf{k}}-\mu$, $h$ is the external staggered
magnetic field. %\begin{align}
The transverse susceptibility is given by \cite{SDW1,SDW2} 
\begin{align}
\chi_{\mathbf{qq},\omega}^{+-} & =\frac{\chi_{\mathbf{qq},\omega}^{+-,0}(1-U%
\chi_{\mathbf{q}+\mathbf{Q},\mathbf{q}+\mathbf{Q},\omega}^{+-,0})+U(\chi_{%
\mathbf{q},\mathbf{q}+\mathbf{Q},\omega}^{+-,0})^{2}}{(1-U\chi_{\mathbf{qq}%
,\omega}^{+-,0})(1-U\chi_{\mathbf{q}+\mathbf{Q},\mathbf{q}+\mathbf{Q}%
,\omega}^{+-,0})-U^{2}(\chi_{\mathbf{q},\mathbf{q}+\mathbf{Q}%
,\omega}^{+-,0})^{2}},  \label{chiMF1} \\
\chi_{\mathbf{q,q}+\mathbf{Q},\omega}^{+-} & =\frac{\chi_{\mathbf{q},\mathbf{%
q}+\mathbf{Q},\omega}^{+-,0}}{(1-U\chi_{\mathbf{qq},\omega}^{+-,0})(1-U\chi_{%
\mathbf{q}+\mathbf{Q},\mathbf{q}+\mathbf{Q},\omega}^{+-,0})-U^{2}(\chi_{%
\mathbf{q},\mathbf{q}+\mathbf{Q},\omega}^{+-,0})^{2}} ,  \label{chiMF2}
\end{align}
where 
\begin{align}
\chi_{\mathbf{qq},\omega}^{+-,0} & =\frac{1}{4}\sum_{\mathbf{k}}\left( 1-%
\frac{\varepsilon_{\mathbf{k}}^{-}\varepsilon_{\mathbf{k+q}}^{-}-\Delta^{2}}{%
E_{\mathbf{k}}^{-}E_{\mathbf{k+q}}^{-}}\right) \left[ \frac {f(E_{\mathbf{k}%
}^{v})-f(E_{\mathbf{k}+\mathbf{q}}^{c})}{i\omega -E_{\mathbf{k}+\mathbf{q}%
}^{c}+E_{\mathbf{k}}^{v}}+\frac{f(E_{\mathbf{k}}^{c})-f(E_{\mathbf{k}+%
\mathbf{q}}^{v})}{i\omega-E_{\mathbf{k}+\mathbf{q}}^{v}+E_{\mathbf{k}}^{c}}%
\right], \\
\chi_{\mathbf{q,q+Q},\omega}^{+-,0} & =\frac{1}{4}\sum_{\mathbf{k}}\frac{%
\Delta(E_{\mathbf{k}}^{-}+E_{\mathbf{k+q}}^{-})}{E_{\mathbf{k}}^{-}E_{%
\mathbf{k+q}}^{-}}\left[ \frac{f(E_{\mathbf{k}}^{c})-f(E_{\mathbf{k}+\mathbf{%
q}}^{v})}{i\omega-E_{\mathbf{k}+\mathbf{q}}^{v}+E_{\mathbf{k}}^{c}}-\frac{%
f(E_{\mathbf{k}}^{v})-f(E_{\mathbf{k}+\mathbf{q}}^{c})}{i\omega-E_{\mathbf{k}+\mathbf{%
q}}^{c}+E_{\mathbf{k}}^{v}}\right],
\end{align}
and $+-$ refer (only in this subsection) to $S^x\pm i S^y$ basis. For $h=0$
we find vanishing susceptibility $\chi_{\mathbf{00},\omega}^{+-}$, which
occurs due to cancellation of the first and second term in the numerator of
Eq. (\ref{chiMF1}). At the same time, for finite staggered field the
cancellation does not occur, and in the limit $\omega\rightarrow 0$ the
first term in the numerator remains finite (being proportional to the
staggered field), while the second term, related to the susceptibility $%
\chi_{\mathbf{q},\mathbf{q}+\mathbf{Q},\omega}^{+-,0}$, vanishes. The
resulting static uniform susceptibility is given by the random phase
approximation (RPA) equation 
\begin{align}
\chi_{\mathbf{00},\omega\rightarrow 0}^{+-} & =\frac{\chi_{\mathbf{00}%
,0}^{+-,0}}{1-U\chi_{\mathbf{00},0}^{+-,0}},  \label{chiMF3}
\end{align}
in agreement with the derivation of the non-linear sigma model in Ref. \onlinecite%
{Dupuis} and previous Ward identity approach \cite{Bonetti,BonettiThesis}.
The latter approach however derived the Eq. (\ref{chiMF3}) for \textit{zero}
external magnetic field, which, as we explain in the discussion above, not
correct. The reason of this discrepancy is in our opinion missing the
off-diagonal terms (like the second terms in the numerator and denominator
of Eq. (\ref{chiMF1})) in the approach of Refs. \onlinecite{Bonetti,BonettiThesis}.

In Fig. \ref{MF_figure} we show field dependence of several components of
dynamical susceptibility for the two-dimensional Hubbard model with $U=7.5t$, $t'/t=0.15$, and their complience to the form, dictated by Ward
identities (\ref{chiQ0AFM}) . Note that apart from explicit field-dependence
in Eq. (\ref{chiQ0AFM}), the static susceptibility $\chi$ also implicitly
depends on magnetic field, saturating, however, at finite value in the limit 
$h\rightarrow$0.
}

\section{Transformation to local coordinate frame}
\label{AppLocal}

\subsection{Transformation of the fermion operators and electron Green's
functions}

%\subsection{Fermion operators and bilinear combinations}
\label{AppLocal1}

The transformation to the local reference frame is obtained by
\cite{OurFirst} 
\begin{equation}
d=\left( 
\begin{tabular}{c}
$d_{\uparrow }$ \\ 
$d_{\downarrow }$%
\end{tabular}%
\ \right) =R^{\theta }\left( 
\begin{array}{c}
c_{\uparrow } \\ 
c_{\downarrow }%
\end{array}%
\ \right) =R^{\theta }c,
\end{equation}%
where 
\begin{equation}
R^{\theta }=\exp \left( i\frac{\theta }{2}\sigma ^{y}\right) =\left( 
\begin{tabular}{ll}
$\cos (\theta /2)$ & $\sin (\theta /2)$ \\ 
$-\sin (\theta /2)$ & $\cos (\theta /2)$%
\end{tabular}%
\ \right) .
\end{equation}%
The corresponding transformation in momentum space reads \cite{OurFirst} 
\begin{align}
c_{{k}\uparrow }& =(D_{k-\mathbf{Q}/2,+}+D_{k+\mathbf{Q}/2,-})/\sqrt{2}, \\
c_{{k}\downarrow }& =(D_{k-\mathbf{Q}/2,+}-D_{k+\mathbf{Q}/2,-})/(\sqrt{2}i),
\end{align}%
where we introduce the operators (Grassmann variables)%
\begin{equation}
D_{{k,}\pm }=\left( d_{{k,}\uparrow }\pm id_{{k}\downarrow }\right) /\sqrt{2}%
.
\end{equation}%
We furthermore define 
\begin{equation*}
c_{{k},\pm }=(c_{{k}\uparrow }\pm ic_{{k}\downarrow })/\sqrt{2}=D_{{k\mp {%
\mathbf{Q}}/}2,\pm }.
\end{equation*}%
The $D$ operators reduce the single electron Green's functions to the $%
2\times 2$ form: 
\begin{align}
\langle \langle c_{{k},+}|c_{{k},+}^{+}\rangle \rangle & =\langle \langle
D_{k-\mathbf{Q}/2,+}|D_{k-\mathbf{Q}/2,+}^{+}\rangle \rangle =\mathcal{G}_{k-%
\mathbf{Q/2},++}, \\
\langle \langle c_{k-\mathbf{Q},-}|c_{k-\mathbf{Q,}-}^{+}\rangle \rangle &
=\langle \langle D_{k-\mathbf{Q}/2,-}|D_{k-\mathbf{Q}/2,-}^{+}\rangle
\rangle =\mathcal{G}_{k-\mathbf{Q}/2,--}, \\
\langle \langle c_{{k},+}|c_{k-\mathbf{Q,}-}^{+}\rangle \rangle & =\langle
\langle D_{k-\mathbf{Q}/2,+}|D_{k-\mathbf{Q}/2,-}^{+}\rangle \rangle =%
\mathcal{G}_{k-\mathbf{Q}/2,+-}, \\
\langle \langle c_{k-\mathbf{Q,}-}|c_{{k,}+}^{+}\rangle \rangle & =\langle
\langle D_{k-\mathbf{Q}/2,-}|D_{k-\mathbf{Q}/2,+}^{+}\rangle \rangle =%
\mathcal{G}_{k-\mathbf{Q}/2,-+}.
\end{align}%
The inverse of the lattice Green's function $\mathcal{G}_{{k},\alpha \alpha
^{\prime }}=-\langle TD_{{k}\alpha }(\tau )D_{{k}\alpha ^{\prime
}}^{+}(0)\rangle $ reads \cite{OurFirst}%
\begin{equation}
\mathcal{G}_{k,\alpha \alpha ^{\prime }}^{-1}=\left( 
\begin{array}{cc}
\phi _{\nu }-\epsilon _{\mathbf{k}+\mathbf{Q}/2} & -(\Sigma _{\nu ,\uparrow
}-\Sigma _{\nu ,\downarrow })/2 \\ 
-(\Sigma _{\nu ,\uparrow }-\Sigma _{\nu ,\downarrow })/2 & \phi _{\nu
}-\epsilon _{\mathbf{k-Q}/2}%
\end{array}%
\right) ,
\end{equation}%
$\phi _{\nu }=i\nu +\mu -(\Sigma _{\nu ,\uparrow }+\Sigma _{\nu ,\downarrow
})/2$. The above mentioned relations allow for obtaining transformation
rules from global to the local coordinate frame: 
\begin{align}
c_{{k}\sigma }^{+}\sigma _{\sigma \sigma ^{\prime }}^{\pm }c_{{k+q,}\sigma
^{\prime }}& =2D_{{k}\pm \mathbf{Q}/2,\mp }D_{{k}\mp \mathbf{Q}/2+{q},\pm
}=d_{{k}\pm \mathbf{Q}/2}^{+}\sigma ^{\pm }d_{{k}\mp \mathbf{Q}/2+{q}},
\label{Tr1} \\
c_{{k}\sigma }^{+}\sigma _{\sigma \sigma ^{\prime }}^{y}c_{{k+q,}\sigma
^{\prime }}& =\sum\limits_{\alpha =\pm }\alpha D_{{k}+\alpha \mathbf{Q}%
/2,\alpha }D_{{k}+\alpha \mathbf{Q}/2+{q},\alpha }  \notag \\
& =(d_{{k}+\mathbf{Q}/2}^{+}(\sigma ^{y}+\sigma ^{0})d_{{k}+\mathbf{Q}/2+{q}%
}+d_{{k}-\mathbf{Q}/2}^{+}\left( \sigma ^{y}-\sigma ^{0}\right) d_{{k}-%
\mathbf{Q}/2+{q}})/2, \\
c_{{k}\sigma }^{+}\sigma _{\sigma \sigma ^{\prime }}^{0}c_{{k+q,}\sigma
^{\prime }}& =\sum\limits_{\alpha =\pm }D_{{k}+\alpha \mathbf{Q}/2,\alpha
}D_{{k}+\alpha \mathbf{Q}/2+{q},\alpha }  \notag \\
& =(d_{{k}+\mathbf{Q}/2}^{+}(\sigma ^{0}+\sigma ^{y})d_{{k}+\mathbf{Q}/2+{q}%
}+d_{{k}-\mathbf{Q}/2}^{+}\left( \sigma ^{0}-\sigma ^{y}\right) d_{{k}-%
\mathbf{Q}/2+{q}})/2,  \label{Tr3}
\end{align}%
where $\sigma ^{\pm }=\sigma ^{z}\pm i\sigma ^{x}$. 
%, $s^\pm=\sigma^x\pm i\sigma^y$.

\subsection{Transformation of the current and spin operators}

Let us consider the transformation of current operators  (\ref{current_operator}).
%\begin{equation}
%    \hat{J}^{\mu \alpha}_{q} = \sum\limits_k T^\mu_{{\mathbf k},{\mathbf q}} c^+_{k} \frac{\sigma^a}{2} c_{k + q},
%    \label{JJa}
%\end{equation}
%where $T^{\mu}_{{\mathbf k},{\mathbf q}} = \frac{t^{\mu}_{k} + t^{\mu}_{k+q}}{2}$. 
%In this notation corresponding hermitian conjugate is
%\begin{equation}
%    \hat{J}^{\mu \alpha}_{-q} = \sum\limits_{k'} c^+_{k'+q} \frac{\sigma^a}{2} c_{k'} T^\mu_{{\mathbf k}',{\mathbf q}}
%\end{equation}
Let us introduce the notation
\begin{equation}
    T^{\mu \pm}_{{\mathbf k},{\mathbf q}} = \frac{T^\mu_{{\mathbf k}-{\mathbf Q}/2,{\mathbf q}} \pm T^\mu_{{\mathbf k}+{\mathbf Q}/2,{\mathbf q}}}{2}.
\end{equation}
We obtain the following transformation rule using Eq. (\ref{Tr1})-(\ref{Tr3})
\begin{align}
    \hat{J}^{\mu 0}_{q} &= \sum_k d^+_{k} \left( T^{\mu +}_{{\mathbf k},{\mathbf q}} \frac{\sigma^0}{2} + T^{\mu -}_{{\mathbf k},{\mathbf q}} \frac{\sigma^y}{2} \right) d_{k+q}, \\
    \hat{J}^{\mu x}_{q} &= 
    \frac{1}{2} 
\sum_{k,\alpha=\pm} T^\mu_{{\mathbf k}+\alpha {\mathbf Q}/2,{\mathbf q}}
%\left( T^{\mu +}_{{\mathbf k},{\mathbf q}} + \alpha T^{\mu -}_{{\mathbf k},{\mathbf q}} \right) 
d^+_{k} 
\left( \frac{\sigma^x}{2} + \alpha i \frac{\sigma^z}{2} \right)  d_{k + q + \alpha {\mathbf Q}},\\
    \hat{J}^{\mu y}_{q} &= \sum_k d^+_{k} \left( T^{\mu -}_{{\mathbf k},{\mathbf q}} \frac{\sigma^0}{2} + T^{\mu +}_{{\mathbf k},{\mathbf q}} \frac{\sigma^y}{2} \right) d_{k+q},\\
    \hat{J}^{\mu z}_{q} &= \frac{1}{2}
    \sum_{k,\alpha=\pm} T^\mu_{{\mathbf k}+\alpha {\mathbf Q}/2,{\mathbf q}}
%    \left( T^{\mu +}_{{\mathbf k},{\mathbf q}} + \alpha T^{\mu -}_{{\mathbf k},{\mathbf q}} \right) 
d^+_{k} \left( \frac{\sigma^z}{2} - \alpha i \frac{\sigma^x}{2} \right) d_{k + q  + \alpha {\mathbf Q}}.
\end{align}
For spin and charge operators we have  $T^{\mu=0}_{{\mathbf k},{\mathbf q}} = T^{\mu=0,+}_{{\mathbf k},{\mathbf q}} = \I$, $T^{\mu=0,-}_{{\mathbf k},{\mathbf q}} = 0$.

\subsection{Transformation of the gauge kernel}

\label{TrKernel}

In the local coordinate frame the momentum is conserved and all correlation
functions become diagonal with respect to the wave vectors $\mathbf{q},%
\mathbf{q}^{\prime }$, so they depend on the only wave vector $\mathbf{q}$,
which greatly simplifies the calculations.
%Full current - current respons
%\begin{equation}
%     = K^{\alpha \beta}_{\mu \nu} (i,j)
%    \label{Kdef}
%\end{equation}
% where $A^{\alpha}_\mu (i)$ - калибровочное поле, cвязанное c токовыми операторами вида
% \begin{align}
% \hat{J}^{\alpha}_\mu (i) &= \sum_j \hat{c}^+_i t^\mu_{ij} \sigma^{\alpha} \hat{c}_j\\
% \hat{J}^{\alpha}_\mu (\mathbf{q}) &= \sum_{\mathbf{q}} \hat{c}^+_{\mathbf{k} + \mathbf{q}} t^\mu_{\mathbf{k}} \sigma^{\alpha} \hat{c}_{\mathbf{k}}
% \end{align}
Introducing the paramagnetic space-time part of the kernel in the local reference frame %\begin{equation}
%    K^{\mu \nu; a b}_{jl}(\tau) = \sum\limits_{j',l'} \gamma^{(1)}_\mu (j,j') \gamma^{(1)}_\nu (l,l') \left\langle \hat{c}^+_j(\tau) \sigma^a \hat{c}_{j'}(\tau) \hat{c}^+_l(0) \sigma^b \hat{c}_{l'}(0) \right\rangle
%\end{equation}
%\begin{equation}
%    K_{q}^{\mu \nu; a b} = -\sum\limits_{k,k'} t^\mu_k \langle\langle \hat{c}^+_{k+q} \sigma^a \hat{c}_{k} \rvert \hat{c}^+_{k'} \sigma^b \hat{c}_{k'+q} \rangle\rangle t^\nu_{k'+q}
%\end{equation}
\begin{align}
\bar{K}_{q;\mu \nu,s s^{\prime }}^{a b} &= \sum\limits_{k,k^{\prime }}
T^{\mu,s}_{\mathbf{k}} \langle\langle \hat{d}^+_{k} \sigma^a \hat{d}_{k+q}
\rvert \hat{d}^+_{k^{\prime }+q} \sigma^b \hat{d}_{k^{\prime }}
\rangle\rangle T^{\nu,s^{\prime }}_{{\mathbf{k}}^{\prime
}},\;\;\;\;\;\;\;\;\;\;\;\;\;\;\;\;\;\;\;\;\;\;\;\;\;\;\;\; a,b=0,y,
\label{Corr1} \\
\bar{K}_{q;\mu \nu,ss}^{a b} &= \sum\limits_{k,k^{\prime }} T^{\mu}_{{%
\mathbf{k}}+s {\mathbf{Q}}/2} \langle\langle \hat{d}^+_{k} \sigma^a 
\hat{d}_{k+q+\alpha {\mathbf{Q}}} \rvert \hat{d}^+_{k^{\prime }+q+\alpha {%
\mathbf{Q}}} \sigma^b \hat{d}_{k^{\prime }} \rangle\rangle T^{\nu}_{{\mathbf{%
k}}^{\prime }+s {\mathbf{Q}}/2},\,\,\,\,\, a,b=x,z  \label{Corr2}
\end{align}
($s,s^{\prime }=\pm$), the components of the kernel in the global reference frame can be expressed as given by the Eqs. (\ref{Kxx}) of the main text. On the other hand, the diamagnetic part is expressed as given by the Eq. (\ref{Kdg}) of the main text with
%\begin{equation}
%    \sum\limits_{k} c^+_{k+q} \sigma^y c_k t_k^\mu = \sum_k d^+_{k+q} \sigma^0 d_k t^{\mu -}_k + d^+_{k+q} \sigma^y d_k t^{\mu +}_k
%\end{equation}
%\begin{equation}
%    \sum\limits_{k} c^+_{k+q} \sigma^0 c_k t_k^\mu = \sum_k d^+_{k+q} \sigma^0 d_k t^{\mu +}_k + d^+_{k+q} \sigma^y d_k t^{\mu -}_k
%\end{equation}
%The Eq. (\ref{Kcorr}) of the main text is proven by using ``duality" property of $\bar{K}^{\phi}$ bubbles 
%\begin{align}
%\bar{K}^{\phi}_{q;0\rho} =   \phi_{q,\rho}-  K^0_{q,0\rho}, \,\,\,  \bar{K}^{\phi}_{q;\mu 0} &=  \phi^{\rm t}_{q,\mu}-  K^0_{q,\mu0},\,\,\,
%    \bar{K}^{\phi}_{q;00} =    \phi_{q}- K^0_{q,00},
%\label{zero_both_simplification}
%\end{align}
%which yields the following representation for the current-spin and spin-spin correlation functions
%\comAK{How this helps to prove the identity?}
\begin{equation}
\bar{K}_{ \mu\nu,s}^{d,ab} = -\delta_{ab} \sum_k T^{\mu\nu,s}_k \langle \hat{d%
}^+_{k} \sigma^a \hat{d}_{k} \rangle,
\end{equation}%
which is equivalent to the Eq. (\ref{Kd}).

%In Fig. \ref{FigChiQ}d,e we show the comparison of the components of the
%susceptibility $\chi _{\mathbf{0},\omega }^{\mathrm{loc}}$ and $\chi _{\pm 
%\mathbf{Q},\omega }^{\mathrm{loc}}$ (including the off-diagonal
%contributions to the local susceptibility $\chi ^{\mathrm{loc},\pm y}$,
%which have parameter-free analytic form $\pm {m}/{\omega }$) in the the
%local reference frame, computed within DMFT, with the corresponding
%analytical results (\ref{chiINCOMlocal}). One can see that the structure of
%the computed quantities match the derived analytical results.

\section{Application of Ward identities for obtaining susceptibilities in the incommensurate phase}
\label{AppSusc1}

In the case of spiral incommensurate order explicit
form of Ward identities is conveniently written in the local coordinate
frame (see Appendix \ref{AppLocal}%
), where the spins are aligned along the local $z$ axis (see also Refs. \onlinecite{OurFirst,Bonetti,BonettiThesis}). The
susceptibilities in the local coordinate frame are diagonal with respect to
momenta. The corresponding ${\mathbf{q}}=0$ block of the susceptibilities in the local reference frame, determined by Ward identity (\ref{Wnu0}) in the basis $S_{\mathbf{q}%
}^{x},S_{\mathbf{q}}^{y},S_{\mathbf{q}}^{z}$ takes the form 
\begin{equation}
\bar{\chi}_{{\mathbf{q}}={\mathbf{0}},\omega }^{ab}=\frac{1}{\omega ^{2}}%
\left( 
\begin{array}{ccc}
\omega ^{2}\bar{\chi}_{\mathbf{0},\omega }^{{xx}} & \omega \bar{r}^x_{\mathbf{0},\omega} & \omega ^{2}\bar{\chi}_{\mathbf{0},\omega }^{{xz%
}} \\ 
-\omega \bar{r}^x_{\mathbf{0},\omega} & h \bar{r}^x_{\mathbf{0},\omega} & h\omega \bar{\chi}_{\mathbf{0},\omega }^{{xz}}
\\ 
\omega ^{2}\bar{\chi}_{\mathbf{0},\omega }^{{zx}} & -h\omega \bar{\chi}_{%
\mathbf{0},\omega }^{{zx}} & \omega ^{2}\bar{\chi}_{\mathbf{0},\omega }^{{zz}%
} 
\end{array}%
\right) _{x,y,z},
\end{equation}%
%\begin{equation}
%\chi^{{\rm loc},ab}_{{\mathbf q}={\mathbf 0},\omega}=\frac{1}{2\omega^2} \left(
%\begin{array}{ccc}
% 2\omega ^2 \chi _{{\mathbf 0},\omega}^{\text{loc},++} &  i \omega  (2 m+h \chi _{{\mathbf 0},\omega}^{\text{loc},++}-h \chi _{{\mathbf 0},\omega}^{\text{loc},+-}) & 2\omega ^2
%   \chi _{{\mathbf 0},\omega}^{\text{loc},+-} \\
%   -i\omega  (2m+h \chi _{{\mathbf 0},\omega}^{\text{loc},++}
%  -h \chi _{{\mathbf 0},\omega}^{\text{loc},-+}) 
%& 
%   2 h (m-h \chi _{{\mathbf 0},\omega}^{\text{loc},xx}) & i \omega  (2
%   m+h \chi
%   _{{\mathbf 0},\omega}^{\text{loc},--}-h \chi _{{\mathbf 0},\omega}^{\text{loc},+-}) \\
% 2\omega ^2 \chi _{{\mathbf 0},\omega}^{\text{loc},-+} & -i \omega  (2 m+h \chi _{{\mathbf 0},\omega}^{\text{loc},--}-h \chi _{{\mathbf 0},\omega}^{\text{loc},-+}) & \omega ^2
%   \chi _{{\mathbf 0},\omega}^{\text{loc},--} \\
%\end{array}
%\right)
%\end{equation}
where $\bar{r}^a_{\mathbf{q},\omega}=m-h\bar{\chi}_{{\mathbf{q}},\omega }^{aa}$ and the bars stand for the local coordinate frame and the index $x,y,z$
refers to the respective spin reference frame. The blocks at the momenta $%
\mathbf{q}=\pm {\mathbf{Q}}$ are written in the basis $S_{%
\mathbf{q}}^{+},S_{\mathbf{q}}^{y},S_{\mathbf{q}}^{-}$ ($S_{\mathbf{q}}^{\pm
}=S_{\mathbf{q}}^{z}\pm iS_{\mathbf{q}}^{x}$) as 
\begin{equation}
\bar{\chi}_{{\mathbf{q}}=-{\mathbf{Q}},\omega }^{ab}=\frac{1}{\omega
^{2}}\left( 
\begin{array}{ccc}
\I h \omega \bar{\chi}^{+y}_{-\mathbf{Q},\omega}& \omega^2 \bar{\chi}^{+y}_{-\mathbf{Q},\omega} & \omega^2 \bar{\chi}^{+-}_{-\mathbf{Q},\omega}\\ 
- \I \omega \bar{r}^y_{-\mathbf{Q},\omega} & \omega
^{2}\bar{\chi}_{-{\mathbf{Q}},\omega }^{yy} & \omega^2 \bar{\chi}^{y-}_{-\mathbf{Q},\omega}  \\ 
h \bar{r}^y_{-\mathbf{Q},\omega} & - \I \omega \bar{r}^y_{-\mathbf{Q},\omega} & \I h \omega \bar{\chi}^{y-}_{-\mathbf{Q},\omega}
\end{array}%
\right) _{+,y,-}.
\end{equation}
%\begin{align}
%    + \I p^+_{\omega} \bar{r}^y_{-\mathbf{Q},\omega}\comAK{=\omega\bar{\chi}^{+y}_{-\mathbf{Q},\omega}}\\
%    + \I p^-_{\omega} \bar{r}^y_{-\mathbf{Q},\omega}\comAK{=\omega\bar{\chi}^{y-}_{-\mathbf{Q},\omega}}  
%    \\
%    \frac{4 \omega^2}{\bar{\kappa}_{-{\mathbf Q},\omega}^{+-}} + h p^+_{\omega} p^-_{\omega} \bar{r}^y_{-\mathbf{Q},\omega}\comAK{=\omega^2\bar{\chi}^{--}_{-\mathbf{Q},\omega}}  
%\end{align}
%where $\bar{\kappa}_{\mathbf{q},\omega}$ is the inverse susceptibility in the local coordinate frame and  $ p^{\pm}_{\omega}$ are defined as $p^{+}_{\omega} = \bar{\kappa}_{-{\mathbf Q},\omega}^{++} / \bar{\kappa}_{-{\mathbf Q},\omega}^{+-}$ and $p^{-}_{\omega} = \bar{\kappa}_{-{\mathbf Q},\omega}^{--} / \bar{\kappa}_{-{\mathbf Q},\omega}^{+-}$.  
The susceptibility $\bar{\chi}_{{\mathbf{q}}={\mathbf{Q}},\omega }$ can be obtained via hermitian conjugate as  $\bar{\chi}_{{\mathbf{q}}={\mathbf{Q}},\omega } = \bar{\chi}_{{\mathbf{q}}=-{\mathbf{Q}},\omega }^+$. 
%Quantities  $ p^{\pm}_{\omega}$, $\frac{1}{\bar{\kappa}_{-{\mathbf Q},\omega}^{+-}}$, t
The components of the susceptibility $\bar{\chi}_{{\mathbf{0}},\omega }^{ab}$ with $a,b=x,z$ and $\bar{\chi}_{\alpha {\mathbf{Q}},\omega }^{yy}$ are not fixed by Ward identities, and have to be determined from the microscopic theory. The form of the $\bar{\chi}_{{\mathbf{0}},\omega }^{xx}$ and $\bar{\chi}_{\alpha {\mathbf{Q}},\omega }^{yy}$ components is parameterized by the frequency dependence of the temporal stiffnesses in Eq. (\ref{chiQQQ}) of the main text. The frequency dependence of the component $\bar{\chi}_{{\mathbf{0}},\omega }^{zz}$ is determined directly from the microscopic approach, since it is non-singular in the ordered phase, {but has a finite discontinuity at $\omega = 0$}. Finally, the components $\bar{\chi}_{{\mathbf{0}},\omega }^{xz,zx}$ vanish by symmetry.

To determine momentum dependencies of susceptibilities, we consider general form of the susceptibilities, Eq.  (\ref{kq0}) and (\ref{kq1}) of the main text, which
includes the zeroth (charge) components allowed by the symmetry \cite%
{Bonetti,BonettiThesis}. 
%\begin{equation}
%\bar{\bar{\kappa}}_{{\mathbf{q}},\omega }^{ab}=\frac{1}{m}\left( 
%\begin{array}{cccc}
%{h}+A_{nl}q_{n}q_{l} & %\omega  & C_{n}q_{n} & D_{n}q_{n} \\ 
%-\omega  & {m}{\chi _{2,\omega }^{-1}} & 0 & 0 \\ 
%-C_{n}q_{n} & 0 & {d^{0z}}{\bar{\chi}_{0}^{00}} & -{d^{0z}}{\bar{\chi}%
%_{0}^{0z}} \\ 
%-D_{n}q_{n} & 0 & -{d^{0z}}{\bar{\chi}_{0}^{0z}} & {d^{0z}}{\bar{\chi}_{0}^{{%
%zz}}} 
%\end{array}%
%\right) _{x,y,z,0},\,\,\,\,\,{\bar{\kappa}}_{-{\mathbf{Q}},\omega }^{ab}=%
%\frac{1}{m}\left( 
%\begin{array}{ccc}
%\ast  & \ast  & \ast  %\\ 
%{2i\omega } & {h+B_{nl}q_{n}q_{l}} & \ast  \\ 
%{4m}{\chi _{1,\omega }^{-1}} & {2i\omega } & \ast  
%\end{array}%
%\right) _{+,y,-}.
%\end{equation}%
%where $d^{0z}=m/(\bar{\chi}_{0}^{00}\bar{\chi}_{0}^{{zz}}-(\bar{\chi}%
%_{0}^{0z})^{2})$, such that the inversion of this form at ${\mathbf{q}}=0$
%yields the result (\ref{chiINCOMlocal}) in the spin sector, double overline
%stands for the quantities in the local reference frame which include charge
%component.
Relating the corresponding coefficients to those in the momentum-dependent $%
\bar{\kappa}_{{\mathbf{q}},\omega}$ and using again Ward identities (\ref%
{Gamma_Ward2}) for the functional $\Gamma$, written in the local coordinate
frame, we find equations for the coefficients $A_{nl}$ and $B_{nl}$
%\begin{align}
%q_n q_l M_{q;nl}^{yy}&={m^2}\kappa^{{\rm loc},xx}_{\mathbf{q},\omega}
%%{1/4}\sum_{\alpha,\beta=\pm 1}(-1)^{\alpha\beta}\kappa^{-\alpha,\beta}_{\mathbf{q}+\alpha\mathbf{Q},\mathbf{q}+\beta\mathbf{Q},\omega}
%-m h \label{ExplicitGammaQ2}\\
%q_n q_l M_{q;nl}^{xx,zz}(q)&=
%%-\frac{m^2}{4}\sum_{\alpha,\beta=\pm 1}\kappa^{yy}_{\mathbf{q}+\alpha\mathbf{Q},\mathbf{q}+\beta\mathbf{Q}}(\omega)+\frac{m h}{2}\\
%%q_n q_l M_{q;nl}^{zz}=
%\frac{m^2}{4}\sum_{\alpha
%%,\beta
%=\pm 1}
%%^(-1)^{\alpha+\beta}
%\kappa^{{\rm loc},yy}_{\mathbf{q}+\alpha\mathbf{Q},\omega}-\frac{m h}{2} \label{ExplicitGammaQ1}
%\end{align}
%In the first equation we have taken into account the diagonal form of $\kappa^{yy}_{\mathbf{q},\mathbf{q}'}$ in momenta $\mathbf{q},\mathbf{q}'$ and in the last equation we consider the basis $S^{\pm}_{\mathbf q}$. Therefore, at small $q$ we find
\begin{align}
\bar{\kappa}^{yy}_{\pm {\mathbf{Q}}+{\mathbf{q}},\omega}&=\frac{h}{m}+B_{nl} 
\frac{q_n q_l}{m}=\frac{h}{m}+\frac{2q_n q_l}{m^2} M_{q;nl}^{xx,zz},
\label{kq12} \\
\bar{\kappa}^{xx}_{{\mathbf{q}},\omega}&=\frac{h}{m}+\left(A_{nl}+\frac{1}{%
d^{0 z} \chi_0^{zz}}D_n D_l\right)\frac{q_n q_l}{m}= \frac{h}{m}+\frac{q_n
q_l}{m^2} M_{q;nl}^{yy}.  \label{kq21}
\end{align}
%These Ward identities allow computing $\lim_{q \to 0} m^2 \partial^2_{q_n} 
%\frac{1}{ \bar{\kappa}^{xx}_{\mathbf{q}} }$ directly from the computation of 
%$M^{yy}_{\mathbf{0}, n n}$. 
Couplings $C_n$ and $D_n$ can also be computed
from the knowledge of current-spin correlation functions. The non-zero
off-diagonal components of the susceptibilities are fixed by the identity (%
\ref{Wnu0}), which takes the form 
\begin{align}
m \delta_{ax}= \tilde{K}^{ay}_{q;0\nu} q_\nu+h \bar{\chi}^{ax}_q,  \label{Id2}
\\
m \delta_{ax}= {-}\tilde{K}^{ya}_{q;\nu 0} q_\nu+h \bar{\chi}%
^{xa}_q.  \label{Id1}
\end{align}
where in the kernels $\tilde{K}^{ya}_{q,\nu 0}$ and $\tilde{K}^{ay}_{q,0\nu}$ we
pass to the local coordinate frame with respect to the spin index $a$ only.
Using Eq. (\ref{chiU}) of the main text, the identities (\ref{Id2}), (%
\ref{Id1}) can be written as 
\begin{equation}
q_\nu \tilde{\phi}^{\mathrm{t},ya}_{q;\nu 0}= \left( 2 U
m+h\right) \bar{\phi}^{xa}_q{-} m \delta_{a,x},\,\,\,
\tilde{\phi}^{ay}_{q;0\nu}q_\nu =m \delta_{a,x} - \left(2 U m+h\right)\bar{\phi}^{ax}_q.
\end{equation}
Calculating the derivatives $\left(\partial_{q_n} \bar{\phi}^{xz}_q\right)_{q=0}$ and $\left(\partial_{q_n} \bar{\phi}^{x0}_q\right)_{q=0}$ from Eq. (\ref{kq0}) of the main text, we find  %\comIG{Может так чуть компактнее?}
\begin{align}
(h+2 m U)\left(\partial_{q_n} \bar{\phi}^{xz}_q\right)_{q=0} = & %
 \frac{C_n (d^{0z} \bar{\chi}^{zz}_0 - 2 m U ) + D_n
d^{0z} \bar{\chi}^{z0}_0 }{d^{0z} - 4 m U^2 - 2 d^{0z} U ( \bar{\chi}^{00}_0
- \bar{\chi}^{zz}_0 ) } ={\tilde{\phi}^{{\rm t},yz}_{0;n0}},
\label{dphi1} \\
(h+2 m U)\left(\partial_{q_n} \bar{\phi}^{x0}_q\right)_{q=0} = & %
 \frac{C_n d^{0z} \bar{\chi}^{z0}_0 + D_n (d^{0z} \bar{%
\chi}^{00}_0 + 2 m U) }{d^{0z} - 4 m U^2 - 2 d^{0z} U ( \bar{\chi}^{00}_0 - 
\bar{\chi}^{zz}_0 ) } ={\tilde{\phi}^{{\rm t},y0}_{0;n0}} .
\label{dphi2}
\end{align}
%\begin{align}
%(h+2 m U)\left(\partial_{q_n} \phi^{xz}_q\right)_{q=0}= &
%-\frac{C \left(m(1-U \chi _0^{00})+d^{0z} \chi _0^{z0}\chi _0^{0z}\right)+D d^{0z} \chi _0^{00} \chi _0^{0z}}{
%   (1-\chi _0^{00} U) (2 m U+d^{0z} \chi _0^{00})+d^{0z} \chi _0^{z0}\chi _0^{0z} U}
%=-{\phi^{{\rm t},yz}_{0;n0}},\\
%(h+2 m U)\left(\partial_{q_n} \phi^{x0}_q\right)_{q=0}=&
%-\frac{\chi _0^{00} (C d^{0z} \chi _0^{0z}+D m U+D %d^{0z} \chi _0^{00})}{(1-\chi _0^{00} U) (m U+d^{0z} \chi
%   _0^{00})+d^{0z} \chi _0^{z0}\chi _0^{0z} U}
%=-{\phi^{{\rm t},y0}_{0;n0}}
%\label{dphi}
%\end{align}
From this we obtain

%\begin{equation}
%    C_n = \frac{{\phi_{0,n0}^{{\rm t},y0}} d^{0z} \chi _0^{0z}-{\phi_{0,n0}^{{\rm t},yz}} (2 m U+d^{0z} \chi _0^{00})}{m},
%    D= -\frac{{\phi_{0,n0}^{{\rm t},y0}} m
% (1-\chi _0^{00} U) + {\phi_{0,n0}^{{\rm t},y0}} d^{0z} \chi _0^{0z \; 2}-{\phi_{0,n0}^{{\rm t},yz}} d^{0z} \chi _0^{00} \chi _0^{0z}}{m \chi^{00}_0}
%\end{equation}

%\comIG{То же, но чуть в более симметричной форме}

\begin{equation}
C_n = \frac{{\phi_{0,n0}^{\mathrm{t},y0}} d^{0z} \chi _0^{0z} - {%
\phi_{0,n0}^{\mathrm{t},yz}} ( d^{0z} \chi _0^{00} + 2 m U) }{m}, D_n = 
\frac{{\phi_{0,n0}^{\mathrm{t},yz}} d^{0z} \chi _0^{0z} - {\phi_{0,n0}^{%
\mathrm{t},y0}} ( d^{0z} \chi _0^{zz} - 2 m U) }{m}  \label{CD}.
\end{equation}

%\begin{equation}
%    C_n = -\frac{{\phi_{0,n0}^{{\rm t},y0}} \kappa_0^{0z} + {\phi_{0,n0}^{{\rm t},yz}} ( \kappa_0^{zz} + 2 m U) }{m},
%    D_n = -\frac{{\phi_{0,n0}^{{\rm t},yz}} \kappa_0^{0z} + {\phi_{0,n0}^{{\rm t},y0}} ( \kappa_0^{00} - 2 m U) }{m}
%\end{equation}

%\comIG{$\bar{\kappa}$ - 3x3 обратная восприимчивость, чему соответствует $x0$ компонента?..}
%\begin{align}
%\bar{\kappa}^{xz}_q=-\bar{\kappa}^{zx}_q= -\frac{q_n}{m}\left( {\phi_{0,n0}^{y0}} \kappa_{0}^{0z}+{\phi_{0,n0}^{yz}} \frac{1+U {\bar{\chi}_{0}^{zz}}}{{\bar{\chi}_{0}^{zz}}}\right),\label{kq3}\\
%   \bar{\kappa}^{x0}_q=-\bar{\kappa}^{0x}_q= -\frac{q_n}{m}\left({\phi_{0,n0}^{yz}}{\kappa_{0} ^{z0}}+{\phi_{0,n0}^{y0}} \frac{1-U {\bar{\chi}_{0}^{00}}}{{\bar{\chi}_{0}^{00}}}\right)\label{kq4}
%\end{align}
%where the argument $0$ stands for the $\omega\rightarrow 0$, $q\rightarrow 0$ order of the limits. 
Using these results and assuming that the kernels $\phi _{n0}^{yz,y0}$ are non-vanishing only
along one of the directions $q_{x,y}$ corresponding to the incommensurate
direction $Q_{x,y}\neq \pi $, we obtain Eqs. (\ref{Drho1}) and (\ref{r2n}) of the main text.
%We can define the quantity $\Delta \rho _{2,n}$ which is a contribution
%to the spin stiffness $\rho _{2,n}$ originating from the coupling of spin $x$
%and spin $z$, spin $x$ and charge $0$ channels by the following equation 
%\begin{equation}
%\Delta \rho _{2,n}=\frac{m^{2}}{2}\lim_{q\rightarrow 0}\partial
%_{q_{n}}^{2}\left( \frac{1}{\bar{\chi}_{\mathbf{q},0}^{xx}}-\bar{\kappa}_{%
%\mathbf{q},0}^{xx}\right) =\frac{(D_{n}\bar{\chi}_{0}^{0z}+C_{n}\bar{\chi}%
%_{0}^{zz})^{2}}{\bar{\chi}_{0}^{zz}}=\frac{1}{\bar{\chi}_{0}^{zz}}\left[ {%
%\phi _{0,n0}^{yz}}+2U({\phi _{0,n0}^{yz}}\bar{\chi}_{0}^{zz}-{\phi
%_{0,n0}^{y0}}\bar{\chi}_{0}^{0z})\right] ^{2},
%\label{Drho}
%\end{equation}%
%where we assume that the kernels $\phi _{n0}^{yz,y0}$ are non-vanishing only
%along one of the directions $q_{x,y}$ corresponding to the incommensurate
%direction $Q_{x,y}\neq \pi $. The resulting susceptibilities of the chargon sector in the global reference
%frame are obtained by inverting the matrix $\bar{\kappa}_{q}$
%with account of identities (\ref{kq12}) and (\ref{kq21}), which yields Eq. (\ref{chiQQQinc}) of the main text.

%\begin{figure}[tbp]
%\includegraphics[scale=0.7]{incom_div0} %
%\includegraphics[scale=0.7]{incom_dyn}
%\caption{Comparison of the equation (\protect\ref{chiINCOMlocal}) with the
%DMFT results for $x = 0.09$}
%\end{figure}

\renewcommand\thefigure{F\arabic{figure}}
\setcounter{figure}{0}

\section{Determination of stable magnetic configuration in the chargon sector}
\label{Det}
To determine thermodynamically  stable magnetic configuration we require positivity  of the spectrum of magnetic excitations, determined by inverse susceptibility in the local coordinate frame $\bar{\chi}^{-1}_{\mathbf{q},i\omega_n}$, given by Eq. (\ref{chiU}) of the main text. In particular the minimal eigenvalue $\lambda_{\mathbf{q}}^{\text{min}}$ of the matrix $1 - U \phi_{\mathbf{q},0}$  should be non-negative everywhere in the Brillouin zone. The points $\mathbf{q} = \pm\mathbf{Q}$ where this eigenvalue is equal to zero determine the ordering wave vector ${\mathbf Q}$ in chargon sector and correspond to Goldstone modes. To find these points we performed the iterative procedure where at every step we have executed a DMFT calculation for a trial magnetic wave vector $\mathbf{Q}^{(n)}$ and the wave vector of next iteration $\mathbf{Q}^{(n+1)}$ was determined by the ${\mathbf q
}$ point, which provides $\min_{\mathbf{q}} \lambda^{min}_{\mathbf{q}}$. This procedure rapidly converged to a stable magnetic wavevector $\mathbf{Q}$.

\begin{figure}[t]
\includegraphics[scale=1.0]{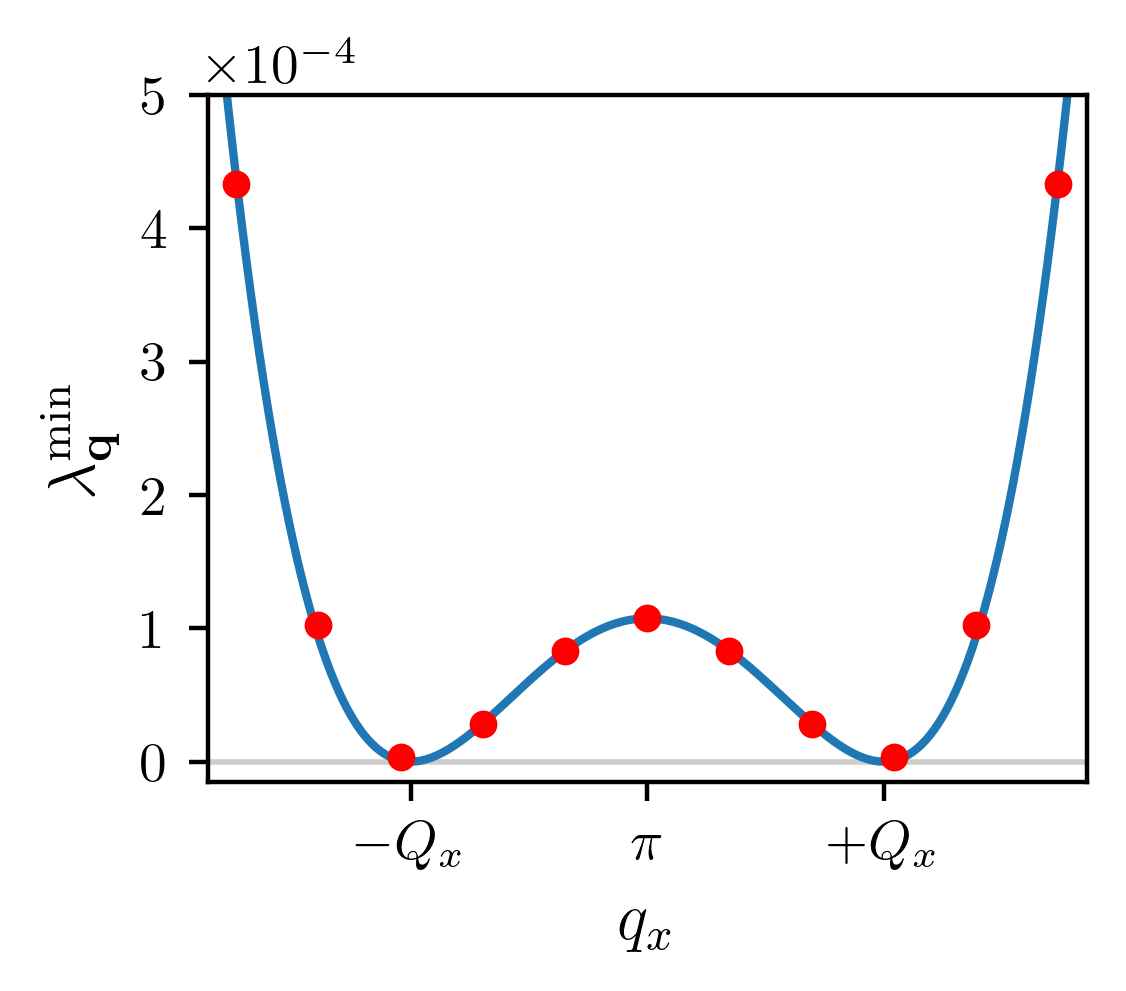}
\caption{(Color online). {The form of the dependence $\lambda^{\text{min}}_{\mathbf{q}}$ along the border of the Brillouin zone $\mathbf{q} = (q_x, \pi)$ 
%according to the
% Approximation of the numerical data for $\lambda_{\mathbf{q}}^{\text{min}}$ with
%empirical function %(\ref{lambda_function}).}}
computed for a state with hole doping level $x=0.09$ (points); the line shows the fit by Eq. (\ref{lambda_function})}.} %\comAK{Please remove points}}
\label{MF_figure}
\end{figure}

In Fig. \ref{MF_figure} we show the typical form of the obtained momentum dependence of eigenvalues $\lambda^{\text{min}}_{\mathbf{q}}$ along the direction $\mathbf{q}=(q_x,\pi)$. The obtained dependencies can be {well} approximated % to a high degree of accuracy 
by
\begin{equation}
    \lambda_{\mathbf{q}}^{\rm min} = C \left( (Q_x - \pi)^2  - (q_x - \pi) ^2 \right)^2,
    \label{lambda_function}
\end{equation}
where $C$ is a numerical constant.
%and $\mathbf{Q}^{\text{AFM}} = (\pi,\pi)$. 
%At zero external magnetic field near the Goldstone mode 
The {out-of-plane} spatial spin stiffness {in $x$ direction $\rho_{1,x}$} is  proportional to the second derivative of minimal eigenvalue
\begin{equation}
    {\rho_{1,x}} \propto \left.\frac{\partial^2 \lambda_{\mathbf{q}}^{\rm min}}{\partial q_x^2}\right\rvert_{q_x = Q_x} = 8 C (\pi - Q_x)^2. 
\end{equation}
For ${Q_x \to \pi}$ (i.e. when the order in the chargon sector approaches  commensurate one) the spin stiffness approaches zero. 

{The argumentation presented above is directly applicable also to the in-plane spatial spin stiffness in $x$ direction $\rho_{2,x}$, for which $\lambda^{\text{min}}_{\mathbf{q}}$ should be considered near $\mathbf{q} = \pm (\pi-Q_x,0)$ points, where the in-plane mode is located in the local coordinate frame. At the incommensurate-commensurate transition full symmetry of the susceptibilities should be restored. Thus all four spin stiffnesses should become equal.
% \comAK{(up to the factor of 2 relating in-plane and out-of-plane spin stiffnesses)}
This is only possible if all four spatial spin stiffness components vanish simultaneously. By continuity of the dependence of spatial stiffnesses  on the doping level $x$, at the commensurate-incommensurate transition point at finite doping the spin stiffnesses therefore vanish approaching the transition also from antiferromagnetic side.}
% Two-dimensional incommensurate and three-dimensional commensurate magnetic order and fluctuations in La2-x Bax CuO4
\end{widetext}

\end{document}